\documentclass[12pt,letterpaper]{article}
\usepackage{putex}
\usepackage{graphicx}
\usepackage{latexsym,amsmath,amsfonts,amssymb,amsthm}
\usepackage{empheq}
\usepackage{bbm}
\usepackage{cite}
\usepackage{indentfirst}
\usepackage{booktabs,array}
\usepackage{placeins}
\usepackage{mathtools}

\DeclarePairedDelimiter\floor{\lfloor}{\rfloor}
\usepackage{colonequals}
\usepackage{cancel}
\usepackage[mode=image|tex]{standalone}
\usepackage[dvipsnames]{xcolor}
\usepackage{tikz}
\usepackage{tikz-cd}
\usepackage{adjustbox}
\usepackage{enumitem}
\usetikzlibrary{decorations.markings}
\usetikzlibrary{decorations.text}
\usepackage{graphicx}
\usepackage[margin=10pt,font=small,labelfont=bf]{caption}
\usepackage{subcaption}
\usepackage{xcolor}
\usepackage{float}
\theoremstyle{definition}

\usepackage{microtype}
\usepackage{hyperref}
\hypersetup{unicode}
\hypersetup{linktoc = all}
\hypersetup{pdfborderstyle={/S/U/W 0.5}}
\hypersetup{linkbordercolor = gray}
\hypersetup{bookmarksnumbered}
\usepackage[T1]{fontenc}
\usepackage[utf8]{inputenc}
\usepackage{lmodern}

\setlist{itemsep=2pt plus 1pt minus 1pt, topsep=2pt plus 1pt minus 1pt}

\renewenvironment{thebibliography}[1]{%
\begin{oldthebibliography}{#1}%
\setlength\itemsep{-2mm}%
}%
{%
\end{oldthebibliography}%
}

\newcommand{\eg}{\textsl{e.g.\@}}

\newcommand{\ie}{\textsl{i.e.\@}}

\numberwithin{equation}{section}

\DeclareMathOperator{\Tr}{Tr}

\DeclareMathOperator{\rank}{rank}

\begin{document}


\title{\begin{LARGE}
The exact Schur index in closed form
\end{LARGE}}

\authors{Yiwen Pan$^1$ and Wolfger Peelaers$^2$
\medskip\medskip\medskip\medskip
 }

\institution{UU}{${}^1$
School of Physics, Sun Yat-Sen University, \cr
$\;\,$ Guangzhou, Guangdong, China}
\institution{Oxford}{${}^2$
Mathematical Institute, University of Oxford, Woodstock Road, \cr
$\;\,$ Oxford, OX2 6GG, United Kingdom}

\abstract{\begin{onehalfspace}{
The Schur limit of the superconformal index of a four-dimensional $\mathcal{N} = 2$ superconformal field theory encodes rich physical information about the protected spectrum of the theory. For a Lagrangian model, this limit of the index can be computed by a contour integral of a multivariate elliptic function. However, surprisingly, so far it has eluded exact evaluation in closed, analytical form. In this paper we propose an elementary approach to bring to heel a large class of these integrals by exploiting the ellipticity of their integrand. Our results take the form of a finite sum of (products of) the well-studied flavored Eisenstein series. In particular, we derive a compact formula for the fully flavored Schur index of all theories of class $\mathcal{S}$ of type $\mathfrak a_1$, we put forward a conjecture for the unflavored Schur indices of all $\mathcal N=4$ super Yang-Mills theories with gauge group $SU(N)$, and we present closed-form expressions for the index of various other gauge theories of low ranks. We also discuss applications to non-Lagrangian theories, modular properties, and defect indices.
}\end{onehalfspace}}

\preprint{}
\setcounter{page}{0}
\maketitle


{
\setcounter{tocdepth}{2}
\setlength\parskip{-0.7mm}
\tableofcontents
}


\section{Introduction}

The superconformal index \cite{Kinney:2005ej}\footnote{See also \cite{Rastelli:2014jja} for a review in the context of theories of class $\mathcal S$.} is of fundamental importance to analyze and characterize four-dimensional $\mathcal N=2$ superconformal field theories (SCFTs) as it encodes succinctly representation theoretical information about the protected spectrum of the theory. It can be defined in terms of a weighted trace over the Hilbert space of states of the radially quantized theory.\footnote{Equivalently, it also admits a definition as a partition function on $S^3\times S^1$. See \cite{Pan:2019bor,Dedushenko:2019yiw,Jeong:2019pzg} for a detailed discussion on such definition for the limit of the index of interest in this paper, namely the Schur limit.} The standard insertion in the trace of minus one raised to the fermion number causes pairwise cancellations and as a result only states of the theory that lie in the kernel of $\{\mathcal Q^\dagger,\mathcal Q\}$, where $\mathcal Q$ is a chosen supercharge of the $\mathcal N=2$ superconformal algebra $\mathfrak{su}(2,2|2)$, contribute nontrivially to the trace \cite{Kinney:2005ej}. Their contribution is weighted by additional fugacities dual to Cartan generators of the commutant of the supercharge $\mathcal Q$ in $\mathfrak{su}(2,2|2)$ . Three such fugacities can be turned on. If the theory has any flavor symmetries, the trace can be further refined by flavor symmetry fugacities.

For generic values of the three conformal fugacities, the superconformal index is a 1/8-BPS object: states preserving the common supercharge $\mathcal Q$ contribute. However, as shown in \cite{Gadde:2011uv}, upon tuning these fugacities appropriately, one can achieve various supersymmetry enhancements. One of these enhanced limits, often referred to as the Schur limit, will be the quantity of interest in this paper. It is a quarter-BPS object defined concretely as \cite{Gadde:2011uv}
\begin{equation}\label{SchurIndexDef}
\mathcal I(q,\vec b) = \Tr\, (-1)^F q^{E-R} \prod_{j=1}^{\rank G_F} b_j^{f_j}\;.
\end{equation}
Here $F$, $E$ and $R$ are respectively the fermion number, the conformal dimension and the $SU(2)_R$ Cartan generator, while $f_j$ are the Cartan generators of the flavor symmetry group $G_F$ of the theory. The thus-defined Schur limit of the superconformal index is often simply referred to as the Schur index.

The Schur index is a remarkable quantity, most notably because of the central role it plays in the SCFT/VOA correspondence of \cite{Beem:2013sza}. (See, for example, \cite{Beem:2014kka,Beem:2014rza,Lemos:2014lua,Rastelli:2014jja,Buican:2015ina,Cordova:2015nma,Song:2015wta,Xie:2016evu,Cordova:2016uwk,Cordova:2017ohl,Cordova:2017mhb,Song:2017oew,Wang:2020oxs,Buican:2019kba,Xie:2019vzr,Xie:2019zlb,Xie:2021omd}.) Indeed, the correspondence associates with every superconformal field theory a vertex operator algebra and the Schur index of the former equals the vacuum character of the latter. What's more, exploiting this map, it was argued in \cite{Beem:2017ooy} that the unflavored Schur index solves a modular differential equation and thus transforms as an element of a vector-valued (quasi)-modular form of weight zero.\footnote{For early work on the modular properties of the Schur index see \cite{Razamat:2012uv}.} However, in all but a handful of instances we currently lack the closed-form, analytical expressions for the Schur index needed to establish these transformation properties directly. In this paper we will develop technology to dramatically improve on this situation. What's more, we will retain all flavor symmetry fugacities and study modular transformations of fully flavored Schur indices. We find these to behave as quasi-Jacobi forms as defined in \cite{Krauel:2013lra}.

By and large, three approaches have been pursued in the literature to compute Schur indices:
\begin{enumerate}
\item For Lagrangian theories, exploiting the independence of the superconformal index of exactly marginal couplings, one can straightforwardly evaluate the index in the zero-coupling limit. The result takes the form of an integral over the gauge group, which implements the projection onto gauge-invariant states, of an integrand reflecting the matter content of the theory. While evaluating these integrals in a series expansion in $q$ to an arbitrary (finite) order is easy, performing them analytically has not yet been achieved in the literature. The goal of this paper is to do just that. 
\item For theories of class $\mathcal S$, the superconformal index has been shown to be computed by a topological field theory correlator on the UV curve describing the theory. The topological field theory has been identified as the zero-area limit of two-dimensional, $q$-deformed Yang-Mills theory and the wavefunctions of the states being correlated have been constructed \cite{Gadde:2011ik,Gadde:2011uv,Buican:2015ina,Lemos:2012ph,Lemos:2014lua,Mekareeya:2012tn,Song:2015wta,Xie:2016evu}. The resulting expressions for the index involve an infinite sum over the irreducible representations of a simple, simply-laced Lie algebra. Exact evaluation of this sum is currently not feasible.
\item For theories of which the associated vertex operator algebra has been identified (or conjectured) through alternative means, it is sometimes possible to compute its vacuum character and thus, indirectly, the Schur index.
\end{enumerate}  
Some further sporadic results and conjectures for Schur indices can be found in, \eg{}, \cite{Bourdier:2015wda,Bourdier:2015sga,Zafrir:2020epd}. Of course, these three strategies intersect and inform one another. What's more, our Lagrangian results allow analytical control over an interesting class of non-Lagrangian theories as well by leveraging S-duality and inversion formulas for integral transforms. In particular, we find a closed-form expression for the Minahan-Nemeschansky theory with $E_6$ flavor symmetry, which is the basic building block in the family of $\mathfrak a_2$ class $\mathcal S$ theories. Furthermore our computations allow access to a wealth of fully flavored vacuum characters of highly nontrivial vertex operator algebras. Examples include, among many others, the vacuum characters of the small $\mathcal N=4$ chiral algebra at $c=-9$, $\widehat{\mathfrak{so}}(8)_{-2}$, $(\widehat{\mathfrak{e}}_6)_{-3}$, and $(\widehat{\mathfrak{e}}_7)_{-4}$.

The technology we develop in this paper was motivated by the observation that the residue of a class of poles of the integrand of the contour integrals defining the Schur index of $\mathcal N=4$ super Yang-Mills theories with simply-laced gauge group carries physical meaning \cite{Pan:2021ulr}. Namely, it is exactly equal to the character of the collection of free fields proposed in \cite{Adamovic:2014lra,Bonetti:2018fqz} that can be used to economically realize the vertex operator algebra associated with the $\mathcal N=4$ theory. What's more, the vertex operator algebra corresponding to the $\mathcal N=4$ theory is a subalgebra of the free field algebra: it is obtained as the kernel of some screening charge. This strongly suggests that one ought to be able to compactly evaluate the index of the original theory in terms of its residues by effectively implementing the projection onto the kernel of the screening charge. As we will see, our techniques realize this expectation in a computationally concrete sense and can be generalized far beyond $\mathcal N=4$ super Yang-Mills theories.

In detail, the methods we introduce exploit the double periodicity of the integrand of the contour integrals defining the Schur index. Concretely, for a Lagrangian $\mathcal N=2$ superconformal gauge theory with hypermultiplets transforming in $\rank \mathfrak{g}_F$ irreducible representations $\mathcal R_l$ of the gauge group $G$ (with corresponding gauge algebra $\mathfrak g$) the Schur index is computed by
\begin{small}
\begin{equation}\label{SchurIndexIntegral}
\mathcal I(q,\vec b) = \frac{(-i)^{\rank \mathfrak g - \dim \mathfrak g}\ \eta(\tau)^{3 \rank \mathfrak g-\dim \mathfrak g}}{|W|} \oint \prod_{j=1}^{\rank \mathfrak g} \frac{da_j}{2\pi i a_j}\ \prod_{\alpha\neq 0} \vartheta_1(\alpha(\mathfrak{a})|\tau)\prod_{l=1}^{\rank \mathfrak{g}_F} \prod_{\rho\in \mathcal{R}_l}\frac{\eta(\tau)}{\vartheta_4(\rho(\mathfrak a)+\mathfrak b_l|\tau)}\;. 
\end{equation}
\end{small}%
Here, $|W|$ denotes the order of the Weyl group of $\mathfrak g$, $\alpha\neq 0$ are the nonzero roots of the gauge algebra, $\rho\in \mathcal R_l$ denote the weights of the representation $\mathcal R_l$, and we expressed the integrand in terms of the standard Dedekind eta function and Jacobi theta functions. Finally, $q=e^{2\pi i \tau}$, $a=e^{2\pi i \mathfrak a}$ and similarly for $b$. It is easy to verify that the integrand of \eqref{SchurIndexIntegral} is elliptic in each gauge fugacity separately, \ie{}, $\mathfrak a_j \sim \mathfrak a_j + 1\sim \mathfrak a_j + \tau$ for all $j$. As a result, the residues of the integrand as a function of one integration variable, say, $\mathfrak{a}_1$ is in fact an elliptic function with respect to the remaining $\mathfrak{a}$'s.

The integral \eqref{SchurIndexIntegral} can be evaluated by performing the contour integrals one after another. We will see that the first integral is the simplest, thanks to the integral formula we derive in section \ref{sec:integrals}
\begin{align}\label{elliptic-integral-intro}
	\oint_{|a| = 1} \frac{da}{2\pi i a} f(\mathfrak{a})
	= f(\mathfrak{a}_0)
	  + \sum_{\text{real } \mathfrak a_j}R_j\ E_1\left[\begin{matrix}
			-1 \\ \frac{a_j}{a_0} q^{\frac{1}{2}}
		\end{matrix}\right]
	  + \sum_{\text{imag. } \mathfrak a_j}R_j\ E_1\left[\begin{matrix}
			-1 \\ \frac{a_j}{a_0} q^{ - \frac{1}{2}}
		\end{matrix}\right] \ ,
\end{align}
where $a_{j = 1, 2, \ldots} = e^{2\pi i \mathfrak{a}_j}$ denotes the (finitely many, simple) poles of the integrand within the \emph{fundamental parallelogram} bounded by the vertices $0, \tau, 1, 1 + \tau$, and the $R_j$ are the corresponding residues $R_j \equiv \operatorname{Res}_{a_j}\frac{1}{a} f(a)$. Throughout this paper we distinguish two types of poles, \emph{real} if $\mathfrak{a}_j \in \mathbb{R}$, and \emph{imaginary} if $\mathfrak{a}_j = \text{real} + \lambda \tau$ with a positive $\lambda$. Finally, $\mathfrak{a}_0$ is an arbitrary (regular) reference point in the parallelogram.

After the first integral, the presence of the Eisenstein series renders the integrand of the subsequent integrals non-elliptic, since $a_j/a_0$ may contain the remaining integration variables. Fortunately, the residues $R_j$ still enjoy ellipticity, and we therefore develop computational techniques to deal with integrals of the form
\begin{align}
	\oint \frac{da}{2\pi i a}f(a)E_k \left[\begin{matrix}
		\pm 1 \\ a b
	\end{matrix}\right] \ .
\end{align}
where $f(a = e^{2\pi i \mathfrak{a}})$ is elliptic with respect to $\mathfrak{a}$. For example, we have
\begin{align}
	& \ \oint_{|z| = 1} \frac{dz}{2\pi i z}f(\mathfrak{z})E_k\left[
	\begin{matrix}
		-1 \\ za
	\end{matrix}\right] \nonumber\\
	= & \ - \mathcal{S}_{k} \left(f(\mathfrak{z}_0)
	  + \sum_{\text{real/imag } \mathfrak{z}_i}R_i  E_{1}\left[\begin{matrix}
			-1 \\ \frac{z_i}{z_0}q^{\pm \frac{1}{2}}
		\end{matrix}\right]
	\right)
	- \sum_{\text{real/imag } \mathfrak{z}_i} R_i  \sum_{\ell = 0}^{k - 1} \mathcal{S}_{\ell} E_{k - \ell + 1}\left[
		\begin{matrix}
			1 \\ z_i a q^{\pm \frac{1}{2}}
		\end{matrix}\right] \ , 
\end{align}
where $\mathcal{S}_{2k}$ is the constant term of the Eisenstein series $E_{2k}\big[\substack{-1 \\ z}\big]$. With the help of the these formulas, we are able to compute the Schur index of a vast set of theories exactly in closed-form as a finite sum of Eisenstein series.

This paper is organized as follows. In section \ref{sec:integrals}, we summarize the integral formula that we will apply to evaluate Schur indices. We apply these formulas to write down in closed form the indices of Lagrangian rank-one theories in section \ref{sec:rankone}. Next, in section \ref{sec:a1-index} we evaluate the Schur index of all theories of class $\mathcal{S}$ of type $\mathfrak a_1$: we derive a universal, compact formula for these indices, see \eqref{Ign}. In section \ref{sec:N=4} and \ref{sec:other}, we further consider Schur indices of $\mathcal{N} = 4$ theories and $SU(N)$ superconformal QCD. In particular, in subsection \eqref{N=4unfl} we conjecture closed-form expressions for the unflavored indices of $\mathcal N=4$ super Yang-Mills theories with gauge group $SU(N).$ In section \ref{applications}, we discuss some applications of our closed-form expressions, including closed-form expressions for several non-Lagrangian theories including the $E_6$ and $E_7$ Minahan-Nemeschansky theory, modular properties, and defect indices. A couple of appendices contain helpful properties and results on elliptic functions and their Fourier series.


\section{Integrating almost elliptic functions\label{sec:integrals}}
Our main goal is to analytically evaluate the Schur index of Lagrangian four-dimensional $\mathcal N=2$ superconformal field theories. In other words, we aim to calculate in closed-form contour integrals of multivariate elliptic functions, see \eqref{SchurIndexIntegral}. As the fully flavored integrand of the Schur index has only simple poles, in this section we develop general techniques and derive widely applicable results to compute multi-integrals of doubly-periodic multivariate functions with simple poles. Our strategy will be to perform these multiple integrals one by one, but, as we will see, ellipticity is typically lost after a single integration. Nevertheless, we overcome this difficulty and present integration formulas to deal with the resulting almost elliptic integrals in a large class of cases.

\subsection{Integrating elliptic functions}

An elliptic function is a meromorphic functions $f(\mathfrak{z})$ that is doubly periodic, \ie{},
\begin{align}
	f(\mathfrak{z}) = f(\mathfrak{z} + 1) = f(\mathfrak{z} + \tau) \;,
\end{align}
where we choose $\tau \in \mathbb C$ to have a positive imaginary part. Equivalently, one can view an elliptic function as a meromorphic function on a torus $T^2$ with complex structure specified by $\tau$. Let us call the parallelogram with vertices $0, 1, \tau, \tau + 1$ the \emph{fundamental parallelogram}. An elliptic function is obviously completely determined by its values in this parallelogram. In fact, up to an additive constant, it is completely determined by its poles and their residues within the parallelogram. Before continuing, let us note that in appendix \ref{specialfuctions} we collect various useful special functions relevant for our purposes, and let us also introduce a handy notational convention: we will relate Latin alphabet letters in normal math font, like $a, b, \ldots, z$, to symbols in fraktur font, like $\mathfrak{a}, \mathfrak{b}, \ldots, \mathfrak{z}$, by
\begin{equation}
a = e^{2\pi i \mathfrak{a}}, \qquad b = e^{2\pi i \mathfrak{b}}, \qquad\ldots \qquad z = e^{2\pi i \mathfrak{z}}.
\end{equation}
Furthermore, as is standard, we also have $q = e^{2\pi i \tau}$.

For our purposes, it is sufficient to focus on elliptic functions possessing only simple poles. Note that such functions necessarily must have vanishing total residue.\footnote{The more rigorously correct statement is that any meromorphic 1-form $f(\mathfrak{z})d\mathfrak{z}$ has vanishing total residue on any compact Riemann surface; on $T^2$ we can construct a single periodic coordinate $z$ and the statement reduces to one about meromorphic functions.} As a consequence, elliptic functions with just a single simple pole (inside the fundamental parallelogram) don't exist. The Weierstrass zeta function, $\zeta(\mathfrak{z})$, which has relatively simple behavior under shifts of its argument by $1$ and $\tau$ -- the failure to be doubly periodic depends only on the nome $\tau$ but not on $\mathfrak z$ --, comes very close though. We refer the reader to appendix \ref{specialfuctions}, and in particular around \eqref{weierstrasszetadef}, for the precise definition of the $\zeta$-function and some helpful properties. Nevertheless, although non-elliptic, the $\zeta$-function plays a crucial role in our analysis, as it is the source of one unit of residue and can therefore be used to construct any elliptic function possessing only simple poles. In detail, let $f(\mathfrak{z})$ be an elliptic function with simple poles located at $\mathfrak{z}_j$ in the fundamental parallelogram and whose residues are given by,\footnote{Note that $R_j = 2\pi i \operatorname{Res}_{\mathfrak{z} \to \mathfrak{z}_j}f(\mathfrak{z})$.}
\begin{equation}
R_j \colonequals \operatorname{Res}_{z_j} \frac{1}{z} f(\mathfrak{z}) = \oint_{\mathfrak{z}_j}\frac{dz}{2\pi i z} f(\mathfrak{z}) \ .
\end{equation}
Here, per our convention, $z=e^{2\pi i \mathfrak z}$. Then we can write the elliptic function $f$ as
\begin{equation}\label{elliptic-expansion}
f(\mathfrak{z}) = C_f(\tau) + \frac{1}{2\pi i}\sum_{j} R_j\ \zeta(\mathfrak{z} - \mathfrak{z}_j) \ ,
\end{equation}
where $C_f$ is independent of $\mathfrak z$ or the pole positions but can depend on the nome $\tau$. It is straightforward to check that the right-hand side is indeed elliptic, using the behavior of the $\zeta$-function under shifts by a period and the fact that the sum of the residues vanishes, \ie{}, $\sum_{j} R_j = 0$. Moreover, the pole positions and their residues on both sides of the equation manifestly match.

We are interested in the contour integral
\begin{align}
	\oint_{|z| = 1} \frac{dz}{2\pi i z} f(\mathfrak{z}) = \int_0^1d \mathfrak{z} f(\mathfrak{z})\ . 
\end{align}
Rewriting the elliptic function $f$ in terms of the $\zeta$ function as in \eqref{elliptic-expansion}, we have
\begin{align}
	\oint_{|z| = 1} \frac{dz}{2\pi i z} f(\mathfrak{z})
	= C_f(\tau) + \frac{1}{2\pi i} \sum_j R_j \int_0^1 d \mathfrak{z} \zeta(\mathfrak{z} - \mathfrak{z}_j) \label{int1}\ .
\end{align}
Our task is thus to evaluate the integral of the Weierstrass zeta functions $\zeta(\mathfrak{z} - \mathfrak{z}_j)$. This can be easily achieved by observing that the $\zeta$-function can be Fourier-expanded (where $'$ ignores $n = 0$)\cite{abramowitz+stegun},\footnote{An alternative approach to evaluate these integrals is as follows. Note that $\zeta$ is the derivative of the Weierstrass $\sigma$-function,
\begin{align}
	\zeta(\mathfrak{z}) = \frac{d}{d\mathfrak{z}} \widehat{\sigma}(\mathfrak{z}), \qquad \widehat{\sigma}(\mathfrak{z}) = \ln \vartheta_1(\mathfrak{z}) - \frac{(2\pi)^2}{2}  \mathfrak{z}^2 E_2(\tau) \ .
\end{align}
Therefore, we have the integral (choosing $\ln (-1) = - \pi i$)
\begin{align}
	\int_0^1d \mathfrak{z} \zeta(\mathfrak{z} - \mathfrak{b}) = \ln \frac{\vartheta_1(1 - \mathfrak{b})}{\vartheta_1( - \mathfrak{b})}  - 4\pi^2 \left(\frac{1}{2} - \mathfrak{b}\right)E_2(\tau) \ = - \pi i - 4\pi^2 \left(\frac{1}{2} - \mathfrak{b}\right)E_2 \ .
\end{align}
.}
\begin{align}\label{fourier-zeta}
	\zeta(\mathfrak{z}) = & \ - 4\pi^2 \mathfrak{z} E_2(\tau) - \pi i + \pi \sum_{n}' \frac{1}{\sin n \pi \tau}q^{- \frac{n}{2}}e^{2 \pi i n \mathfrak{z}} \ , \qquad \text{if}\quad \operatorname{Im}\mathfrak{z} = 0 \ , \\
	\zeta(\mathfrak{z}) = & \ - 4\pi^2 \mathfrak{z} E_2(\tau) + \pi i + \pi \sum_{n}' \frac{1}{\sin n \pi \tau}q^{+ \frac{n}{2}}e^{2 \pi i n \mathfrak{z}} \ , \qquad \text{if}\quad -1 < \frac{\operatorname{Im}\mathfrak{z}}{\operatorname{Im}\tau} < 0 .\label{fourier-zeta2}
\end{align}
For $\mathfrak{z}$ with $\operatorname{Im}\mathfrak{z}$ outside of the above ranges, one simply applies the shift formula (\ref{shift-formula-zeta}). In the integral of interest \eqref{int1}, the variables $\mathfrak{z} - \mathfrak{z}_j$ belong precisely to either one of these ranges. To distinguish these two cases, we will call $\mathfrak{z}_j$ \emph{real} if $\operatorname{Im}\mathfrak{z}_j = 0$, or \emph{imaginary} if $\operatorname{Im}\mathfrak{z}_j > 0$.

Now we simply compute the integral of the Fourier series and obtain
\begin{equation}
	\oint_{|z| = 1} \frac{dz}{2\pi i z} f(\mathfrak{z})
	= C_f(\tau)
	  + \frac{1}{2\pi i} \sum_{\text{real } \mathfrak{z}_j} R_j \, (4\pi^2 \mathfrak{z}_jE_2(\tau) - \pi i)
	  + \frac{1}{2\pi i} \sum_{\text{imag. } \mathfrak{z}_j}  R_j\, (4\pi^2 \mathfrak{z}_jE_2(\tau) + \pi i)\ .
\end{equation}
This result can be simplified more by observing that the Weierstrass zeta function is related to the Eisenstein series as (see appendix \ref{app:usefulidentities} for more details)
\begin{equation}
	\zeta(\mathfrak{z})
	= 2\pi i\, E_1\left[\begin{matrix}
		- 1 \\ zq^{\frac{1}{2}}
	\end{matrix}
	\right] + \pi i - 4\pi^2 \mathfrak{z}E_2 
  = 2\pi i\, E_1\left[\begin{matrix}
		- 1 \\ zq^{ - \frac{1}{2}}
	\end{matrix}
	\right] - \pi i - 4\pi^2 \mathfrak{z}E_2\ ,
\end{equation}
and by writing $C_f$ as
\begin{equation}
	C_f(\tau) = f(\mathfrak{z}_0) - \frac{1}{2\pi i}\sum_{j} R_j\ \zeta(\mathfrak{z}_0 - \mathfrak{z}_j) \ .
\end{equation}
where $\mathfrak{z}_0$ is an \textit{arbitrary} reference value. One then finds the explicit, analytic evaluation of the contour integral of the elliptic function $f$ to be:
\begin{equation}\label{elliptic-integral}
\boxed{
	\oint_{|z| = 1} \frac{dz}{2\pi i z} f(\mathfrak{z})
	= f(\mathfrak{z}_0)
	  + \sum_{\text{real } \zeta_j}R_j\ E_1\left[\begin{matrix}
			-1 \\ \frac{z_j}{z_0} q^{\frac{1}{2}}
		\end{matrix}\right]
	  + \sum_{\text{imag. } \zeta_j}R_j\ E_1\left[\begin{matrix}
			-1 \\ \frac{z_j}{z_0} q^{ - \frac{1}{2}}
		\end{matrix}\right] \ ,
		}
\end{equation}
We can choose $\mathfrak{z}_0$ at will, and we will sometimes exploit this freedom to simplify expressions on a case-dependent basis. Most often though, we simply choose $\mathfrak{z}_0 = 0$. 

A couple of remarks are in order. First, the evaluation formula \eqref{elliptic-integral} allows one to compute the Schur index of Lagrangian rank-one theories, \ie{}, $\mathcal N=4$ super Yang-Mills with gauge group $SU(2)$ and an $SU(2)$ gauge theory with four hypermultiplets transforming in the fundamental representation of the gauge group. We will do so in the next section. Second, we notice that after a single integral, the result stops being elliptic in a very manifest manner, as the Eisenstein series $E_1$ is not doubly periodic (see (\ref{Eisenstein-shift-1}) for its behavior under a full period shift). To evaluate the indices of higher-rank theories, we will thus need evaluation formulas involving the product of an elliptic function and an Eisenstein series. We will refer to such integrands as almost elliptic.

\subsection{Integrating almost elliptic functions\label{section:integrating-almost-elliptic-functions}}\label{integratinfalmostelliptic}
We now turn to the task of evaluating contour integrals whose integrand is the product of an elliptic function (possessing only simple poles) and an Eisenstein series,
\begin{equation}\label{integral-ell-times-eis}
\oint_{|z| = 1} \frac{dz}{2\pi i z}f(\mathfrak{z})E_k\left[\begin{matrix}\pm 1 \\ za\end{matrix}\right] \;,
\end{equation}
where $a$ is an arbitrary complex number different from $a$ and $q$ (or powers thereof). Without further ado, we immediately present our results for the integration formulas. First of all, we have for $k \ge 0$ (when $k = 0$ it reduces to (\ref{elliptic-integral}))
{\begin{empheq}[box=\fbox]{align}\label{integral-formula-1}
	& \ \oint_{|z| = 1} \frac{dz}{2\pi i z}f(\mathfrak{z})E_k\left[
	\begin{matrix}
		-1 \\ za
	\end{matrix}\right] \\
	= & \ - \mathcal{S}_{k} \left(f(\mathfrak{z}_0)
	  + \sum_{\text{real/imag } \mathfrak{z}_j}R_j  E_{1}\left[\begin{matrix}
			-1 \\ \frac{z_j}{z_0}q^{\pm \frac{1}{2}}
		\end{matrix}\right]
	\right)
	- \sum_{\text{real/imag } \mathfrak{z}_j} R_j\,  \sum_{\ell = 0}^{k - 1} \mathcal{S}_{\ell}\, E_{k - \ell + 1}\left[
		\begin{matrix}
			1 \\ z_j a q^{\pm \frac{1}{2}}
		\end{matrix}\right] \ ,\nonumber 
\end{empheq}}%
{{\vspace{-2cm}

\noindent As in the previous subsection, $\mathfrak{z}_j$ for $j>0$ are the positions of the simple poles of $f$ and they are called real or imaginary depending on whether their imaginary part is zero or strictly positive, and $R_j$ are their residues, \ie{}, $\operatorname{Res}_{z \to z_j}\frac{1}{z} f(z)$. Depending on the reality of the $\mathfrak{z}_j$, the argument of the Eisenstein series involves a positive or negative power of $\sqrt{q}$. Furthermore, $\mathfrak{z}_0$ is an arbitrary reference value. Finally, $\mathcal{S}_{\ell}$ are rational numbers defined as
\begin{equation}\label{S2k_maintext}
\frac{1}{2}\frac{y}{\sinh \frac{y}{2}} \equiv \sum_{\ell \ge 0} \mathcal{S}_\ell\, y^\ell \qquad \text{for $y<1$}\;.
\end{equation}
More explicitly,
\begin{center}
	\begin{tabular}{c|c|c|c|c|c|c|c|c|c|c|c}
		$\ell$ & 0 & 1 & 2 & 3 & 4 & 5 & 6 & 7 & 8 & 9 & 10 \\
		\hline
		$\mathcal{S}_\ell$ & 1 & 0 & $- \frac{1}{24}$ & 0 & $\frac{7}{5760}$ & 0 & $ - \frac{31}{967680}$ & 0 & $\frac{127}{154828800}$ & 0 & $- \frac{73}{3503554560}$
	\end{tabular}
\end{center}

Similarly, we have established the following integration formula
{\begin{empheq}[box=\fbox]{align}\nonumber
	& \ \oint_{|z| = 1} \frac{dz}{2\pi i z} f(\mathfrak{z}) E_k\left[\begin{matrix}
		+ 1 \\ za
	\end{matrix}
	\right]\\
	= & \ - \mathcal{A}_k\left(f(\mathfrak{z}_0) + \sum_{\text{real/imag } \mathfrak{z}_j}R_j\, E_1\left[\begin{matrix}
			- 1\\ \frac{z_j}{z_0}q^{\pm \frac{1}{2}}
		\end{matrix}\right]\right)  \nonumber\\
	  & \ - \sum_{\text{real/imag } \mathfrak{z}_j} R_j \left(
	    - \mathcal{B}_k\, E_1\left[\begin{matrix}
	  	  -1 \\ z_jaq^{\pm \frac{1}{2}}
	    \end{matrix}\right]
	  + \sum_{\ell = 0}^{k - 1} \mathcal{S}_{\ell}\, E_{k + 1 - \ell} \left[\begin{matrix}
	  	  		-1 \\ z_j a q^{\pm  \frac{1}{2}}
	  	  	\end{matrix}\right]\right) \; , \label{integral-formula-2}
\end{empheq}}%

\noindent where $\mathcal{A}_k$ and $\mathcal{B}_k$ are rational numbers given by
\begin{align}
	\mathcal{A}_{2n} = \frac{B_{2n}}{(2n)!}, \qquad \mathcal{A}_{2n + 1} = \frac{\delta_{n, 0}}{2}, \qquad \mathcal{B}_{2n} = \frac{B_{2n}}{(2n)!} - \mathcal{S}_{2n}, \qquad \mathcal{B}_{2n + 1} = \frac{\delta_{n,0}}{2} \ .
\end{align}
Here $B_{2n}$ are the Bernoulli numbers.

Let us make some brief comments about the derivation of \eqref{integral-formula-1} and \eqref{integral-formula-2}, referring the reader to the appendices for more details. Integrals of the type \eqref{integral-ell-times-eis} can be analyzed by first expanding the elliptic function $f(\mathfrak{z})$ and the Eisenstein series in Fourier series. The former Fourier series has been obtained in the previous subsection, while we propose Fourier expansions of the Eisenstein series in appendix \ref{app:fourier}. The contour integral of products of Fourier series is easily evaluated and results in a novel Fourier series. The nontrivial task is to recognize the resulting series as combinations of Jacobi theta functions or Eisenstein series. We have done so explicitly for relatively small $k$ by hand, while for $k$ up to 11, we considered a suitable Ansatz based on the predicted structure of the result and checked the integral formula in series expansions in $q$ to very high order.


\section{Rank-one theories}\label{sec:rankone}
A first class of theories whose Schur index we can evaluate in closed form are Lagrangian superconformal field theories with a one-dimensional Coulomb branch. In other words, gauge theories with gauge group of rank one. Two such theories exist: first, $\mathcal N=4$ super Yang-Mills theory with gauge group $SU(2)$, and second, $SU(2)$ superconformal QCD, \ie{}, an $\mathcal N=2$ supersymmetric gauge theory with four hypermultiplets transforming in the fundamental representation of the gauge group $SU(2)$.

\subsection{\texorpdfstring{$\mathcal N=4$}{N=4} super Yang-Mills theory with gauge group \texorpdfstring{$SU(2)$}{SU(2)}}\label{N=4rankone}

Using \eqref{SchurIndexIntegral}, it is easy to write down the contour integral computing the Schur index of $\mathcal N=4$ super Yang-Mills theory with gauge group $SU(2)$:
\begin{equation}
	\mathcal{I}_{\mathcal{N} = 4 \ SU(2)}
	= -\frac{1}{2} \oint \frac{da}{2\pi i a} \frac{\eta(\tau)^3}{\vartheta_4(\mathfrak{b})} \prod_{\pm} \frac{\vartheta_1(\pm 2 \mathfrak{a})}{\vartheta_4(\pm 2 \mathfrak{a} + \mathfrak{b})} \ .
\end{equation}
Here $b=e^{2\pi i \mathfrak b}$ is a fugacity for the $SU(2)_F$ flavor symmetry of the theory when viewed as an $\mathcal N=2$ theory; in other words, it rotates the adjoint hypermultiplet. For simplicity, we start by rescaling the integration variable as $\mathfrak{a} \to \mathfrak{a}/2$ and using the periodicity properties of the integrand find
\begin{equation}\label{N=4resc}
	\mathcal{I}_{\mathcal{N} = 4 \ SU(2)}
	= \frac{1}{2} \oint \frac{da}{2\pi i a} \frac{\eta(\tau)^3}{\vartheta_4(\mathfrak{b})} \prod_{\pm} \frac{\vartheta_1(\pm \mathfrak{a})}{\vartheta_4(\pm \mathfrak{a} + \mathfrak{b})} \ .
\end{equation}

The integrand of \eqref{N=4resc} is elliptic with respect to $\mathfrak{a}$ and has two simple poles in the fundamental parallelogram at
\begin{equation}
	\mathfrak{a} = \pm \mathfrak{b} + \frac{\tau}{2} \ .
\end{equation}
As classified in the previous section (see below \eqref{fourier-zeta2}), these are poles of ``imaginary type''. Their respective residues are of course opposite and are given explicitly by
\begin{equation}\label{resN=4}
	\mathop{\operatorname{Res}}\limits_{\mathfrak{a} \to \pm \mathfrak{b} + \frac{\tau}{2}} \frac{1}{2}\frac{\eta(\tau)^3}{\vartheta_4(\mathfrak{b})} \prod_{\pm} \frac{\vartheta_1(\pm \mathfrak{a})}{\vartheta_4(\pm \mathfrak{a} + \mathfrak{b})} = \pm \frac{1}{2 i} \frac{\vartheta_4(\mathfrak{b})}{\vartheta_1(2 \mathfrak{b})} \ .
\end{equation}

Using formula \eqref{elliptic-integral}, it is straightforward to compute the contour integral,
\begin{align}
	\mathcal{I}_{\mathcal{N} = 4 \ SU(2)}
	= & \ - \frac{1}{2 i} \frac{\vartheta_4(\mathfrak{b})}{\vartheta_1(2 \mathfrak{b})}E_1\left[\begin{matrix}
		-1 \\ bq^{\frac{1}{2}}q^{- \frac{1}{2}}
	\end{matrix}\right]
	+ \frac{ 1}{2 i} \frac{\vartheta_4(\mathfrak{b})}{\vartheta_1(2 \mathfrak{b})}E_1\left[\begin{matrix}
			-1 \\ b^{-1}q^{\frac{1}{2}}q^{- \frac{1}{2}}
		\end{matrix}\right] \\
	= & \ - \frac{1}{2i} \frac{\vartheta_4(\mathfrak{b})}{\vartheta_1(2 \mathfrak{b})} \left(E_1\left[\begin{matrix}
				-1 \\ b
			\end{matrix}\right]
			- E_1\left[\begin{matrix}
				-1 \\ b^{-1}
			\end{matrix}\right]
	\right)\;.
\end{align}
Here we used the reference value $\mathfrak z_0 = 0.$

One can choose to further simplify the above result for the Schur index of $SU(2)$ $\mathcal N=4$ super Yang-Mills by using the symmetry property of Eisenstein series and rewriting the Eisenstein series in terms of Jacobi-theta functions. One then finds
\begin{equation}\label{N=4SU2}
	\mathcal{I}_{\mathcal{N} = 4 \ SU(2)}
	= \frac{i \vartheta_4(\mathfrak{b})}{\vartheta_1(2 \mathfrak{b})}E_1\left[\begin{matrix}
		-1 \\ b
	\end{matrix}\right]
	= \frac{1}{2\pi} \frac{\vartheta'_4(\mathfrak{b})}{\vartheta_1(2 \mathfrak{b})} \ .
\end{equation}

When including an additional free hypermultiplet and identifying its $SU(2)$ flavor symmetry with $SU(2)_F$ identified above, the theory has a class $\mathcal{S}$ description of type $\mathfrak a_1$ in terms of a one-punctured torus. Denoting the Schur index of the theory associated with an $s$-punctured genus $g$ surface as $\mathcal I_{g,s}$, we thus find
\begin{equation}\label{I11}
	\mathcal{I}_{1,1}(\mathfrak{b})
	= \frac{1}{2\pi} \frac{\vartheta_4'(\mathfrak{b})}{\vartheta_1(2 \mathfrak{b})} \frac{\eta(\tau)}{\vartheta_4(\mathfrak{b})}
	= \frac{i \eta(\tau)}{2\vartheta_1(2 \mathfrak{b})}
	\sum_{\alpha = \pm}
	\alpha E_1\left[\begin{matrix}
		- 1 \\ b^\alpha
	\end{matrix}\right] \ .
\end{equation}
The rewriting we performed in the second equality foreshadows our general results for the indices $\mathcal I_{g,s}$ of section \ref{sec:a1-index}. More generally, the flavor symmetry of the additional free hypermultiplet need not be identified with $SU(2)_F$. In that case, we have
\begin{equation}\label{I11general}
	\mathcal{I}_{1,1}(\mathfrak{b}, \mathfrak{b}') = \frac{1}{2\pi} \frac{\vartheta_4'(\mathfrak{b})}{\vartheta_1(2 \mathfrak{b})} \frac{\eta(\tau)}{\vartheta_4(\mathfrak{b}')} \ ,
\end{equation}
where $\mathfrak{b}'$ is the fugacity of the flavor symmetry of the free hypermultiplet.

\subsection{\texorpdfstring{$SU(2)$}{SU(2)} superconformal QCD} \label{SU2SCQCD}
We now consider an $SU(2)$ gauge theory with four hypermultiplets transforming in the fundamental representation of the gauge group. We denote its Schur index as $\mathcal{I}_{0, 4}$, as it has a class $\mathcal{S}$ description of type $\mathfrak a_1$ in terms of a four-punctured sphere. It is computed by the following contour integral
\begin{equation}
	\mathcal{I}_{0, 4} = - \frac{1}{2}\oint \frac{da}{2\pi i a} \vartheta_1(2 \mathfrak{a})\vartheta_1(- 2 \mathfrak{a}) \prod_{j = 1}^4\prod_{\pm}\frac{\eta(\tau)}{\vartheta_4(\pm\mathfrak{a} + \mathfrak{m}_j)} \ .
\end{equation}
The poles $\mathfrak{a} = \pm \mathfrak{m}_j + \frac{\tau}{2}\ , j = 1, \ldots, 4$ in the fundamental parallelogram are all of imaginary type with residues
\begin{equation}\label{SQCDresidues}
	R_{j,\pm} \equiv \mathop{\operatorname{Res}}\limits_{a \to m_j^{\pm 1}q^{\frac{1}{2}}} (\text{integrand}) = \pm \frac{i}{2} 
  \frac{\vartheta_1(2\mathfrak{m}_j)}{\eta(\tau)}
  \prod_{\ell \ne j} \frac{\eta(\tau)}{\vartheta_1(\mathfrak{m}_j + \mathfrak{m}_\ell)}
  \frac{ \eta(\tau)}{\vartheta_1(\mathfrak{m}_j - \mathfrak{m}_\ell)}\ .
\end{equation}
We pick $\mathfrak{a} = 0$ as the arbitrary reference value, which happens to be a zero of the integrand. Consequently, using \eqref{elliptic-integral}, the index reads
\begin{equation}\label{I04-1}
	\mathcal{I}_{0, 4}
	= \sum_{j = 1}^4 E_1\left[
    \begin{matrix}
      - 1 \\ m_j
    \end{matrix}
    \right]
    \frac{ i \vartheta_1(2\mathfrak{m}_j) }{\eta(\tau)}\prod_{\ell \ne j}\frac{\eta(\tau)}{\vartheta_1(\mathfrak{m}_j + \mathfrak{m}_\ell)}
    \frac{ \eta(\tau)}{\vartheta_1(\mathfrak{m}_j - \mathfrak{m}_\ell)} \ ,
\end{equation}
where we have used the symmetry
\begin{equation}
	E_1\left[\begin{matrix}
		-1 \\ m_j^{-1}
	\end{matrix}\right] = - E_1\left[\begin{matrix}
		-1 \\ m_j
	\end{matrix}\right]\ .
\end{equation}

Note that the four fugacities $\mathfrak{m}_j$ are combinations of those associated to the four punctures in the class $\mathcal{S}$ description. Denoting the latter as $b_{s}, s=1,2,3,4$, they are related to $m_j$ by
\begin{equation}
  m_1 =  b_1 b_2 , \qquad m_2 = \frac{b_1}{b_2}, \quad
  m_3 =  b_3 b_4, \qquad m_4 = \frac{b_3}{b_4}\ .
\end{equation}
Notice, however, that after this change of variables S-duality is not yet manifest, as the expression is not explicitly permutation invariant in the flavor fugacities. In section \ref{sec:a1-index}, we will revisit this theory, and all other theories of class $\mathcal S$ of type $\mathfrak a_1$, and uncover an alternative expression which is explicitly invariant under (generalized) S-dualities.


\section{Theories of class \texorpdfstring{$\mathcal S$}{S} of type \texorpdfstring{$\mathfrak a_1$}{a1}}\label{sec:a1-index}

The next class of theories whose Schur indices we set out to evaluate in closed form are those of class $\mathcal S$ of type $\mathfrak a_1$. Such theory can be engineered by compactifying the six-dimensional $\mathcal N=(2,0)$ theory of type $\mathfrak a_1$ on a Riemann surface $\Sigma_{g,n}$ of genus $g$ and with $n$ punctures. The punctures mark the location of (regular) codimension-two defects spanning the four non-compact spatial dimensions. For type $\mathfrak a_1$, a unique such defect is available. The complex structure moduli of the Riemann surface encode the exactly marginal couplings of the resulting four-dimensional theory. We denote this theory $\mathcal T_{g,n}$. Different degeneration limits of the Riemann surface correspond to different (generalized) S-dual descriptions of the theory \cite{Gaiotto:2009we}.  Because the Schur index is independent of exactly marginal couplings, it's independent of the choice of duality frame. Consequently, it is computed by a TQFT correlator on $\Sigma_{g,n}$ \cite{Gadde:2009kb,Gadde:2011ik,Beem:2014rza}. In particular, the index should be invariant under all permutations of the flavor symmetry fugacities associated with the punctures.

In this section, we show that the Schur index of the theory $\mathcal T_{g,n}$ is given by the closed-form expression
\begin{align}\label{Ign}
	\mathcal{I}_{g, n}(q; \vec b) = & \ \frac{i^n}{2} \frac{\eta(\tau)^{n + 2g - 2}}{\prod_{j = 1}^{n}\vartheta_1(2 \mathfrak{b}_j)}
	\sum_{\vec\alpha = \pm}\Big(  \prod_{j = 1}^{n}\alpha_j  \Big)\sum_{k = 1}^{n + 2g - 2}\lambda_k^{(n + 2g - 2)} E_k\left[\begin{matrix}
		(-1)^n \\ \prod_{j = 1}^{n}b_j^{\alpha_j}
	\end{matrix}\right]\ ,\\
	\mathcal{I}_{g, n = 0}(q) = & \ \frac{1}{2}\eta(\tau)^{2g - 2} \sum_{k = 1}^{g - 1} \lambda_{2k}^{(2g - 2)} \left(E_{2k} + \frac{B_{2k}}{(2k)!}\right)\;, \qquad \text{for } g>0\ ,
\end{align}
where in the first line we sum over all signs $\alpha_j = \pm 1$ independently. The numerical coefficients $\lambda$ are determined recursively by the equations
\begin{align}\label{lambda-def}
	\lambda^{(\text{even})}_0 = & \ \lambda^{(\text{odd})}_\text{even} = \lambda^{(\text{even})}_\text{odd} = 0\;, \qquad \lambda^{(1)}_1 = \lambda^{(2)}_2 = 1 \; , \\
	\lambda^{(2k + 1)}_1 = & \ \sum_{\ell = 1}^{k} \lambda_{2 \ell}^{2k} \left( \mathcal{S}_{2\ell} - \frac{B_{2\ell}}{(2\ell)!} \right)  \;,  \\
	\lambda^{(2k + 1)}_{2m + 1} = & \ \sum_{\ell = m}^{k}\lambda_{2\ell}^{(2k)} \mathcal{S}_{2(\ell - m)} \;, \quad
	\lambda^{(2k + 2)}_{2m + 2} = \sum_{\ell = m}^{k}\lambda^{(2k + 1)}_{2\ell + 1}\mathcal{S}_{2(\ell - m)}\; ,
\end{align}
where $\mathcal S$ was defined in \eqref{S2k_maintext} and $B$ are again the Bernouilli numbers. Note that $\mathcal{I}_{1, 1}$ has already been proved to take the form \eqref{Ign} in \eqref{I11}. It is quite pleasing to note that in \eqref{Ign} all fugacities appear on equal footing, directly reflecting the generalized S-duality invariance of the index.

To prove this statement, we first show that the Schur index of the trinion theory $\mathcal T_{0,3}$ takes the form \eqref{Ign}. Subsequent gaugings to add punctures and handles can then be easily performed using the formulas presented in section \ref{sec:integrals}.

\subsection{Trinion theory}
The $\mathfrak{a}_1$ trinion theory $\mathcal T_{0,3}$ is a theory of eight half-hypermultiplets. Breaking their $\mathfrak{usp}(8)$ flavor symmetry as $\mathfrak{usp}(8)\rightarrow \mathfrak{su}(2) \oplus \mathfrak{so}(4) = \mathfrak{su}(2) \oplus \mathfrak{su}(2) \oplus \mathfrak{su}(2)$, its Schur index is easily written down:
\begin{equation}\label{I03_orig}
\mathcal I_{0,3}\left(q; \vec{b}\right) = \frac{\eta(\tau)^4}{\prod_{\pm, \pm}\vartheta_4(\mathfrak b_1 \pm \mathfrak b_2 \pm \mathfrak b_3)}\;.
\end{equation}

For our current purposes, a more useful expression for $\mathcal I_{0,3}\left(q; \vec{b}\right)$ is as follows:
\begin{equation}\label{I03_eis}
\mathcal I_{0,3}\left(q; \vec{b}\right) = \frac{1}{2i} \frac{\eta(\tau)}{\prod_{i = 1}^3 \vartheta_1(2 \mathfrak{b}_i)}\sum_{\vec\alpha = \pm} \Big(\prod_{i = 1}^3\alpha_i\Big) E_1\left[\begin{matrix}
		- 1 \\ \prod_{i = 1}^3 b_i^{\alpha_i}
	\end{matrix}\right]\ .
\end{equation}
Note that this expression indeed conforms to \eqref{Ign}. The proof of the equality of \eqref{I03_orig} and \eqref{I03_eis} is essentially contained in appendix E of \cite{Gadde:2011uv}. There it was proved that the trinion index of type $\mathfrak a_1$ given in \eqref{I03_orig} can be written in terms of a TQFT three-point function \cite{Gadde:2011ik,Gadde:2011uv}:\footnote{We have re-organized this expression slightly compared to \cite{Gadde:2011ik,Gadde:2011uv} and have expressed the prefactor in terms of elliptic functions.}
\begin{equation}\label{I03_TQFT}
\mathcal I_{0,3}\left(q; \vec{b}\right) = i \eta(\tau)\left(\prod_{i = 1}^{3} \frac{b_i - b_i^{-1}}{\vartheta_1(2 \mathfrak{b}_i)}\right)\sum_{j \in \frac{1}{2}\mathbb{N}} \frac{\prod_{i = 1}^{3}\chi^{(j)}(b_i)}{q^{-\frac{1}{2}(2j + 1)} - q^{\frac{1}{2}(2j + 1)}} \ ,
\end{equation}
where $\chi^{(j)}$ denotes the spin-$j$ character of $SU(2)$: $\chi^{(j)}(a) = (a^{2j+1} - a^{-2j-1})/(a-a^{-1}).$ To establish the equality of this expression and \eqref{I03_eis}, one can apply the identity
\begin{equation}
	\sum_{n = 1}^{+\infty}\frac{x^n}{q^{- \frac{n}{2}} - q^{\frac{n}{2}}} = \sum_{n = 0}^{+\infty} \frac{x q^{n + \frac{1}{2}}}{1 - x q^{n + \frac{1}{2}}} \ ,
\end{equation}
to the first Eisenstein series
\begin{equation}
	E_1\left[\begin{matrix}
		-1 \\ z
	\end{matrix}\right] = - B_1(1/2)
	+ \sum_{r = 0}^{+\infty} \frac{z^{-1} q^{r + \frac{1}{2}}}{1 - z^{-1}q^{r + \frac{1}{2}}}
	- \sum_{r = 0}^{+\infty} \frac{z q^{r + \frac{1}{2}}}{1 - z q^{r + \frac{1}{2}}} \ ,
\end{equation}
to deduce that
\begin{equation}
	\sum_{n = 1}^{+\infty} \frac{\prod_{i = 1}^{3}(b_i^n - b_i^{-n})}{q^{- \frac{n}{2}} - q^{\frac{n}{2}}} = - \frac{1}{2} \sum_{\vec \alpha = \pm} (\alpha_1\alpha_2\alpha_3)E_1\left[\begin{matrix}
		-1 \\ b_1^{\alpha_1}b_2^{\alpha_2}b_3^{\alpha_3}
	\end{matrix}\right] \ .
\end{equation}

\subsection{Induction on number of punctures}
In the previous subsection we proved that $\mathcal I_{0,3}$ is of the form \eqref{Ign}. To increase the number of punctures $n$, we perform induction on $n$. We thus assume that $\mathcal I_{g,n}$ is of the form \eqref{Ign} and aim to prove that then also $\mathcal I_{g,n+1}$ is. Adding a puncture requires the evaluation of the following contour integral:
\begin{equation}
\mathcal I_{g,n+1} = - \frac{1}{2}\oint \frac{da}{2\pi i a} \vartheta_1(2 \mathfrak{a})\vartheta_1(- 2 \mathfrak{a})\ \mathcal I_{g,n}\left(q;b_1,\ldots,b_{n-1},a\right)\ \frac{\eta(\tau)^4}{\prod_{\pm, \pm}\vartheta_4(\mathfrak a \pm \mathfrak  b_n \pm \mathfrak b_{n+1})}\;.
\end{equation}
Substituting \eqref{Ign} for $\mathcal I_{g,n}$, this integral is exactly of the type \eqref{integral-ell-times-eis} for which we have presented integration formulas in \eqref{integral-formula-1} and \eqref{integral-formula-2}. To use these formulas, we need to evaluate the residues of the poles of the elliptic part of the integrand. These poles only arise from the additional hypermultiplet and are located (inside the fundamental parallelogram) at $a = b_n^{\alpha_n} b_{n+1}^{\alpha_{n + 1}} q^{\frac{1}{2}},$ where $\alpha_n, \alpha_{n + 1}$ are independent signs. These poles are of imaginary type. Now we observe the interesting fact that
\begin{multline}
\mathop{\operatorname{Res}}_{a = b_n^{\alpha_n} b_{n+1}^{\alpha_{n + 1}} q^{\frac{1}{2}}} \frac{1}{a}
	\frac{\eta(\tau)^{n + 2 g -2}}{\vartheta_1(2 \mathfrak{a})\prod_{i = 1}^{n - 1}\vartheta_1(2 \mathfrak{b}_i)}
	\left(-\frac{1}{2} \vartheta_1(2\mathfrak a) \vartheta_1(-2\mathfrak a) \right) \left( \frac{\eta(\tau)^4}{\prod_{\pm, \pm}\vartheta_4(\mathfrak a \pm \mathfrak  b_n \pm \mathfrak b_{n+1})} \right)\\
	= \frac{-i \alpha_n \alpha_{n + 1} \eta(\tau)^{n + 2g - 1}}{2\prod_{i = 1}^{n + 1}\vartheta_1(2 \mathfrak{b}_i)} \ .
\end{multline}
These residues can now be plugged into the formulas \eqref{integral-formula-1} and \eqref{integral-formula-2} and we can use the reference value $\mathfrak{a}_0 = 0$, which is a zero of the integrand. Explicitly, we treat $n = \text{even}$ and $n = \text{odd}$ separately. For $n = \text{even}$, we first notice that
\begin{align}
	E_k\left[\begin{matrix}
		+ 1 \\ a^\pm b
	\end{matrix}\right]
	= (\pm 1)^k E_k\left[\begin{matrix}
		+ 1 \\ a b^{\pm}
	\end{matrix}\right] \ .
\end{align}
Performing the integral over $a$ using \eqref{integral-formula-2} we obtain
\begin{small}
\begin{align}
	&\mathcal I_{g,n+1} = \ \sum_{\vec\alpha = \pm} \frac{i^n}{2} \frac{\alpha_{n}\alpha_{n + 1}(- i) \eta(\tau)^{n + 2g - 1}}{2 \prod_{j = 1}^{n + 1}\vartheta_1(2 \mathfrak{b}_j)}\left(\prod_{j = 1}^{n - 1} \alpha_j\right)
	 \nonumber\\
	& \qquad\qquad \times \sum_{k = 1}^{n + 2g - 2}\lambda_k^{(n + 2g - 2)}\left(
		\mathcal{B}_k E_1\left[\begin{matrix}
			-1 \\ \prod_{j = 1}^{n + 1}b_j^{\alpha_j}
		\end{matrix}\right]
		- \sum_{\ell = 0}^{\floor{\frac{k - 1}{2}}}\mathcal{S}_{2\ell}E_{k + 1 - 2\ell}\left[\begin{matrix}
			-1 \\ \prod_{j = 1}^{n + 1}b_j^{\alpha_j}
		\end{matrix}\right]
	\right)\nonumber\\
	&-  \ \sum_{\vec\alpha = \pm} \frac{i^n}{2}\frac{\alpha_n \alpha_{n + 1}(-i)\eta(\tau)^{n + 2g - 1}}{2 \prod_{j = 1}^{n + 1}\vartheta_1(2 \mathfrak{b}_j)}  \left(\prod_{j = 1}^{n - 1}\alpha_j\right) \nonumber\\
	&\quad\; \times \sum_{k = 1}^{n + 2g - 2} \lambda^{(n + 2g - 2)}_k\left(
		 \mathcal{B}_k E_1\left[\begin{matrix}
			-1 \\ b_n^{\alpha_n}b_{n + 1}^{\alpha_{n + 1}}\prod_{j = 1}^{n - 1}b_j^{-\alpha_j}
		\end{matrix}\right]
		- \sum_{\ell = 0}^{\floor{\frac{k - 1}{2}}}\mathcal{S}_{2\ell}E_{k + 1 - 2\ell}\left[\begin{matrix}
			-1 \\ b_n^{\alpha_n}b_{n + 1}^{\alpha_{n + 1}}\prod_{j = 1}^{n - 1}b_j^{-\alpha_j}
		\end{matrix}\right]
	\right) \ .\label{Ign+1itnermediate}
\end{align}
\end{small}%
The terms with coefficients $\mathcal{A}_k$ arising from the application of (\ref{integral-formula-2}) vanish thanks to the sum over $\alpha_n, \alpha_{n + 1}$ and the simple identity $\sum_{\alpha, \beta = \pm} \alpha \beta E_\text{odd}\Big[\substack{-1 \\ a^\alpha b^\beta}\Big] = 0$. Moreover, the two sums in \eqref{Ign+1itnermediate} are actually identical thanks to the symmetry properties of $E_k$. Hence, it simplifies to
\begin{align}
	\mathcal I_{g,n+1} = & \ \sum_{\vec\alpha = \pm} \frac{i^{n + 1}}{2} \frac{ \eta(\tau)^{n + 2g - 1}}{\prod_{j = 1}^{n + 1}\vartheta_1(2 \mathfrak{b}_j)}\left(\prod_{j = 1}^{n + 1} \alpha_j\right)
	\sum_{k = 1}^{n + 2g - 2}\lambda_k^{(n + 2g - 2)} \nonumber\\
	& \qquad\qquad \times \left(
		+ \sum_{\ell = 0}^{\floor{\frac{k - 1}{2}}}\mathcal{S}_{2\ell}E_{k + 1 - 2\ell}\left[\begin{matrix}
			-1 \\ \prod_{j = 1}^{n + 1}b_j^{\alpha_j}
		\end{matrix}\right]
		- \mathcal{B}_k E_1\left[\begin{matrix}
			-1 \\ \prod_{j = 1}^{n + 1}b_j^{\alpha_j}
		\end{matrix}\right]
	\right)\nonumber \\
	= & \ \sum_{\vec\alpha=\pm}\frac{i^{n + 1}}{2} \frac{ \eta(\tau)^{n + 2g - 1}}{\prod_{j = 1}^{n + 1}\vartheta_1(2 \mathfrak{b}_j)}\left(\prod_{j = 1}^{n + 1} \alpha_j\right) \sum_{k = 1}^{n + 2g - 1} \lambda_k^{(n + 2g - 1)} E_k\left[\begin{matrix}
		-1 \\ \prod_{j = 1}^{n + 1}b_j^{\alpha_j}
	\end{matrix}\right] \ .
\end{align}
Comparing the coefficients of $E_k$ on both sides, one deduces the recursion relations\footnote{Note that $n +2g - 1$ is odd, and $k + 1 - 2\ell \ge 2$ for $\ell \le \floor{\frac{k - 1}{2}} \le \frac{k - 1}{2}$.}
\begin{equation}
	\lambda_1^{n + 2g - 1} = \sum_{\substack{k = 1 \\ k \text{ even}}}^{n + 2g -2} \lambda_k^{(n + 2g -2)} ( - \mathcal{B}_k) = \sum_{\substack{k = 1 \\ k \text{ even}}}^{n + 2g -2} \lambda_k^{(n + 2g - 2)}(\mathcal{S}_{k} - \frac{B_k}{k!}) \ ,
\end{equation}
and, noting that $\lambda^{(n - 2g - 2)}_\text{odd} = \lambda^{(n - 2g - 1)}_\text{even} = 0$,
\begin{equation}
	\lambda_{k > 1}^{(n + 2g - 1)} = \sum_{k' = k - 1}^{n + 2g - 2}\lambda_{k'}^{(n + 2g - 2)} \mathcal{S}_{k' - (k - 1)} \ .
\end{equation}

An almost identical computation can be carried out for odd $n$. Combining both cases, we recover the relations in (\ref{lambda-def}). Starting from the trinion index $\mathcal{I}_{0,3}$, we can claim rigorously the validity of the formula (\ref{Ign}) for all $\mathcal{I}_{0, n \ge 3}$. The remaining task is to apply induction on the genus $g$ as well.

\subsection{Induction on genus}
Having shown that \eqref{Ign} is correct for all genus zero theories, we now increase the genus. Once again, we proceed inductively by first assuming the validity of (\ref{Ign}) at some $g$ and $n \ge 2$. One can easily add an additional handle by gluing two punctures, say, the two associated with $b_{n - 1}$ and $b_n$. Then the index of the genus-$(g+1)$ theory with $n - 2$ punctures is given by the contour integral
\begin{align}
	\mathcal{I}_{g + 1, n-2}
	= & \ \oint \frac{da}{2\pi i a} \mathcal{I}_{g, n}(b_1, \ldots, b_{n -2}, a) \left(- \frac{1}{2}\vartheta_1(2 \mathfrak{a})\vartheta_1( - 2 \mathfrak{a})\right) \nonumber\\
	= & \ - \frac{1}{2} \oint\frac{da}{2\pi i a}  \frac{i^n}{2}
	\frac{\eta(\tau)^{n + 2g - 2}}{\prod_{j = 1}^{n - 2}\vartheta_1(2 \mathfrak{b}_j)}
	\sum_{\alpha_j} \Big(\prod_{j = 1}^{n}\alpha_i\Big) \sum_{k = 1}^{n + 2g - 2} \lambda^{(n + 2g - 2)}_k E_k\left[\begin{matrix}
		(-1)^n \\ a^{\alpha_{n - 1} - \alpha_n} \prod_{j = 1}^{n - 2} b_j^{\alpha_j}
	\end{matrix}\right] \ , 
\end{align}
where we made the identification $\mathfrak{b}_{n - 1} \to \mathfrak{a}$, $\mathfrak{b}_{n} \to - \mathfrak{a}$ upon gauging. Note the cancellations between the $\vartheta_1(\pm 2 \mathfrak{a})$. For $\alpha_{n - 1} = - \alpha_n$, the corresponding terms are completely independent of the $a$ variable. For $\alpha_{n - 1} = \alpha_{n}$, the terms depend on $a$ through the Eisenstein series and the integral over $a$ extracts their constant terms. The integral can be evaluated using the integral formula \eqref{integral-formula-1}, \eqref{integral-formula-2}, and it gives
\begin{align}\label{I2n-21}
	\mathcal{I}_{g + 1, n - 2}(b)
	= & \ - \frac{i^n}{2} \frac{\eta(\tau)^{n + 2g - 2}}{\prod_{j = 1}^{n - 2}\vartheta_1(2 \mathfrak{b}_i)}\sum'_{\vec\alpha = \pm} \Big(  \prod_{j = 1}^{n - 2}\alpha_j  \Big)\sum_{k = 1}^{n + 2g - 2}\lambda_k^{(n + 2g - 2)}E_k\left[\begin{matrix}
		(-1)^n \\ \prod_{j = 1}^{n - 2}b_j^{\alpha_j}
	\end{matrix}\right] \nonumber\\
	& \ - \frac{i^n}{4} \frac{\eta(\tau)^{n + 2g -2}}{\prod_{j = 1}^{n - 2}\vartheta_1(2 \mathfrak{b}_i)}\sum'_{\vec\alpha = \pm} \Big(  \prod_{j = 1}^{n - 2}\alpha_j  \Big)\left(+ \lambda_1^{(n + 2g - 2)} + 2\sum_{\substack{k = 1 \\ k = \text{even}}}^{n + 2g - 2}\lambda_k^{(n + 2g - 2)} \frac{B_{k}}{k!}\right) \ , 
\end{align}
where $\sum'$ implies a smaller sum over signs $\alpha_1, \ldots, \alpha_{n - 2}$ and we applied the results about constant terms collected in (\ref{const-terms}). To obtain the second line, we note that only when $k = \text{even}$ can $E_k$ have a non-vanishing constant term (with the exception of $k = 1$), while $\lambda_{k \text{ even}}^{(n + 2g -2)}$ is non-zero only when $n$ is also even. Moreover, when $n \ge 3$, the sum over $\alpha_i = \pm$ kills the second line. Therefore, given $\mathcal{I}_{g, n \ge 3}$ and in particular the rational numbers $\lambda_k^{(n + 2g - 2)}$, the index $\mathcal{I}_{g + 1, n - 2}$ is completely determined,
\begin{equation}
	\mathcal{I}_{g + 1, n - 2}(b)
	=  \ - \frac{i^n}{2} \frac{\eta(\tau)^{n + 2g - 2}}{\prod_{j = 1}^{n + 2g - 2}\vartheta_1(2 \mathfrak{b}_i)}\sum'_{\alpha_j = \pm} \Big(  \prod_{j = 1}^{n - 2}\alpha_j  \Big)\sum_{k = 1}^{n + 2g - 2}\lambda_k^{(n + 2g - 2)}E_k\left[\begin{matrix}
		(-1)^n \\ \prod_{j = 1}^{n - 2}b_j^{\alpha_j}
	\end{matrix}\right] \ .
\end{equation}
Renaming $g + 1 \to g, n - 2 \to n$, we then recover the proposed formula for $g \ge 1, n \ge 1$,
\begin{equation}\label{I2n1}
  \mathcal{I}_{g, n}(b)
  =  \ + \frac{i^{n}}{2} \frac{\eta(\tau)^{n + 2g - 2}}{\prod_{i = 1}^{n}\vartheta_1(2 \mathfrak{b}_i)}\sum_{\alpha_i = \pm} \Big(  \prod_{i = 1}^{n}\alpha_i  \Big)\sum_{k = 1}^{n + 2g - 2}\lambda_k^{(n + 2g - 2)}E_k\left[\begin{matrix}
    (-1)^n \\ \prod_{i = 1}^{n}b_i^{\alpha_i}
  \end{matrix}\right] \ .
\end{equation}

On the other hand, when $n = 2$ in \eqref{I2n-21}, $\lambda_1^{(2g)} = 0$ and we have
\begin{equation}
	\mathcal{I}_{g + 1, 0} = + \frac{1}{2} \eta(\tau)^{2g}
	\sum_{\substack{k = 1 \\ k \text{ even}}}^{2g} \lambda_k^{(2g)}\left(
	  E_k\left[\begin{matrix}
			+ 1 \\ 1
		\end{matrix}\right] + \frac{B_k}{k!}
	\right)\ .
\end{equation}
In other words,
\begin{equation}
	\mathcal{I}_{g, n = 0} =  \ + \frac{1}{2}\eta(\tau)^{2g - 2} \sum_{\ell = 1}^{g - 1} \lambda_{2\ell}^{(2g - 2)} \left(E_{2\ell} + \frac{B_{2\ell}}{(2\ell)!}\right)\ .
\end{equation}

Finally, combining the induction on $n$ and on $g$, one can start with $g = 0, n = 3$ and obtain the index for all other values of $g, n$.

\subsection{Flavoring index of genus two class \texorpdfstring{$\mathcal S$}{S} theory without punctures}
As $\mathcal T_{2,0}$ is of class $\mathcal S$, its Schur index has been computed already above; it reads
\begin{equation}
\mathcal{I}_{2,0}(q) = \frac{1}{2}\eta(\tau)^{2} \left(E_{2} + \frac{1}{12}\right) \ .
\end{equation}
However, $\mathcal T_{2,0}$ possesses a $U(1)$ flavor symmetry which is not manifest in the class $\mathcal S$ description.\footnote{It is relatively common that a class $\mathcal S$ description of a theory doesn't make manifest the full flavor symmetry group, but only a subgroup, often a maximal one. A simple example is $\mathcal T_{0,4}$ whose full flavor symmetry is $SO(8)$ but of which only an $SU(2)^4$ subgroup is manifest. The theory $\mathcal T_{2,0}$ is somewhat more unusual in that no flavor symmetry is visible at all.} In this subsection, we aim to include it. What's more, we perform the computation in two different degeneration limits of $\Sigma_{2,0}$, thus allowing us to perform a refined test of generalized $S$-duality. Finally, also note that $\mathcal T_{2,0}$ has recently received attention in the context of the VOA/SCFT correspondence \cite{Kiyoshige:2020uqz,Beem:2021jnm}.

\subsubsection{Duality frame I}
One duality frame of the genus-two theory is given in terms of two one-punctured tori $\Sigma_{1,1}^{(i)}, i=1,2$ glued together by a long tube. The theory associated to each $\Sigma_{1,1}^{(i)}$ has been discussed already in subsection \ref{N=4rankone}: it is an $\mathcal{N} = 4$ super Yang-Mills theory together with an additional doublet of free half-hypermultiplets $Q^{(i)}$.\footnote{Our notation suppresses the $SU(2)$ flavor index of the free half-hypermultiplets.} The $SU(2)$ flavor symmetries of the Yang-Mills theory and the free hypermultiplet are identified. To form the genus-two theory $\mathcal T_{0,2}$, it is this symmetry that will be gauged amongst the two copies -- we will denote its associated fugacity as $a$. Furthermore, the hypermultiplets $Q^{(1)}$ and $Q^{(2)}$  can be combined in a pair of complex conjugate combinations that transform under the $U(1)$ flavor symmetry with opposite charges, $\phi = \frac{1}{\sqrt{2}}(Q^{(1)} + i Q^{(2)})$ and $\bar\phi = \frac{1}{\sqrt{2}}(Q^{(1)} - i Q^{(2)})$. The flavor fugacity associated with this $U(1)$ symmetry is denoted $b$.

The corresponding flavored Schur index can thus be written as
\begin{align}
	\mathcal{I}_{2,0}(\mathfrak{b})
	= & \ -\frac{1}{2} \oint \frac{da}{2\pi i a} \vartheta_1(2 \mathfrak{a})\vartheta_1( - 2 \mathfrak{a}) \mathcal{I}_{1,1}(\mathfrak{a},\mathfrak{a} + \mathfrak{b})\mathcal{I}_{1,1}(\mathfrak{a}, \mathfrak{a} - \mathfrak{b})  \nonumber\\
	= & \  \frac{\eta(\tau)^2 }{8 \pi^2} \oint \frac{da}{2\pi i a} \frac{\vartheta'_4(\mathfrak{a})^2}{\vartheta_4(\mathfrak{a} + \mathfrak{b})\vartheta_4(\mathfrak{a} - \mathfrak{b})}
	=  \frac{\eta(\tau)^2}{8 \pi^2} \oint \frac{da}{2\pi i a} \frac{\vartheta_4(\mathfrak{a})\vartheta_4(\mathfrak{a})}{\vartheta_4(\mathfrak{a} + \mathfrak{b})\vartheta_4(\mathfrak{a} - \mathfrak{b})} \frac{\vartheta'_4(\mathfrak{a})^2}{\vartheta_4(\mathfrak{a})^2}\ ,
\end{align}
where $\mathcal{I}_{1,1}$ is given by \eqref{I11general}. The integral can be computed by first observing that
\begin{equation}
	\left(\frac{\vartheta_4'(\mathfrak{a})}{\vartheta_4(\mathfrak{a})}\right)^2
	= - \partial_\mathfrak{a} \zeta(\mathfrak{a} + \frac{\tau}{2}) + 8\pi^2 E_2\left[\begin{matrix}
		-1\\ a
	\end{matrix}\right]\ ,
\end{equation}
where $\zeta$ is the Weierstrass zeta function, as before. Next one can note that $\vartheta_4(\mathfrak{a})^2/(\vartheta_4(\mathfrak{a} + \mathfrak{b}) \vartheta_4(\mathfrak{a} - \mathfrak{b}))$ is elliptic with respect to $\mathfrak{a}$. The relevant poles and residues are
\begin{equation}
	\mathfrak{a} = \pm \mathfrak{b} + \frac{\tau}{2}, \qquad \mathop{\operatorname{Res}}\limits_{\pm \mathfrak{b} + \frac{\tau}{2}} = \mp \frac{i \vartheta_1(\mathfrak{b})^2}{8\pi^2 \eta(\tau) \vartheta_1(2 \mathfrak{b})} \ .
\end{equation}

Using the Fourier expansion (\ref{fourier-zeta2}) of $\zeta(\mathfrak{a} + \frac{\tau}{2})$ and the integration formula \eqref{integral-formula-1}, we have
\begin{align}\label{I201}
	\mathcal{I}_{2,0}(\mathfrak{b}) = \frac{ i \vartheta_1(\mathfrak{b})^2}{\eta(\tau) \vartheta_1(2 \mathfrak{b})}\bigg( & \ 
	  E_3\left[\begin{matrix}
	  	+ 1 \\ b
	  \end{matrix}\right]
   +E_1\left[\begin{matrix}
   	  + 1 \\ b
   \end{matrix}\right]E_2\left[\begin{matrix}
   	  + 1 \\ b
   \end{matrix}\right]
   - E_2(\tau) E_1\left[\begin{matrix}
   	  +1 \\ b
   \end{matrix}\right]
   + E_2(\tau) E_1\left[\begin{matrix}
      	  -1 \\ b
      \end{matrix}\right] \nonumber\\
   & \ 
   + \frac{1}{12} E_1\left[\begin{matrix}
   	  -1 \\ b
   \end{matrix}\right]
	\bigg) + \frac{\eta(\tau)^2}{2} \left(E_2 + \frac{1}{12}\right)\frac{\vartheta_4(0)^2}{\vartheta_4(\mathfrak{b})^2}\ .
\end{align}
Here we have chosen the reference value $\mathfrak{a} = 0$ when applying the integration formula.

\subsubsection{Duality frame II}
Another gauge theory description of the genus-two theory can be obtained by gluing two three-puncture spheres together via three long tubes. In this frame, the flavored Schur index is given by the contour integral
\begin{equation}
	\mathcal{I}_{2, 0} = - \frac{1}{8}\oint \prod_{i = 1}^3 \left[\frac{da_i}{2\pi i a_i}\vartheta_1(2\mathfrak{a}_i)\vartheta_1(- 2\mathfrak{a}_i)\right]
	\prod_{\pm, \pm, \pm} \frac{\eta(\tau)}{\vartheta_4(\pm \mathfrak{a}_1 \pm \mathfrak{a}_2 \pm \mathfrak{a}_3 + \mathfrak{b})} \ ,
\end{equation}
where $\mathfrak{b}$ is the fugacity associated with the $U(1)$-flavor symmetry.

The $a_1$ integral involves eight poles of imaginary type,
\begin{equation}
	\mathfrak{a}^{\alpha \beta \gamma}_1 = \alpha \mathfrak{a}_2 + \beta \mathfrak{a}_3 + \gamma \mathfrak{b} + \frac{\tau}{2}\ ,
	\qquad
	\alpha, \beta, \gamma = \pm 1\ ,
\end{equation}
with residues
\begin{equation}
	R^{(1)}_{\alpha \beta \gamma}
	= - i \eta(\tau)^5 \frac{\vartheta_1(2 \alpha \mathfrak{a}_2 + 2 \beta \mathfrak{a}_3 + 2 \gamma \mathfrak{b})}{
	  \vartheta_1(2 \alpha \mathfrak{a}_2 + 2 \beta \mathfrak{a}_3)
	  \vartheta_1(2 \alpha \mathfrak{a}_2 + 2 \gamma \mathfrak{b})
	  \vartheta_1(2 \beta \mathfrak{a}_3 + 2 \gamma \mathfrak{b})
	}
	\frac{\vartheta_1(2 \alpha \mathfrak{a}_2)\vartheta_1(2 \beta \mathfrak{a}_2)}{\vartheta_1(2 \gamma \mathfrak{b})} \ .
\end{equation}
Choosing the reference value to be $\mathfrak{a}_1 = 0$, which happens to be a zero of the integrand, we find the result of the first integral
\begin{equation}
	\mathcal{I}_{2,0} = \frac{1}{8} \oint \prod_{i = 2}^3 \frac{da_i}{2\pi i a_i} \sum_{\alpha, \beta, \gamma = \pm}
	R^{(1)}_{\alpha \beta \gamma}(a_2, a_3, b) E_1\left[\begin{matrix}
		-1 \\ a_2^\alpha a_3^\beta b^\gamma
	\end{matrix}\right]\ .
\end{equation}
Note that all the $E_1$ factors in the sum over $\alpha, \beta, \gamma$ depend on $a_2$.

To proceed, we note that the function $R^{(1)}_{\alpha \beta \gamma}$ has poles in $\mathfrak{a}_2$ at
\begin{align}
	\mathfrak{a}_2 = \mathfrak{a}^{\alpha\beta\gamma|1, k, \ell}_2 \equiv - \alpha \beta \mathfrak{a}_3 + \frac{k}{2} \tau + \frac{\ell}{2} , 
	\qquad \text{and}\qquad
	\mathfrak{a}_2 = \mathfrak{a}_2^{\alpha\beta\gamma|2, k, \ell} \equiv - \alpha \gamma \mathfrak{b} + \frac{k}{2} \tau + \frac{\ell}{2}\ ,
\end{align}
with $k, \ell = 0, 1$. Here, poles with $k = 0$ are of real type and those with $k = 1$ are of imaginary type. Their residues are
\begin{align}
	R^{(2)}_{\alpha\beta\gamma|1,k,\ell} 
	\equiv & \ \oint_{\mathfrak{a}^{\alpha\beta\gamma|1, k, \ell}_2 } \frac{da_2}{2\pi i a_2}R^{(1)}_{\alpha\beta\gamma}
	= \frac{\alpha}{2}\eta(\tau)^2 \frac{
	  \vartheta_1(2\beta \mathfrak{a}_3)
	  \vartheta_1(-2\beta \mathfrak{a}_3)
	}{
	  \vartheta_1(2 \beta \mathfrak{a}_3 + 2 \gamma \mathfrak{b})
	  \vartheta_1(- 2 \beta \mathfrak{a}_3 + 2 \gamma \mathfrak{b})
	}\\
R^{(2)}_{\alpha\beta\gamma|2,k,\ell} \equiv & \ \oint_{\mathfrak{a}^{\alpha\beta\gamma|2, k, \ell}_2 } \frac{da_2}{2\pi i a_2}R^{(1)}_{\alpha\beta\gamma} = - R^{(2)}_{\alpha\beta\gamma|1,k,\ell} \ .
\end{align}
After the $a_2$ integral, and simplifying the resulting integrand by explicitly performing various half-period shifts, the index reads

\begin{align}
\mathcal{I}_{2,0} = \oint \frac{da_3}{2\pi i a_3}\frac{\eta(\tau)^2 \vartheta_1(2 \mathfrak{a}_3)^2}{2\prod_\pm\vartheta_1(2\mathfrak{a}_3 \pm 2 \mathfrak{b})}
	\sum_{\pm \pm}\left(
	E_2\left[\begin{matrix}
		\pm 1 \\ \pm b
	\end{matrix}\right]
	- E_2\left[\begin{matrix}
		\pm 1 \\ \pm a_3
	\end{matrix}\right]
	\right) \ .
\end{align}
The factor in front of the sum is again an elliptic function in $\mathfrak{a}_3$ with periodicities $1$ and $\tau$. The relevant poles and residues are
\begin{align}
	\mathfrak{a}_3 = \pm \mathfrak{b} + \frac{ k}{2} \tau + \frac{\ell}{2},
	\qquad
	\mathop{\operatorname{Res}}\limits_{\pm \mathfrak{b} + \frac{ k}{2} \tau + \frac{\ell}{2}} = \pm \frac{2i \vartheta_1(2\mathfrak{b})^2}{\eta(\tau) \vartheta_1(4 \mathfrak{b})} \ ,
	\qquad
	k, \ell = 0, 1\ .
\end{align}
The final result of the Schur index then reads
\begin{align}
	\mathcal{I}'_{2,0}(\mathfrak{b})
	= & \ \frac{i \vartheta_1(2 \mathfrak{b})^2}{24 \eta(\tau) \vartheta_1(4 \mathfrak{b})}\left(
	\sum_{\pm, \pm}E_1\left[\begin{matrix}
		\pm 1\\ \pm b
	\end{matrix}\right]
 + 12 \sum_{\pm, \pm, \pm, \pm}E_1\left[
	\begin{matrix}
		\pm 1\\ \pm b
	\end{matrix}\right]
	E_2\left[
		\begin{matrix}
			\pm 1\\ \pm b
		\end{matrix}\right]
	+ 48 \sum_{\pm, \pm} E_3\left[\begin{matrix}
		\pm 1\\ \pm b
	\end{matrix}\right]
	\right) \nonumber\\
	= & \ \frac{i \vartheta_1(2 \mathfrak{b})^2}{\eta(\tau) \vartheta_1(4 \mathfrak{b})}\left(
		E_3\left[\begin{matrix}
			+ 1\\ b^2
		\end{matrix}\right]
	  + E_1\left[
		\begin{matrix}
			+ 1\\ b^2
		\end{matrix}\right]
		E_2\left[
		\begin{matrix}
			+ 1\\ b^2
		\end{matrix}\right]
		+ \frac{1}{12}E_1\left[\begin{matrix}
			1\\ b^2
		\end{matrix}\right]
		\right)\ .
\end{align}
To obtain this result, we also used \eqref{duplication-Eisenstein}.

If generalized S-duality holds, the two descriptions of the genus-two theory should have identical Schur index. In other words,
\begin{equation}
	\mathcal{I}_{2,0}(2\mathfrak{b}) = \mathcal{I}'_{2,0}(\mathfrak{b}) \ ,
\end{equation}
where we rescaled the $U(1)$ charges appropriately. The difference of the two indices is given by
\begin{equation}
\mathcal{I}'_{2,0}(\mathfrak{b}) - \mathcal{I}_{2,0}(2\mathfrak{b}) = \left(E_2 + \frac{1}{12}\right)\left[\frac{i \vartheta_1(2 \mathfrak{b})^2}{\eta(\tau)\vartheta_1(4 \mathfrak{b})}  \left(
				E_1\left[\begin{matrix}
					+ 1 \\ b^2
				\end{matrix}\right]
				- E_1\left[\begin{matrix}
					-1 \\ b^2
				\end{matrix}\right]
			\right) - \frac{\eta(\tau)^2}{2} \frac{\vartheta_4(0)^2}{\vartheta_4(2\mathfrak{b})^2}\right] = 0\ ,
\end{equation}
where the right-hand side vanishes thanks to (\ref{Eisenstein-identity-1}).

\subsection{Unflavoring}

Suppose we have an index $\mathcal{I}_{g, n}(q;\vec b)$ with $n > 0$ and we would like to unflavor one of its flavor fugacities. We consider sending $b_n \to 1$ in the compact formula for $\mathcal{I}_{g, n}(b)$. To do so, we invoke the simple limit
\begin{equation}
	\lim_{\mathfrak{b} \to 0}
	\sum_{\beta = \pm}\frac{\beta}{\vartheta_1(2 \mathfrak{b})}\frac{\vartheta^{(k)}_i(\mathfrak{a} + \beta \mathfrak{b})}{\vartheta_i(\mathfrak{a} + \beta \mathfrak{b})}
	= \frac{1}{\vartheta'_1(0)}\left(- \frac{\vartheta'_i(\mathfrak{a})}{\vartheta_i(\mathfrak{a})} \frac{\vartheta^{(k)}_i(\mathfrak{a})}{\vartheta_i(\mathfrak{a})}
	+ \frac{\vartheta^{(k + 1)}_i(\mathfrak{a})}{\vartheta_i(\mathfrak{a})}
	\right) \ ,
\end{equation}
or equivalently,
\begin{footnotesize}
\begin{equation}
	\lim_{\mathfrak{b} \to 0}
	\sum_{\beta = \pm}\frac{\beta}{\vartheta_1(2 \mathfrak{b})}
	E_k\left[\begin{matrix}
		\pm 1 \\ a b^\beta
	\end{matrix}\right] = \frac{2\pi i}{\vartheta_1'(0)}
	\Bigg(
	- E_1\left[\begin{matrix}
	 		\pm 1\\ a
  \end{matrix}\right]  \
	E_k\left[\begin{matrix}
	  \pm 1\\ a
	\end{matrix}\right]
	- (k + 1)E_{k + 1}\left[\begin{matrix}
	  \pm 1 \\ a
	\end{matrix}\right]+ \sum_{\ell = 1}^{\floor{\frac{k + 1}{2}}}
	  E_{2\ell}(\tau) E_{k + 1 - 2\ell}\left[\begin{matrix}
	  	\pm 1 \\ a
	  \end{matrix}\right]
	\Bigg) \ .
\end{equation}
\end{footnotesize}%

As a result, the $\mathcal{I}_{g, n}$ indices with one fugacity unflavored is given by
\begin{align}
&\mathcal I_{g,n}(b_n=1) =  \frac{i^n}{2} \frac{2\pi i \eta(\tau)^{2 + 2g - 2}}{\vartheta_1'(0)\prod_{i = 1}^{n - 1}\vartheta_1(2 \mathfrak{b}_i)}
	\sum'_{\alpha_i} \left(\prod_{i = 1}^{n - 1}\alpha_i \right)
	\sum_{k = 1}^{n + 2g - 2} \lambda_k^{(n + 2g - 2)} \nonumber\\
	& \ \times \left(- E_1\left[\begin{matrix}
		(-1)^n \\ \mathbf{b}_{n-1}^\alpha
	\end{matrix}\right]
	E_k\left[\begin{matrix}
			(-1)^n \\ \mathbf{b}_{n-1}^\alpha
		\end{matrix}\right]
	- (k + 1)E_{k + 1}\left[\begin{matrix}
			(-1)^n \\ \mathbf{b}_{n-1}^\alpha
		\end{matrix}\right]
		+ \sum_{\ell = 1}^{\floor{\frac{k + 1}{2}}}E_{2\ell}(\tau)E_{k + 1 - 2\ell}\left[\begin{matrix}
			(-1)^n \\ \mathbf{b}_{n - 1}^\alpha
		\end{matrix}\right]
	\right) \ , 
\end{align}
where $\mathbf{b}^\alpha_{n - 1}$ is short-hand for $\prod_{i = 1}^{n-1}b_i^{\alpha_i}$. One can repeat this computation to further unrefine the indices. 

We have not pursued this logic in all generality to arrive at a compact formula for the fully unrefined limit.\footnote{We have been informed that \cite{beemetal} has a proposal for such formula.} However, as an example of an unflavored index, we consider the unflavored limit of $\mathcal{I}_{0,4}$. This result can be compared against the unflavored vacuum character of $(\widehat{\mathfrak{d}}_4)_{-2}$ found in \cite{Arakawa:2016hkg}. The index reads
\begin{equation}
	\mathcal{I}_{0,4}\left(q;\vec b\right) = \frac{1}{2} \frac{\eta(\tau)^2}{\prod_{j = 1}^{4}\vartheta_1(2 \mathfrak{b}_j)}\sum_{\alpha_j} \left(\prod_{j = 1}^{4}\alpha_j\right) E_2\left[\begin{matrix}
		+ 1 \\ \prod_{j = 1}^{4}b_j^{\alpha_j}
	\end{matrix}\right] \ .
\end{equation}
Recall that
\begin{align}
	E_2\left[\begin{matrix}
		+ 1 \\ z
	\end{matrix}\right]
	= \frac{1}{8\pi^2} \frac{\vartheta_1''(\mathfrak{z})}{\vartheta_1(\mathfrak{z})} - \frac{1}{2}E_2 \ ,
	\qquad
	E_2 = \frac{1}{12\pi^2} \frac{\vartheta_1'''(0)}{\vartheta_1'(0)} \ .
\end{align}
Taking the $b_1, \ldots, b_4 \to 1$ limit carefully one after another, we obtain
\begin{equation}
	\lim_{\vec b \to 1}\mathcal{I}_{0, 4}(q;\vec b) = \frac{
	  10 \vartheta_1^{(3)}(0)^3
	  - 13 \vartheta'_1(0)\vartheta^{(3)}_1(0)\vartheta_1^{(5)}(0)
	  + 3 \vartheta'(0)^2 \vartheta^{(7)}(0)
	}{
	  240 \times 2^{2/3}\pi^{8/3}\vartheta_1'(0)^{19/3}
	}\ .
\end{equation}

On the other thand, $E_4$ is related to $E_2$ and Jacobi theta function by
\begin{align}
	E_4(\tau) = q \partial_q E_2(\tau) + E_2(\tau)^2 \ ,
	\qquad
	E_2 = \frac{1}{12\pi^2} \frac{\vartheta_1'''(0)}{\vartheta_1'(0)}\ ,
\end{align}
and also $\vartheta'_1(0) = 2\pi \eta(\tau)^3$. Putting everything together, it is then straightforward to show that
\begin{align}
	\lim_{\vec b \to 1}\mathcal{I}_{0, 4}(q;\vec b) = 3\frac{q \partial_q E_4(\tau)}{\eta(\tau)^{10}}  \ .
\end{align}
Note that the normalization conventions in \cite{Arakawa:2016hkg} are different from ours.

\subsection{Resumming TQFT formula}

From the identification of the trinion index \eqref{I03_eis} with the TQFT formula \eqref{I03_TQFT}, it is natural to expect that the compact, closed-form expressions in \eqref{Ign} are a resummation of the TQFT formulas for all theories of class $\mathcal S$ of type $\mathfrak a_1$. The TQFT formula reads
\begin{equation}
	\mathcal{I}_{g,n}
	= q^{\frac{1}{12}(13(g-1)+5n)} \sum_{j \in \frac{1}{2}\mathbb{N}} C_{\mathcal{R}_j}(q)^{n + 2g - 2} \prod_{i = 1}^{n} \psi_{\mathcal{R}_j}(b_i,q) \ ,
\end{equation}
where
\begin{equation}
	C_{\mathcal{R}_j}(q) = \ - \frac{q^{\frac{1}{2}}(q;q)}{q^{j + \frac{1}{2}} - q^{- j - \frac{1}{2}}} \ , \quad
	\psi_{\mathcal{R}_j}(b,q)
	= \frac{- i q^{\frac{1}{8}}}{\vartheta_1(2 \mathfrak{b})} \sum_{\alpha = \pm} \alpha b^{\alpha( 2j + 1)} \ .
\end{equation}
We have also included the central charge factor $q^{-c/24}$, where
\begin{equation}
	- \frac{c}{24} = \frac{1}{12}(13(g - 1) + 5n) \ .
\end{equation}

More explicitly, the TQFT formula can be written as
\begin{equation}
	\mathcal{I}_{g,n} =  \frac{i^n \eta(\tau)^{n + 2g - 2} }{\prod_{i = 1}^{n}\vartheta_1(2 \mathfrak{b}_1)} \sum_{\vec\alpha=\pm}\left(\prod_{i = 1}^{n} \alpha_i\right)\sum_{k = 1}^{+\infty}\frac{(\prod_{i = 1}^{n}  b^{\alpha_i} )^k}{(q^{\frac{k}{2}} - q^{ - \frac{k}{2}})^{n + 2g - 2}}\ .
\end{equation}
Comparing with \eqref{Ign}, one immediately notices the strong similarlities and one derives the following resummation formula:
\begin{equation}
	\frac{1}{2}\sum_{\alpha_i = \pm}\Big(  \prod_{i = 1}^{n}\alpha_i  \Big)\sum_{\ell = 1}^{n + 2g - 2}\lambda_\ell^{(n + 2g - 2)} E_\ell\left[\begin{matrix}
		(-1)^n \\ \prod_{i = 1}^{n}b_i^{\alpha_i}
	\end{matrix}\right]
	= \sum_{\alpha_i}\left(\prod_{i = 1}^{n} \alpha_i\right)\sum_{k = 1}^{+\infty}\frac{(\prod_{i = 1}^{n}  b^{\alpha_i} )^k}{(q^{\frac{k}{2}} - q^{ - \frac{k}{2}})^{n + 2g - 2}}\ .
\end{equation}


\section{\texorpdfstring{$\mathcal{N} = 4$}{N=4} theories \label{sec:N=4}}

Another series of Lagrangian theories whose Schur index can be evaluated exactly are the $\mathcal{N} = 4$ super-Yang-Mills models. In subsection \ref{N=4rankone}, we have already computed the index of the $\mathcal N=4$ theory with gauge group $SU(2)$. In this section we look at rank-two cases of classical gauge groups in flavored detail, and present results for unflavored indices of all $SU(N)$ theories.

\subsection{\texorpdfstring{$SU(3)\ \mathcal N=4$}{SU(3) N=4} super Yang-Mills theory}

The Schur index of $SU(3)\ \mathcal N=4$ super Yang-Mills theory is given by the double contour integral
\begin{equation}\label{N=4SU3}
	\mathcal{I}_{\mathcal{N} = 4 \ SU(3)}
	 = \frac{1}{3!} \oint \prod_{j = 1}^2\frac{da_j}{2\pi i a_j} \frac{\eta(\tau)^6}{\vartheta_4(\mathfrak{b})^2} \prod_{\substack{i,j=1\\ i\neq j}}^{3}\frac{\vartheta_1(\mathfrak{a}_i - \mathfrak{a}_j)}{\vartheta_4(\mathfrak{a}_i - \mathfrak{a}_j + \mathfrak{b})}
\end{equation}
where $\mathfrak{a}_3 = - \mathfrak{a}_1 - \mathfrak{a}_2$ and $a_3 = (a_1a_2)^{-1}$. It is easy to verify that the integrand is elliptic in both $\mathfrak{a}_1$ and $\mathfrak{a}_2$. Its poles are determined by the following six equations, which define the zeroes of the $\vartheta_4$'s in the denominator,
\begin{equation}
	\mathfrak{a}_i - \mathfrak{a}_j = \pm \mathfrak{b} + \frac{\tau}{2}, \qquad i,j=1,2,3, \quad i \ne j \  .
\end{equation}

To evaluate the integral \eqref{N=4SU3}, we first perform the integration over $a_1$. We thus have to take into account six poles, located at
\begin{align}
	& \mathfrak{a}_1 = \mathfrak{a}_2 \pm \mathfrak{b} + \frac{\tau}{2}\\
	&\mathfrak{a}_1 = - 2\mathfrak{a}_2 \pm \mathfrak{b} + \frac{\tau}{2}, \\
	& \mathfrak{a}_1 = - \frac{\mathfrak{a}_2}{2} \pm \frac{\mathfrak{b}}{2} + \frac{\tau}{4} + \frac{k \tau}{2} + \frac{\ell}{2},
	\qquad
	k, \ell = 0, 1 \ .
\end{align}
Note that the set of poles in the last line arise from the pole equation in $\mathfrak{a}_1 - \mathfrak{a}_3 = 2 \mathfrak{a}_1 + \mathfrak{a}_2$. Due to the factor $2$ in front of $\mathfrak{a}_1$, a larger number of poles lies within the fundamental parallelogram as compared to the cases in the first two lines. The respective residues of these poles are given by
\begin{align}
	R_{1,\pm}^{(1)} = & \ \frac{i}{6}\eta(\tau)^3
  \frac{
    \vartheta_4(3 \mathfrak{a}_2 \pm \mathfrak{b})
    \vartheta_1(3 \mathfrak{a}_2 \pm 2 \mathfrak{b})
  }{
    \vartheta_1(\pm 2 \mathfrak{b})
    \vartheta_1(3 \mathfrak{a}_2)
    \vartheta_4(3 \mathfrak{a}_2 \pm 3 \mathfrak{b})
  } \ , \\
  R_{2,\pm}^{(1)} = & \ \frac{i}{6} \eta(\tau)^3
    \frac{
      \vartheta_4(3 \mathfrak{a}_2 \mp \mathfrak{b})
      \vartheta_1(3 \mathfrak{a}_2 \mp 2 \mathfrak{b})
    }{
      \vartheta_1(\pm 2 \mathfrak{b})
      \vartheta_1(3 \mathfrak{a}_2)
      \vartheta_4(3 \mathfrak{a}_2 \mp 3 \mathfrak{b})} \ ,\\ 
  R^{(1)}_{3,\pm, k\ell} = & \ \frac{i}{12} \frac{\eta(\tau)^3}{\vartheta_1(\pm 2 \mathfrak{b})} \prod_{\gamma = \pm} \frac{\vartheta_1(\frac{3}{2} \gamma \mathfrak{a}_2 \pm \frac{1}{2}\mathfrak{b} + \frac{1}{4}\tau + \frac{k}{2}\tau + \frac{\ell}{2})^2}{
    \vartheta_4(\frac{3}{2} \gamma \mathfrak{a}_2 \pm \frac{3}{2}\mathfrak{b} + \frac{1}{4} \tau + \frac{k}{2}\tau + \frac{\ell}{2})
    \vartheta_4(\frac{3}{2} \gamma \mathfrak{a}_2 \mp \frac{1}{2}\mathfrak{b} + \frac{1}{4} \tau + \frac{k}{2}\tau + \frac{\ell}{2})
  } \ . 
\end{align}

The poles in $\mathfrak a_1$ are all of imaginary type, hence the result of the $a_1$ integral is given by
\begin{align}
	& \ \sum_{\pm}R^{(1)}_{1,\pm} E_1\left[\begin{matrix}
		-1 \\ b^{\pm 1}
	\end{matrix}
	\right]
	+ \sum_{\pm} R^{(1)}_{2, \pm}E_1\left[\begin{matrix}
		-1 \\ a_2^{-3}b^{\pm}
	\end{matrix}\right] + \sum_{\pm, k,\ell}R^{(1)}_{3, \pm, k\ell}E_1\left[\begin{matrix}
		-1 \\ a_2^{-\frac{3}{2}}b^{\pm \frac{1}{2}} q^{\frac{1}{4}} q^{\frac{k - 1}{2}} e^{\pi i \ell}
	\end{matrix}\right] \ ,
\end{align}
where we have chosen the reference value $\mathfrak{a}_1 = \mathfrak{a}_2$, which is a zero of the integrand. To integrate these terms with respect to $a_2$, one can first rescale (where applicable) $3\mathfrak{a}_2 \to \mathfrak{a}_2$ and $\frac{3}{2}\mathfrak{a}_2 \to \mathfrak{a}_2$ without affecting the integral. Next, to be able to apply the integration formulas of subsection \ref{integratinfalmostelliptic}, we list the relevant poles and residues in the following table:
{
\renewcommand{\arraystretch}{1.5}
\begin{table}[h!]
\centering
	\begin{tabular}{c|c|c}
		factor & poles & residues\\
		\hline
		$R^{(1)}_{1,\pm}$ & $\mathfrak{a}_2 = 0$ & $ - \frac{i}{6\eta(\tau)} \frac{\vartheta_4( \mathfrak{b})}{\vartheta_4( 3 \mathfrak{b})}$\\
		                  & $\mathfrak{a}_2 = \mp 3 \mathfrak{b} + \frac{\tau}{2}$ & $ + \frac{i}{6\eta(\tau)} \frac{\vartheta_4( \mathfrak{b})}{\vartheta_4( 3 \mathfrak{b})}$\\
    \hline
		$R^{(1)}_{2,\pm}$ & $\mathfrak{a}_2 = 0$ & $ + \frac{i}{6\eta(\tau)} \frac{\vartheta_4( \mathfrak{b})}{\vartheta_4( 3 \mathfrak{b})}$\\
		                  & $\mathfrak{a}_2 = \pm 3 \mathfrak{b} + \frac{\tau}{2}$ & $ - \frac{i}{6\eta(\tau)} \frac{\vartheta_4( \mathfrak{b})}{\vartheta_4( 3 \mathfrak{b})}$\\
		\hline
		$R^{(1)}_{3,\pm,k\ell}$ & $\mathfrak{a}_2 = \mp \frac{3}{2} \gamma \mathfrak{b} + \frac{\tau}{2} + \frac{1}{4}(2k - 1)\gamma \tau + \frac{\ell}{2}$, $\gamma = \pm 1$ & $ \gamma \frac{\vartheta_4(\mathfrak{b})}{12 \vartheta_4(3 \mathfrak{b})}$
	\end{tabular}
\end{table}
}

Finally, putting everything together and simplifying, we obtain the Schur index of $\mathcal N=4$ super Yang-Mills with gauge group $SU(3)$
\begin{align}\label{N4SU3index}
	\mathcal{I}_{\mathcal{N} = 4 \ SU(3)} = - \frac{1}{8} \frac{\vartheta_4(\mathfrak{b})}{\vartheta_4(3 \mathfrak{b})}\left(
	  - \frac{1}{3} + 4 E_1 \left[\begin{matrix}
	  	-1 \\ b
	  \end{matrix}\right]^2 - 4 E_2\left[
	  \begin{matrix}
	  	+ 1 \\ b^2
	  \end{matrix}\right]
	\right) \ .
\end{align}
It is noteworthy that the prefactor is precisely equal to the simultaneous residue of the original integrand.

\subsection{\texorpdfstring{$SO(4)\ \mathcal N=4$}{SO(4) N=4} super Yang-Mills theory}
The gauge group $SO(4)$ is not a simple Lie group: its Lie algebra is isomorphic to $\mathfrak{su}(2) \oplus \mathfrak{su}(2)$. The physical theory is a product theory and its Schur index equals the square of $\mathcal{I}_{\mathcal{N} = 4\ SU(2)}$. This can be easily shown by a change of variables and exploiting the periodicity with respect to the integration variables.

The integral that computes the index reads
\begin{equation}
\mathcal{I}_{\mathcal{N} = 4 \ SO(4)} = \frac{1}{4} \oint_{|a_i| = 1} \prod_{i = 1}^2\frac{da_i}{2\pi i a_i}
 	  \prod_{\alpha, \beta = \pm}\prod_{i < j}  \ \frac{\vartheta_1(\alpha \mathfrak{a}_i + \beta \mathfrak{a}_j)}{\vartheta_4(\alpha \mathfrak{a}_i + \beta \mathfrak{a}_j + \mathfrak{b})} \equiv \oint_{|a_i| = 1} \prod_{i = 1}^2\frac{da_i}{2\pi i a_i} Z(\mathfrak{a}_1, \mathfrak{a}_2)\ .
 \end{equation}
 Changing the variables to
 \begin{equation}
   \mathfrak{b}_1 \equiv \mathfrak{a}_1 + \mathfrak{a}_2, \qquad \mathfrak{b}_2 = \mathfrak{a}_1 - \mathfrak{a}_2 \ ,
 \end{equation}
 the integral now reads
 \begin{equation}
 	\mathcal{I}_{\mathcal{N} = 4 \ SO(4)} = \int_0^2  d \mathfrak{b}_1 \int_{\mathfrak{b}_2 - 2}^{2 - \mathfrak{b}_2} d \mathfrak{b}_2 \left(\frac{1}{2}\right)Z\left(\frac{1}{2} (\mathfrak{b}_1 + \mathfrak{b}_2), \frac{1}{2} (\mathfrak{b}_1 - \mathfrak{b}_2)\right) \ ,
 \end{equation}
where the factor $1/2$ is the Jacobian. Using the periodicity with respect to both $\mathfrak{a}_1, \mathfrak{a}_2$, the integrals decouple
\begin{equation}
	\mathcal{I}_{\mathcal{N} = 4 \ SO(4)}
	= \frac{1}{2^2} \left[\int_0^2 d \mathfrak{b}_1 \frac{\eta(\tau)^3 \vartheta_1(\pm \mathfrak{b}_1)}{2\vartheta_4(\pm \mathfrak{b}_1 + \mathfrak{b})}\right]
	  \left[\int_{-1}^1 d \mathfrak{b}_2\frac{\eta(\tau)^3 \vartheta_1(\pm \mathfrak{b}_2)}{2\vartheta_4(\pm \mathfrak{b}_2 + \mathfrak{b})}\right] \ ,
\end{equation}
from which we recognize that indeed
\begin{equation}
	\mathcal{I}_{\mathcal{N} = 4 \ SO(4)} = \mathcal{I}_{\mathcal{N} = 4 \ SU(2)}(\mathfrak{b})^2 = \frac{1}{4\pi^2} \frac{\vartheta'_4(\mathfrak{b})^2}{\vartheta_1(2 \mathfrak{b})^2}\ .
\end{equation}

\subsection{\texorpdfstring{$SO(5)\ \mathcal N=4$}{SO(5) N=4} super Yang-Mills theory}
Next, we consider $\mathcal N=4$ super Yang-Mills theory with gauge group $SO(5)$. Since $\mathfrak{so}(5)\cong \mathfrak{usp}(4)$, the theory is trivially S-dual to the $USp(4)$ theory. Manipulations similar to the ones in the previous subsection make sure that this expectation is borne out at the level of the index as well. Let's thus focus on the $SO(5)$ theory.

Its Schur index is computed by
\begin{equation}
\mathcal{I}_{\mathcal{N}  = 4 \ SO(5)} = \frac{1}{8} \oint \prod_{A = 1}^2\frac{da_A}{2\pi i a_A} \frac{\eta(\tau)^6}{\vartheta_4(\mathfrak{b})^2} \frac{
    - \vartheta_1(\mathfrak{a}_1)^2
    \vartheta_1(\mathfrak{a}_2)^2
    \vartheta_1(\mathfrak{a}_1 + \mathfrak{a}_2)^2
    \vartheta_1(\mathfrak{a}_1 - \mathfrak{a}_2)^2
  }{
    \vartheta_4(\mathfrak{a}_1 \pm \mathfrak{b})
    \vartheta_4(\mathfrak{a}_2 \pm \mathfrak{b})
    \vartheta_4(\mathfrak{a}_1 + \mathfrak{a}_2\pm \mathfrak{b})
    \vartheta_4(\mathfrak{a}_1 - \mathfrak{a}_2\pm \mathfrak{b})
  }\ .
\end{equation}

Here we simply list the poles and residues relevant for the computation. The poles and residues in the $\mathfrak{a}_1$-integrals are,
{
\renewcommand{\arraystretch}{1.5}
\begin{table}[h!]
\centering
	\begin{tabular}{c|c}
	  poles & residues \\
	  \hline
		$\mathfrak{a}_1 = \pm \mathfrak{b} + \frac{\tau}{2}$, $\alpha = \pm 1$ & $ \frac{i}{8}\eta(\tau)^3 \frac{\vartheta_4(\mathfrak{a}_2 + \alpha \mathfrak{b})\vartheta_4(\mathfrak{a}_2 - \alpha \mathfrak{b})}{
		  \vartheta_1(2 \alpha \mathfrak{b})
		  \vartheta_1(\mathfrak{a}_2 + 2 \alpha \mathfrak{b})
		  \vartheta_1(\mathfrak{a}_2 - 2 \alpha \mathfrak{b})}$\\
		$\mathfrak{a}_1 = \beta \mathfrak{a}_2 + \gamma \mathfrak{b} + \frac{\tau}{2}$, $\beta, \gamma = \pm 1$ & $\frac{i}{8} \eta(\tau)^3 \frac{
		  \vartheta_4(\mathfrak{a}_2 + \beta \gamma \mathfrak{b})
		  \vartheta_1(\mathfrak{a}_2)
		  \vartheta_4(2 \mathfrak{a}_2 + \beta \gamma \mathfrak{b})^2
		}{
		  \vartheta_1(\mathfrak{a}_2 + 2 \beta \gamma \mathfrak{b})
		  \vartheta_4(\mathfrak{a} - \beta \gamma \mathfrak{b})
		  \vartheta_1(2\mathfrak{a} )
		  \vartheta_1(2 \gamma \mathfrak{b}) \vartheta_1(2 \mathfrak{a}_2 + 2 \beta \gamma \mathfrak{b})
		}$
	\end{tabular}
\end{table}
}

\noindent while the poles and residues in the subsequent $\mathfrak{a}_2$-integral are
{
\renewcommand{\arraystretch}{1.5}
\begin{table}[h!]
\centering
	\begin{tabular}{c|c|c}
		factor & poles & residues\\
		\hline
		$ \frac{i}{8} \frac{\eta(\tau)^3\vartheta_4(\mathfrak{a}_2 + \alpha \mathfrak{b})\vartheta_4(\mathfrak{a}_2 - \alpha \mathfrak{b})}{
				  \vartheta_1(2 \alpha \mathfrak{b})
				  \vartheta_1(\mathfrak{a}_2 + 2 \alpha \mathfrak{b})
				  \vartheta_1(\mathfrak{a}_2 - 2 \alpha \mathfrak{b})}$
		& $\mathfrak{a}_2 = 2 \alpha \delta \mathfrak{b}$, $\delta = \pm$ &
		$- \frac{\delta}{8} \frac{\vartheta_4(3 \mathfrak{b})\vartheta_4(\mathfrak{b})}{\vartheta_1(2 \mathfrak{b})\vartheta_1(4 \mathfrak{b})}$\\
		$\frac{i}{8} \frac{
		  \eta(\tau)^3
		  \vartheta_4(\mathfrak{a}_2 + \beta \gamma \mathfrak{b})
		  \vartheta_1(\mathfrak{a}_2)
		  \vartheta_4(2 \mathfrak{a}_2 + \beta \gamma \mathfrak{b})^2
		}{
		  \vartheta_1(\mathfrak{a}_2 + 2 \beta \gamma \mathfrak{b})
		  \vartheta_4(\mathfrak{a} - \beta \gamma \mathfrak{b})
		  \vartheta_1(2\mathfrak{a} )
		  \vartheta_1(2 \gamma \mathfrak{b}) \vartheta_1(2 \mathfrak{a}_2 + 2 \beta \gamma \mathfrak{b})
		}$& $\mathfrak{a}_2 = - 2\beta \gamma \mathfrak{b}$ & $ + \frac{\beta}{8} \frac{\vartheta_4( \mathfrak{b})\vartheta_4(3 \mathfrak{b})}{\vartheta_1(2  \mathfrak{b})\vartheta_1(4  \mathfrak{b})}$\\
		& $\mathfrak{a}_2 = \beta \gamma \mathfrak{b} + \frac{\tau}{2}$ & $- \frac{\beta}{8} \frac{\vartheta_4( \mathfrak{b})\vartheta_4(3  \mathfrak{b})}{\vartheta_1(2 \mathfrak{b})\vartheta_1(4 \mathfrak{b})}$\\
		& $\mathfrak{a}_2 = \frac{\tau}{2}$ & $\frac{\beta \vartheta_4(\mathfrak{b})^2 \vartheta_4(0)}{16 \vartheta_1(2 \mathfrak{b})^2 \vartheta_4(2 \mathfrak{b})}$\\
		& $\mathfrak{a}_2 = \frac{1}{2}$ & $ - \frac{\beta \vartheta_4(\mathfrak{b})
		^2 \vartheta_2(0)}{16 \vartheta_1(2 \mathfrak{b})^2 \vartheta_2(2 \mathfrak{b})}$\\
		& $\mathfrak{a}_2 = \frac{1}{2} + \frac{\tau}{2}$ & $ - \frac{\beta \vartheta_4(\mathfrak{b})^2 \vartheta_3(0)}{16 \vartheta_1(2 \mathfrak{b})^2 \vartheta_3(2 \mathfrak{b})}$\\
		& $\mathfrak{a}_2 = - \beta \gamma \mathfrak{b}$ & $ - \frac{\beta \vartheta_4(\mathfrak{b})^2 \vartheta_4(0)}{16 \vartheta_1(2 \mathfrak{b})^2 \vartheta_4(2 \mathfrak{b})}$\\
		& $\mathfrak{a}_2 = - \beta \gamma \mathfrak{b} + \frac{1}{2}$ & $ + \frac{\beta \vartheta_4(\mathfrak{b})^2 \vartheta_3(0)}{16 \vartheta_1(2 \mathfrak{b})^2 \vartheta_3(2 \mathfrak{b})}$\\
		& $\mathfrak{a}_2 = - \beta \gamma \mathfrak{b} + \frac{1}{2} + \frac{\tau}{2}$ & $ + \frac{\beta \vartheta_4(\mathfrak{b})^2 \vartheta_2(0)}{16 \vartheta_1(2 \mathfrak{b})^2 \vartheta_2(2 \mathfrak{b})}$
	\end{tabular}
\end{table}
}

After performing both integrals, we find
\begin{align}
  \mathcal{I}_{\mathcal{N} = 4 \ SO(5)} = & \ \frac{
    \vartheta_4(\mathfrak{b})\vartheta_4(3\mathfrak{b})
  }{
    16 \vartheta_1(2 \mathfrak{b}) \vartheta_1(4 \mathfrak{b})
  }\left(
    1 + \frac{\vartheta'_4(\mathfrak{b})^2}{\pi^2 \vartheta_4(\mathfrak{b})^2}
    + \frac{\vartheta'_4(\mathfrak{b})\vartheta'_4(3\mathfrak{b})}{\pi^2\vartheta_4(\mathfrak{b})\vartheta_4(3\mathfrak{b})}
    + \frac{\vartheta''_1(2 \mathfrak{b})}{\pi^2 \vartheta_1(2 \mathfrak{b})}
    - \frac{\vartheta''_4(\mathfrak{b})}{\pi^2 \vartheta_4(\mathfrak{b})}
  \right) \nonumber\\
  & \ + \frac{1}{16} \frac{\vartheta_4(\mathfrak{b})^2}{\vartheta_1(2 \mathfrak{b})^2}\left(
    \frac{i}{\pi} \frac{\vartheta'_3(0)}{\vartheta_3(2 \mathfrak{b})}
    - \frac{i}{\pi} \frac{\vartheta'_4(0)}{\vartheta_4(2 \mathfrak{b})}
    + \frac{1}{2\pi^2}\left(
      - \frac{\vartheta''_2(0)}{\vartheta_2(2 \mathfrak{b})}
      - \frac{\vartheta''_3(0)}{\vartheta_3(2 \mathfrak{b})}
      + \frac{\vartheta''_4(0)}{\vartheta_4(2 \mathfrak{b})}
    \right)
  \right) \nonumber\\
  & \ + \frac{1}{32}\vartheta_4(0) \frac{\vartheta_4(\mathfrak{b})^2}{\vartheta_1(2 \mathfrak{b})^2 \vartheta_4(2 \mathfrak{b})}\left(- 1 - \frac{\vartheta''_1(\mathfrak{b})}{\pi^2 \vartheta_1( \mathfrak{b})}\right) \nonumber\\
  & \ + \frac{1}{32}\vartheta_3(0) \frac{\vartheta_4(\mathfrak{b})^2}{\vartheta_1
  (2 \mathfrak{b})^2 \vartheta_3(2 \mathfrak{b})}\left(+ 1 + \frac{\vartheta''_2(\mathfrak{b})}{\pi^2 \vartheta_2( \mathfrak{b})}\right) \nonumber\\
  & \ + \frac{1}{32}\vartheta_2(0) \frac{\vartheta_4(\mathfrak{b})^2}{\vartheta_1(2 \mathfrak{b})^2 \vartheta_2(2 \mathfrak{b})}\left(- 1 + \frac{\vartheta''_3(\mathfrak{b})}{\pi^2 \vartheta_3( \mathfrak{b})}\right)\ . 
\end{align}
Here we chose to express the final result in terms of Jacobi theta functions. Note that unlike the $SU(N)$ cases, different prefactors appear, corresponding to different independent simultaneous residues of the original integral. These are given by (up to signs)
\begin{align}\label{SO5residues}
	\frac{8\vartheta_4(3 \mathfrak{b})\vartheta_4(\mathfrak{b})}{\vartheta_1(2 \mathfrak{b} )\vartheta_1(4 \mathfrak{b})}, \qquad \frac{\vartheta_4(\mathfrak{b})^2 \vartheta_{i = 2,3,4}(0)}{16\vartheta_1(2 \mathfrak{b})^2\vartheta_{i = 2,3,4}(2 \mathfrak{b})} \ .
\end{align}

\subsection{Unflavored indices for \texorpdfstring{$SU(N)\ \mathcal N=4$}{SU(N) N=4} theories}\label{N=4unfl}
Unflavored indices of $\mathcal{N} = 4$ super Yang-Mills theories with gauge group $SU(N)$ were obtained in terms of elliptic integrals in \cite{Bourdier:2015wda}. In \cite{Kang:2021lic}, it was further pointed out that when the gauge groups are $SU(2N + 1)$, the unflavored indices are given by the generating function $\mathfrak{M}_N$ of MacMahon's generalized ``sum-of-divisor'' function,
\begin{align}
	\mathcal{I}_{\mathcal{N} = 4 \ SU(2N + 1)} = q^{- \frac{N(N+1)}{2}} \mathfrak{M}_N\ , \qquad \mathfrak{M}_N \equiv \sum_{0 < n_1 < \ldots < n_N} \frac{q^{n_1 + \ldots + n_N}}{(1 - q^{n_1})^2 \ldots (1 - q^{n_N})^2} \ .
\end{align}

On the other hand, the unflavored limit of the Schur indices computed via our integral formulas are written in terms of Eisenstein series or Jacobi theta functions. By making suitable Ans\"atze, we can increase the rank beyond the rank-two cases for which we have performed analytic computations in the previous subsections. We collect our results for $SU$ gauge groups of low ranks in table \ref{unflavored-indices-N4}. 
{\renewcommand{\arraystretch}{1.5}
\begin{table}[t]
\centering
	\begin{tabular}{c|l}
		$G$ &  Schur index\\
		\hline
		$SU(2)$ & $\frac{1}{4\pi} \frac{\vartheta_4''(0)}{\vartheta'_1(0)}$\\
		$SU(3)$ & $\frac{1}{24} + \frac{1}{2}E_2$\\
		$SU(4)$ & $\frac{\vartheta''_4(0)}{48 \pi \vartheta_1'(0)} + \frac{\vartheta_4''''(0)}{192 \pi^3 \vartheta_1'(0)}$\\
		$SU(5)$ & $\frac{3}{640} + \frac{1}{16}E_2 - \frac{1}{4} [E_4 - \frac{1}{2} (E_2)^2]$\\
		$SU(6)$ & $\frac{1}{360\pi} \frac{\vartheta^{(2)}_4(0)}{\vartheta'_1(0)}
		+ \frac{1}{1152\pi^3} \frac{\vartheta_4^{(4)}(0)}{\vartheta'_1(0)}
		+ \frac{1}{23040\pi^5}\frac{\vartheta_4^{(6)}(0)}{\vartheta'_1(0)}$\\
		$SU(7)$& $\frac{5}{7168} + \frac{37}{3840}E_2 - \frac{5}{96} [E_4-\frac{1}{2} (E_2)^2] + \frac{1}{6} [E_6 - \frac{3}{4} E_4 E_2 + \frac{1}{8} E_2^3] $\\
		$SU(8)$ & $\frac{1}{2240\pi} \frac{\vartheta^{(2)}_4(0)}{\vartheta'_1(0)}
		+ \frac{1}{46080\pi^3} \frac{\vartheta_4^{(4)}(0)}{\vartheta'_1(0)}
		+ \frac{1}{92160\pi^5}\frac{\vartheta_4^{(6)}(0)}{\vartheta'_1(0)}
		+ \frac{1}{5160960\pi^7}\frac{\vartheta_4^{(8)}(0)}{\vartheta'_1(0)}$ \\
		$SU(9)$ & $ \frac{35}{294912} + \frac{3229}{1935360}E_2 -\frac{47}{4608} [E_4-\frac{1}{2} (E_2)^2]+\frac{7}{144} [E_6 -\frac{3}{4} E_4 E_2 + \frac{1}{8} (E_2)^3] $\\
		& $-\frac{1}{8} [E_8 - \frac{2}{3} E_6 E_2 - \frac{1}{4} (E_4)^2 - \frac{1}{48} (E_2)^4 + \frac{1}{4} E_4 (E_2)^2 ] $ \\
		$SU(11)$ & $\frac{63}{2883584} + \frac{10679}{34406400}E_2 - \frac{1571}{774144}(E_4 - \frac{1}{2}(E_2)^2) + \frac{133}{11520}(E_6 - \frac{3}{4}E_4 E_2 + \frac{1}{8})$\\
		& $- \frac{3}{64} (E_8 - \frac{2}{3}E_6E_2- \frac{1}{4}(E_4)^2 + \frac{1}{4}E_4(E_2)62 - \frac{1}{48}(E_2)^4)$\\
		& $ + \frac{1}{10} (E_{10} - \frac{5}{8}E_8 E_2 - \frac{5}{12}E_6E_4 + \frac{5}{24}E_6(E_2)^2 + \frac{5}{32}(E_4)^2 E_2 - \frac{5}{96}E_4 (E_2)^3 + \frac{1}{384} (E_2)^5)$
	\end{tabular}
	\caption{Unflavor indices for various low-rank $\mathcal{N} = 4$ theories with $SU$ gauge groups. In \eqref{SU2N+1 N=4} and \eqref{SU2N N=4} we present conjectures generalizing these closed-form expressions to all ranks. \label{unflavored-indices-N4}}
\end{table}}%
From these results, we observe some clear patterns. For $SU(\text{odd})$ gauge groups, the indices can be organized into the form
\begin{equation}\label{SU2N+1 N=4}
\mathcal{I}_{\mathcal{N} = 4 \ SU(2N + 1)} = (-1)^N\sum_{k = 0}^{N} \frac{\tilde \lambda^{(2N + 3)}_{2k + 2}(2)}{\max(2k, 1)}  \widetilde{\mathbb{E}}_{2k} \ ,
\end{equation}
where $\tilde{\mathbb{E}}$'s are defined in terms of the Weierstrass elliptic $P$-function \eqref{P2}\footnote{It is curious to observe that almost identical expressions $\mathbb{E}_{2k}$ appear in the translation (\ref{EisensteinToTheta-2}) from the twisted Eisenstein series $E_{2k}\Big[\substack{\#\\\#}\Big]$ to Jacobi-theta functions, and that the same numbers $\tilde \lambda$ will appear in the residues of $A_1$-indices, see \eqref{A1-residue-2}.}
\begin{equation}
\widetilde{\mathbb{E}}_0 = 1, \quad \widetilde{\mathbb{E}}_{2k} \colonequals 2k \oint\frac{dy}{2\pi i y} y^{-2k} e^{P_2(y)} = \sum_{\substack{\vec n \\ \sum_{j \ge 1}j  n_j = k}} \prod_{p\ge 1} \frac{1}{n_p !} \left( - \frac{1}{2p}E_{2p}\right)^{n_p} \ .
\end{equation}
Here the summation is over integer partitions $[1^{n_1},2^{n_2},\ldots]$ of $k$, where $n_j$ is the multiplicity of the part of length $j$. The coefficients $\tilde \lambda$ are rational functions defined in terms of \eqref{lambda-def},
\begin{equation}
\tilde \lambda_\ell^{(n)}(K) \colonequals \sum_{\ell' = \max(\ell, 1)}^{n}\bigg(\frac{K}{2}\bigg)^{\ell' - \ell} \frac{1}{(\ell' - \ell)!}\lambda_{\ell'}^{(n)}\ .
\end{equation}

Similarly, the $SU(2N)$ indices with $N = 1, 2, 3, 4$ can be written as
\begin{equation}\label{SU2N N=4}
	\mathcal{I}_{\mathcal{N} = 4 \ SU(2N)}
	=  (-1)^N \sum_{\ell = 1}^{N} \frac{(-1)^{\ell} 2^{2\ell} \tilde \lambda^{(2N+2)}_{2
	\ell+1} (2)}{2(2\ell)!} \left(\frac{1}{4\pi}\right)^{2\ell - 1} \frac{\vartheta_4^{(2\ell)}(0)}{\vartheta_1'(0)} \ .
\end{equation}
We conjecture that this expression captures the unflavored Schur indices of all $\mathcal N=4$ super Yang-Mills theories with $SU(\text{even})$ gauge groups.


\section{Superconformal QCD \label{sec:other}}
A third class of Lagrangian theories we consider is $\mathcal N=2$ superconformal QCD. The case with gauge group $SU(2)$ has been considered in subsection \ref{SU2SCQCD} and was revisited from the point of view of class $\mathcal S$ in section \ref{sec:a1-index}. While by now the computational are familiar, their concrete application tends to be technical, so we will be a bit more schematic in this section.  

\subsection{\texorpdfstring{$SU(3)$}{SU(3)} superconformal QCD\label{sec:SU3-SQCD}}

The Schur index of $SU(3)$ superconformal QCD is given by the contour integral
\begin{equation}
	\mathcal{I}_{SU(3) \ \text{SQCD}} = - \frac{1}{3!} \eta(\tau)^{16} \oint \prod_{A = 1}^2 \frac{da_A}{2\pi i a_A}
	\frac{\prod_{A \ne B} \vartheta_1(\mathfrak{a}_A - \mathfrak{a}_B)}{\prod_{A = 1}^3 \prod_{i = 1}^{6} \vartheta_4(\mathfrak{a}_A - \mathfrak{b}_i)}
	\equiv \oint \prod_{A = 1}^2 \frac{da_A}{2\pi i a_A}Z(\mathfrak{a})\ ,
\end{equation}
where $\mathfrak{a}_3 = - \mathfrak{a}_1 - \mathfrak{a}_2$, $a_3 = (a_1 a_2)^{-1}$.

The $a_1$ integral can be easily performed by considering the (imaginary) poles and residues listed in the following table
{
\renewcommand{\arraystretch}{2}
\begin{table}[h!]
\centering
	\begin{tabular}{c|c}
		poles $(j_1 = 1, \ldots,6)$ & residues\\
		\hline
		$\mathfrak{a}_1 = \mathfrak{b}_{j_1} + \frac{\tau}{2}$
		& $\frac{1}{6} \eta(\tau)^{13}q^{\frac{1}{8}}
	\frac{
	  \prod_{A\ne B} \vartheta_1(\mathfrak{a}_A - \mathfrak{a}_B)|_{\mathfrak{a}_1 = \mathfrak{b}_{j_1} + \frac{\tau}{2}}
	}{
	  \prod_i\vartheta_4(\mathfrak{a}_2 - \mathfrak{b}_i)
	  \prod_i\vartheta_4(\mathfrak{a}_2 + \mathfrak{b}_{j_1} + \mathfrak{b}_i + \frac{\tau}{2})
	  \prod_{i \ne j_1}\vartheta_4(\mathfrak{b}_i - \mathfrak{b}_{j_1} - \frac{\tau}{2})
	} $\\
	$\mathfrak{a}_1 = - \mathfrak{a}_2 - \mathfrak{b}_{j_1} + \frac{\tau}{2}$
	& $\frac{1}{6}\eta(\tau)^{13}q^{ \frac{1}{8}} \frac{
		\prod_{A \ne B}\vartheta_1(\mathfrak{a}_A - \mathfrak{a}_B)|_{\mathfrak{a}_1 = - \mathfrak{a}_2 - \mathfrak{b}_{j_1} + \frac{\tau}{2} }
	}{
		\prod_{i}\vartheta_4(\mathfrak{a}_2 - \mathfrak{b}_i)
	  \prod_{i}\vartheta_4(\mathfrak{a}_2 + \mathfrak{b}_{j_1} + \mathfrak{b}_i - \frac{\tau}{2})
	  \prod_{i \ne j_1} \vartheta_4(\mathfrak{b}_i - \mathfrak{b}_{j_1} + \frac{\tau}{2})
	} $
	\end{tabular}
\end{table}
}

\noindent Note that the two residues sharing the same index $j_1$ are opposite to one another. We define
\begin{equation}
	R_{j_1} \colonequals \mathop{\operatorname{Res}}\limits_{\mathfrak{a}_1 = \mathfrak{b}_{j_1} + \frac{\tau}{2}}Z(\mathfrak{a}) = - \mathop{\operatorname{Res}}\limits_{\mathfrak{a}_1 = - \mathfrak{a}_2 - \mathfrak{b}_{j_1} + \frac{\tau}{2}}Z(\mathfrak{a}) \ .
\end{equation}
Choosing the reference value $\mathfrak{a}_1 = 0$, the index becomes after the $a_1$-integration
\begin{equation}
	\mathcal{I}_{SU(3) \ \text{SQCD}} = \oint \frac{da_2}{2\pi i a_2} (\mathfrak{R}_0Z + \mathfrak{R}_1 Z + \mathfrak{R}_2 Z) \ ,
\end{equation}
where
\begin{equation}
	\mathfrak{R}_0 Z \colonequals  Z(\mathfrak{a}_1 = 0), \qquad
	\mathfrak{R}_1 Z \colonequals \sum_{j_1 = 1}^{6} R_{j_1} E_1\left[\begin{matrix}
		-1 \\ b_{j_1}
	\end{matrix}\right] \ , \qquad
	\mathfrak{R}_2 Z \colonequals  \sum_{j_1 = 1}^{6} R_{j_1} E_1\left[\begin{matrix}
		-1 \\ a_2b_{j_1}
	\end{matrix}\right] \ .
\end{equation}

The $a_2$-integral picks up the following poles and residues:

{
\renewcommand{\arraystretch}{1.8}
\begin{table}[h!]
\centering
	\begin{tabular}{c|c|c}
		factor & poles & residues\\
		\hline
		$\mathfrak{R}_0Z$
		& $\mathfrak{a}_2 = \pm \mathfrak{b}_{j_2} + \frac{\tau}{2}$
		& $ \pm R_{0 j_2} \colonequals \pm \frac{
		  i \eta(\tau)^{13}\vartheta_1(2 \mathfrak{b}_{j_2})
		  \vartheta_4(\mathfrak{b}_{j_2})^3
		}{
		  6 \prod_{i \ne j_2}\vartheta_1(\mathfrak{b}_{j_2} - \mathfrak{b}_i)
		  \vartheta_1(\mathfrak{b}_{j_2} + \mathfrak{b}_i)
		  \prod_{i \ne j_2}\vartheta_4(\mathfrak{b}_i)
		}$ \\
		\hline
		$R_{j_1}$
		& $\mathfrak{b}_{j_2} + \frac{\tau}{2}, \  j_2 \ne j_1$
		& $ R_{j_1 j_2} \colonequals \frac{
		  \eta(\tau)^{10}
		  \vartheta_4(2 \mathfrak{b}_{j_1} + \mathfrak{b}_{j_2})
		  \vartheta_4(\mathfrak{b}_{j_1} + 2\mathfrak{b}_{j_2})}{
	  6
	  \prod_{i\ne j_1, j_2}\vartheta_1(\mathfrak{b}_{j_1} - \mathfrak{b}_i)\vartheta_1(\mathfrak{b}_{j_2} - \mathfrak{b}_i)
	  \prod_{i \ne j_1, j_2} \vartheta_4(\mathfrak{b}_{j_1}+ \mathfrak{b}_{j_2} + \mathfrak{b}_i)
	}$\\
		& $ - \mathfrak{b}_{j_1} - \mathfrak{b}_{j_2}, \  j_2 \ne j_1$
		& $ - R_{j_1 j_2}$\\
	\end{tabular}
\end{table}
}

\noindent Note that
\begin{align}
	\mathop{\operatorname{Res}}\limits_{\mathfrak{a}_2 = \mathfrak{b}_{j_2} + \frac{\tau}{2}}R_{j_1}
	= - \mathop{\operatorname{Res}}\limits_{\mathfrak{a}_2 = - \mathfrak{b}_{j_ 1} - \mathfrak{b}_{j_2}} R_{j_1}
	= \left\{ \begin{array}{c c}
		R_{j_1 j_2} & j_2 \ne j_1 \\ 0 & j_2 = j_1
	\end{array}
	\right. \ .
\end{align}
Choosing again $\mathfrak{a}_2 = 0$ as the reference value which happens to be a zero of $Z(\mathfrak{a}_1 = 0)$, the index can be computed by finishing the $\ \mathfrak{a}_2$-integral (with implicitly $R_{j_1 j_2} = 0$ when $j_1 = j_2$), 
\begin{align}
	\mathcal{I}_{SU(3) \ \text{SQCD}} = & \ \sum_{j_2 = 1}^{6}2R_{0j_2}E_1\left[\begin{matrix}
		-1 \\ b_{j_2}
	\end{matrix}\right] \nonumber \\
	& \ + \sum_{j_1}^{6}
	  \left(
	  R_{j_1}(\mathfrak{a}_2 = 0) + \sum_{j_2 = 1}^6
	  \Big(  
	  R_{j_1 j_2}E_1\left[\begin{matrix}
	  	-1 \\ b_{j_2}
	  \end{matrix}\right]
	  + R_{j_1 j_2} E_1\left[\begin{matrix}
	  	-1 \\ b_{j_1}b_{j_2}q^{ - \frac{1}{2}}
	  \end{matrix}\right]  \Big)
	\right)E_1\left[\begin{matrix}
		-1 \\ b_{j_1}
	\end{matrix}\right]\nonumber\\
	& \ + \sum_{j_1, j_2 = 1}^{6} R_{j_1 j_2}\left(
			E_2\left[\begin{matrix}
				1 \\ b_{j_1} b_{j_2}
			\end{matrix}\right]
			- E_2\left[
			\begin{matrix}
				1 \\ b_{j_2}q^{ - \frac{1}{2}}
			\end{matrix}\right]
		\right)  \ ,
\end{align}
where we noticed that
\begin{align}
	R_{0j} \equiv \mathop{\operatorname{Res}}\limits_{\mathfrak{a}_2 \to b_{j} + \frac{\tau}{2}} Z(\mathfrak{a}_1 = 0) = \left[\mathop{\operatorname{Res}}\limits_{\mathfrak{a}_1 \to b_{j} + \frac{\tau}{2}} Z(\mathfrak{a})\right]_{\mathfrak{a}_2 = 0 } \ .
\end{align}

$SU(3)$ superconformal QCD is a theory of class $\mathcal{S}$ of type $\mathfrak a_2$ associated with a four-punctured spheres with two maximal and two minimal punctures. The manifest flavor symmetry is $SU(3)_{c^{(1)}_1, c^{(1)}_2} \times U(1)_{d^{(1)}} \times SU(3)_{c^{(2)}_1, c^{(2)}_2} \times U(1)_{d^{(2)}} $ where we assigned names to the fugacities of the respective flavor symmetry factors in their subscript. They are related to the $b$'s we have used above as
\begin{align}
  c^{(1)}_1 = & \ b_1/d^{(1)}, \qquad c^{(1)}_2 = b_2 / d^{(1)}, \qquad
  d^{(1)} = (b_1 b_2 b_3)^{1/3} \ ,\\
  c^{(2)}_1 = & \ b_4/d^{(2)}, \qquad c^{(2)}_2 = b_5/d^{(2)}, \qquad
  d^{(2)} =  (b_4 b_5 b_6)^{1/3} \ .
\end{align}
With this parametrization, we shall denote the index as $\mathcal{I}_{SU(3) \ \text{SQCD}}(c^{(1)}, c^{(2)}, d^{(1)}, d^{(2)})$, and it will be used to compute the Schur index of the $E_6$ SCFT in the next section.

\subsection{\texorpdfstring{$SU(4)$}{SU(4)} superconformal QCD}

The Schur index of $SU(4)$ superconformal QCD is computed by the integral
\begin{equation}
	\mathcal{I}_{SU(4) \ \text{SQCD}} = + \frac{1}{4!} \eta(\tau)^{26} \oint \prod_{A = 1}^3 \frac{da_A}{2\pi i a_A}
	\frac{\prod_{A \ne B} \vartheta_1(\mathfrak{a}_A - \mathfrak{a}_B)}{\prod_{A = 1}^3 \prod_{i = 1}^{6} \vartheta_4(\mathfrak{a}_A - \mathfrak{b}_i)}
	\equiv \oint \prod_{A = 1}^3 \frac{da_A}{2\pi i a_A}Z(\mathfrak{a})\ ,
\end{equation}
where $\mathfrak{a}_4 = - \mathfrak{a}_1 - \mathfrak{a}_2 - \mathfrak{a}_3$. 

We simply list the relevant poles and residues as we integrate $\mathfrak{a}_1$, $\mathfrak{a}_2$, $\mathfrak{a}_3$ one after another.
{
\renewcommand{\arraystretch}{1.5}
\begin{table}[h!]
\centering
	\begin{tabular}{c|c|c}
		factor & $\mathfrak{a}_1$-poles & residues\\
		\hline
		$Z$ & $\mathfrak{a}_1 = \mathfrak{b}_{j_1} + \frac{\tau}{2}$ & $R_{j_1}$ \\
		& $\mathfrak{a}_1 = - \mathfrak{a}_2 - \mathfrak{a}_3 - \mathfrak{b}_{j_1} + \frac{\tau}{2}$ & $ - R_{j_1}$\\
		\hline
		\hline
		factor & $\mathfrak{a}_2$-poles & residues\\
		\hline
		$Z( \mathfrak{a}_1 = 0)$ & $\mathfrak{a}_2 = \mathfrak{b}_{j_2} + \frac{\tau}{2}$ & $R_{0 j_2}$ \\
		& $\mathfrak{a}_2 = - \mathfrak{a}_3 - \mathfrak{b}_{j_1} + \frac{\tau}{2}$ & $- R_{0 j_2}$\\
		$R_{j_1}$ & $\mathfrak{a}_2 = \mathfrak{b}_{j_2} + \frac{\tau}{2}$ & $R_{j_1 j_2}$ \\
		& $\mathfrak{a}_2 = - \mathfrak{a}_3 - \mathfrak{b}_{j_1} - \mathfrak{b}_{j_2}$ & $- R_{j_1 j_2}$\\
		\hline
		\hline
		factor & $\mathfrak{a}_3$-poles & residues\\
		\hline
		$R_{j_1 j_2}$ & $\mathfrak{a}_3 = \mathfrak{b}_{j_3} + \frac{\tau}{2}$ & $R_{j_1 j_2 j_3}$\\
		              & $\mathfrak{a}_3 = - \mathfrak{b}_{j_1}- \mathfrak{b}_{j_2}- \mathfrak{b}_{j_3} + \frac{\tau}{2}$ & $ - R_{j_1 j_2 j_3}$\\
	\end{tabular}
\end{table}
}

\noindent In the above, $R_{j}, R_{j_1 j_2} $, $ R_{0j} = R_{j0}$ and so forth denotes residues of the integrand. Using also that
\begin{align}
	R_{j 0} \colonequals & \ R_{j}(\mathfrak{a}_2 = 0) = \operatorname{Res}_{\mathfrak{a}_2 = \mathfrak{b}_j + \frac{\tau}{2}}Z(\mathfrak{a}_1 = 0) = R_{0 j} \\
	R_{0j_2 j_3} \colonequals & \ \operatorname{Res}_{\substack{\mathfrak{a}_2 = \mathfrak{b}_{j_2} + \frac{\tau}{2}\\\mathfrak{a}_3 = \mathfrak{b}_{j_3} + \frac{\tau}{2}}}Z(\mathfrak{a}_1 = 0)  \ ,
\end{align}
and similarly for $R_{j_1 0 j_3}$, $R_{j_1 j_2 0}$, we find
\begin{align}
	\mathcal{I}_{SU(4) \ \text{SQCD}} = & \ \sum_{j_1, j_2, j_3 = 0}^{8} R_{j_1j_2j_3}\prod_{A = 1}^{3}E_1\left[\begin{matrix}
		-1 \\ b_{j_A}
	\end{matrix}\right]
	+ \sum_{\substack{j_1, j_2 = 0 \\ j_3 = 1}}^{8} (R_{j_1 j_2 j_3}) E_1\left[\begin{matrix}
		-1 \\ b_{j_1}
	\end{matrix}\right]
	E_1\left[\begin{matrix}
		-1 \\ b_{j_2}
	\end{matrix}\right]
	E_1\left[\begin{matrix}
		-1 \\ b_{j_1}b_{j_2}b_{j_3}
	\end{matrix}\right] \nonumber\\
	& - \sum_{\substack{j_1 = 0\\j_2,j_3 = 1}}^8 R_{j_1 j_2 j_3} \left(
		E_2\left[\begin{matrix}
			1 \\ b_{j_1}b_{j_2}b_{j_3} q^{- \frac{1}{2}}
		\end{matrix}\right]
		+ E_2\left[\begin{matrix}
			1 \\ b_{j_3} q^{ \frac{1}{2}}
		\end{matrix}\right]
	\right) \nonumber\\
	& \ + \sum_{\substack{j_3 = 0\\j_1 j_2 = 1}}^{8} R_{j_1 j_2 j_3}E_1\left[\begin{matrix}
		-1 \\ b_{j_3}
	\end{matrix}\right]E_2\left[\begin{matrix}
		1 \\ b_{j_1}^{-1 }q^{\frac{1}{2}}
	\end{matrix}\right]
	+ \sum_{j_1, j_2, j_3 = 1}^8 R_{j_1 j_2 j_3} E_1\left[\begin{matrix}
		-1 \\ b_{j_1} b_{j_2}b_{j_3}
	\end{matrix}\right]E_2\left[\begin{matrix}
		1 \\ b_{j_1}^{-1 }q^{\frac{1}{2}}
	\end{matrix}\right] \nonumber\\
	& \ + \frac{1}{12} \sum_{j_1, j_2 = 1}^{8}\left(
	  \sum_{j_3 = 0}^8 R_{j_1 j_2 j_3}E_1\left[\begin{matrix}
	  	-1 \\ b_{j_3}
	  \end{matrix}\right]
	  + \sum_{j_3 = 1}^{8} R_{j_1 j_2 j_3} E_1\left[
	  \begin{matrix}
	  	-1 \\ b_{j_1} b_{j_2} b_{j_3}
	  \end{matrix}\right]
	\right) \nonumber \\
	& \ + \sum_{j_1, j_2,j_3 = 1}^{8}R_{j_1 j_2j_3}\left(
    - \frac{1}{8} E_1\left[\begin{matrix}
    	-1 \\ b_{j_1} b_{j_2} b_{j_3}
    \end{matrix}\right]
    - \frac{1}{24} E_3\left[\begin{matrix}
    	-1 \\ b_{j_1}b_{j_2}b_{j_3}
    \end{matrix}\right]
	\right)  \ .
\end{align}

\subsection{\texorpdfstring{$SU(N)$}{SU(N)} superconformal QCD}
So far, we haven't encountered an obstruction to evaluating Schur indices using the integral formulas presented in section \ref{sec:integrals}. However, the formulas given there only allow us to compute integrals whose integrand is an elliptic function times an Eisenstein series. When performing the integrals defining the Schur index one by one, a situation could arise where the integrand contains, apart from an elliptic function, products of Eisenstein series several of which contain the integration variable. The integration formulas of section \ref{sec:integrals} are insufficient to evaluate such integrals. In this section we show that this situation does not occur for $SU(N \ge 4)$ superconformal QCD and thus that all their Schur indices can in principle be evaluated in closed-form using the technology developed in this paper. In section \ref{nonLAgr}, however, we will encounter a situation where more general integration formulas are needed.

The integral we would like to evaluate is
\begin{equation}
	\mathcal{I}_{SU(N) \ \text{SQCD}} =  \frac{(-i)^{(N - N^2)} \eta(\tau)^{-2 + 3 N + N^2}}{N!}  \oint \prod_{A = 1}^N \frac{da_A}{2\pi i a_A}
	\frac{\prod_{A \ne B} \vartheta_1(\mathfrak{a}_A - \mathfrak{a}_B)}{\prod_{A = 1}^N \prod_{i = 1}^{2N} \vartheta_4(\mathfrak{a}_A - \mathfrak{b}_i)}\;,
\end{equation}
where again $\mathfrak{a}_N = - \sum_{A=1}^{N-1} \mathfrak a_A$. We will denote the full integrand as $Z$. If we compute the integrals of $\mathfrak{a}_1, \ldots, \mathfrak{a}_{N - 1}$ one after another, then the first integral picks up two types of poles,
\begin{equation}
  \mathfrak{a}_1 = \mathfrak{p}^{(1)}_{j_1} \colonequals \mathfrak{b}_{j_1} + \frac{\tau}{2}, \qquad \mathfrak{a}_ 1 = \tilde {\mathfrak{p}}^{(1)}_{j_1} \colonequals - \mathfrak{b}_{j_1} - \mathfrak{a}_2 - \ldots - \mathfrak{a}_{N - 1} + \frac{\tau}{2} \ , \qquad j_1 = 1, \ldots, 2N \ .
\end{equation}
It can be verified that the residues of $\mathfrak{p}^{(1)}_{j_1}$ and $\tilde {\mathfrak{p}}^{(1)}_{j_1}$ are opposite. As before, we denote them $R_{j_1}$ and $- R_{j_1}$, and the first integral then equals
\begin{equation}
Z(\mathfrak{a}_1 = 0)
	+ \sum_{j_1 = 1}^{2N} R_{j_1}E_1\left[\begin{matrix}
		-1 \\ b_{j_1}
	\end{matrix}\right]
	+ \sum_{j_1}^{2N} R_{j_1}E_1\left[\begin{matrix}
		-1 \\ a_2 \ldots a_{N - 1}b_{j_1}
	\end{matrix}\right]\;.
\end{equation}
If we define $\mathfrak{p}^{(1)}_0 = 0, R_{0} = Z(\mathfrak{a}_1 = 0), E_1\left[\substack{-1 \\ b_0}\right] = 1$, it can be written more succinctly as 
\begin{equation}
\sum_{j_1 = 0}^{2N} R_{j_1}E_1\left[\begin{matrix}
		-1 \\ b_{j_1}
	\end{matrix}\right]
	+ \sum_{j_1}^{2N} R_{j_1}E_1\left[\begin{matrix}
		-1 \\ a_2 \ldots a_{N - 1}b_{j_1}
	\end{matrix}\right] \ .
\end{equation}
Note that the residues $R_{j_1}$ contain $\mathfrak{a}_2$ in the denominator via two types of factors,
\begin{align}
	\prod_{j = 1}^{2N}\vartheta_4(\mathfrak{a}_2 - \mathfrak{b}_j), \qquad \text{and}, \qquad
	\prod_{j = 1}^{2N} \vartheta_4( - \mathfrak{b}_{j_1} - \frac{\tau}{2} - \mathfrak{a}_2 - \ldots - \mathfrak{a}_{N - 1} - \mathfrak{b}_j) \ ,
\end{align}
where the first factor was present in the original integrand, while the second comes from evaluating $\prod_{j = 1}^{2N}\vartheta_4(\mathfrak{a}_N - \mathfrak{b}_j)$ at $\mathfrak{p}_{j_1}$. These factors lead to two types of $\mathfrak{a}_2$-poles,
\begin{align}
	\mathfrak{a}_2 = \mathfrak{p}^{(2)}_{j_2} \equiv  \mathfrak{b}_{j_2} + \frac{\tau}{2}, \qquad
	\mathfrak{a}_2 = \tilde {\mathfrak{p}}^{(2)}_{j_2} = - \mathfrak{b}_{j_2} - \mathfrak{a}_1 - \mathfrak{a}_3 - \ldots + \frac{\tau}{2} \Big|_{\mathfrak{a}_1 = \mathfrak{p}^{(1)}_{j_1}} \ .
\end{align}
The corresponding residues are
\begin{align}
	R_{j_2j_1} \colonequals \mathop{\operatorname{Res}}_{\mathfrak{a}_2 = \mathfrak{p}^{(2)}_{j_2}} R_{j_1} , \qquad \text{and} \qquad \mathop{\operatorname{Res}}_{\mathfrak{a}_2 = \tilde{ \mathfrak{p}}^{(2)}_{j_2}} R_{j_1} = - R_{j_2j_1}\  .
\end{align}
Furthermore, the variables $\mathfrak{a}_{A > 1}$ appear in a single Eisenstein series inside each term, and they are organized in the product $\prod_{A > 1}a_A$. Therefore $\mathfrak{a}_{A > 1}$, and $\mathfrak{a}_2$ in particular, can be further integrated with our integration formula.

Let us now show that one can carry out all $N - 1$ integrals inductively. Suppose we can perform the $\mathfrak{a}_{1, \ldots, k}$-integrals as above. It is not difficult to see that the residue (and we also replace $\mathop{\operatorname{Res}}_{\mathfrak{a}_{A} = \mathfrak{b}_{j_{A}} + \frac{\tau}{2}}$ by $\lim _{\mathfrak{a}_{A} \to 0}$ when $j_A = 0$)
\begin{align}
	R_{j_k \ldots j_2j_1} \equiv \mathop{\operatorname{Res}}_{\mathfrak{a}_k = \mathfrak{b}_{j_k} + \frac{\tau}{2}} \ldots \mathop{\operatorname{Res}}_{\mathfrak{a}_k = \mathfrak{b}_{j_1} + \frac{\tau}{2}} (\text{integrand})\ ,
\end{align}
has two types of poles in $\mathfrak{a}_{k + 1}$,
\begin{align}
	\mathfrak{p}^{(k+1)}_{j_{k + 1}} \equiv \mathfrak{b}_{j + 1} + \frac{\tau}{2}\ ,\quad
  \tilde{\mathfrak{p}}^{(k+1)}_{j_{k + 1}} \equiv - \mathfrak{a}_1 - \ldots - \mathfrak{a}_{k} + \mathfrak{a}_{k + 2} + \ldots \mathfrak{a}_{N - 1} + \frac{\tau}{2}\ ,
\end{align}
where again the $\mathfrak{a}_1, \ldots, \mathfrak{a}_k$ should be properly substitute. Assume that the result of the first $k$-integrals is a sum of terms of the form
\begin{align}
	\ldots + R_{j_k \ldots j_1} E_*\left[\begin{matrix}
		* \\ *
	\end{matrix}\right] \ldots  E_*\left[\begin{matrix}
		* \\ *
	\end{matrix}\right] + \ldots
\end{align}
where at most only one $E_*$ in each product contains the combination $\prod_{A > k} a_A$, while the remaining $E_*$ factors depend only on $\mathfrak{b}$'s. One can perform the $a_{k + 1}$ integral because
\begin{itemize}
\item If the product of $E_*$'s inside a term is completely independent of $a_{k > A}$, then the integral will produce
\begin{align}
	\left(R_{j_{k + 1} j_k \ldots j_1} E_1\left[\begin{matrix}
	 	 		-1 \\ b_{j_{k + 1}}
	 	 	\end{matrix}\right]
	 	 	+ R_{j_{k + 1} j_k \ldots j_1} E_1\left[\begin{matrix}
	 	 		-1 \\ a_{k + 2} \ldots a_{N - 1} \times (b\text{'s})
	 	 	\end{matrix}\right]
	\right)
	E_*\left[\begin{matrix}
		* \\ *
	\end{matrix}\right]\ldots 
	E_*\left[\begin{matrix}
		* \\ *
	\end{matrix}\right] \ ,
\end{align}
where we also denote the $R_{0 j_k \ldots j_1} = R_{j_k \ldots j_1}(\mathfrak{a}_{k + 1} = 0)$, $E_1\left[\substack{-1 \\ b_0}\right] = 1$.
\item If, instead, one $E_*$ factor depends on some combination $a_{k + 1} \ldots a_{N - 1}$, then the integration of that term leads to terms of the form
\begin{align}
	R_{j_{k + 1}j_k\ldots j_1} E_*\left[\begin{matrix}
		* \\ *
	\end{matrix}\right]\ldots E_{* + 1}\left[\begin{matrix}
		* \\ a_{k + 2}\ldots a_{N - 1}\times b\text{'s}
	\end{matrix}\right], \quad \text{from} \quad \mathfrak{a}_{k + 1} \to \mathfrak{p}^{(k + 1)}_{j_{k + 1}}, \\
	R_{j_{k + 1}\ldots j_1} E_*\left[\begin{matrix}
		* \\ *
	\end{matrix}\right]\ldots E_{* + 1}\left[\begin{matrix}
		* \\ b\text{'s}
	\end{matrix}\right], \quad \text{from} \quad \mathfrak{a}_{k + 1} \to \tilde{\mathfrak{p}}^{(k + 1)}_{j_{k + 1}} \\
	R_{j_{k + 1}\ldots j_1} E_*\left[\begin{matrix}
		* \\ *
	\end{matrix}\right]\ldots E_{1}\left[\begin{matrix}
		-1 \\ b_{j_{k + 1}}
	\end{matrix}\right] \ , \quad \text{from} \quad \mathfrak{a}_{k + 1} \to \mathfrak{p}^{(k + 1)}_{j_{k + 1}}\\
	R_{j_{k + 1}\ldots j_1} E_*\left[\begin{matrix}
		* \\ *
	\end{matrix}\right]\ldots E_{1}\left[\begin{matrix}
		-1 \\ a_{k + 2} \ldots a_{N - 1} \times (b\text{'s})
	\end{matrix}\right] \ , \quad \text{from} \quad \mathfrak{a}_{k + 1} \to \tilde{\mathfrak{p}}^{(k + 1)}_{j_{k + 1}} \ .
\end{align}
\end{itemize}

Now we see that in each term that is generated by the $\mathfrak{a}_{k + 1}$-integral, only one $E_*$ factor depends on the combination $a_{k + 2}\ldots a_{N - 1}$, and one can keep on integrating $\mathfrak{a}_{k + 2}$, and so forth. However, deriving a compact closed-form expression for all the superconformal QCD indices has proved to be challenging. We hope to return to this in the future.


\section{Applications}\label{applications}

In this section we consider three applications of the closed-form expressions for the Schur index derived in the previous sections. In particular, we derive closed-form expressions for several non-Lagrangian theories, we consider the modular properties of the Schur index, which are of particular importance when viewing the Schur index as the vacuum character of a vertex operator algebra via the SCFT/VOA correspondence, and finally we look at defect indices. These applications are meant to illustrate the usefulness of our closed-form expressions, but are not intended as an exhaustive study of each of these topics.

\subsection{Non-Lagrangian theories}\label{nonLAgr}

In the previous sections, we have focused on evaluating the Schur index of Lagrangian theories, as those are naturally computed by integrals of multi-variate elliptic functions. However, the Schur indices of several non-Lagrangian theories are algebraically related to indices of Lagrangian theories and can thus be derived in closed form as well.

The first example is the trinion theory $T_3$, \ie{}, the theory of class $\mathcal{S}$ of type $\mathfrak a_2$ associated with a sphere with three maximal punctures. This theory can be identified with the $E_6$ superconformal field theory of Minahan and Nemeschansky \cite{Minahan:1996fg}. What's more, in \cite{Gadde:2010te,Razamat:2012uv}, it was shown that its Schur index $\mathcal{I}_{T^3}$ can be expressed as a finite sum involving the index of $SU(3)$ superconformal QCD. This comes about as follows: one starts by observing that $SU(3)$ superconformal QCD is of class $\mathcal{S}$ of type $\mathfrak a_2$, associated to a Riemann sphere with two maximal and two minimal punctures. The theory has two interesting S-duality frames -- in fact, their duality is the well-known Argyres-Seiberg duality \cite{Argyres:2007cn}. The first frame is the gauge theory description in terms of an $SU(3)$ gauge theory with six fundamental flavors. The second one involves the $E_6$ superconformal field theory and a hypermultiplet with $SU(2)$ flavor symmetry. These are gauged together along an $SU(2)$ (sub)group of their respective flavor symmetries. At the level of the (Schur) index, the integral implementing this latter gauging can be inverted \cite{2004math.....11044S}. Using the result in section \ref{sec:SU3-SQCD} we then find
\begin{align}
  & \ \mathcal{I}_{E_6}(\vec c^{(1)}, \vec c^{(2)}, (wr, w^{-1}r, r^{-2})) \nonumber\\
  = & \ \frac{\mathcal{I}_{SU(3) \ \text{SQCD}}(\vec c^{(1)}, \vec c^{(2)}, \frac{w^{\frac{1}{3}}}{r}, \frac{w^{- \frac{1}{3}}}{r})_{w \to q^{\frac{1}{2}}w}}{\theta(w^2)} + \frac{\mathcal{I}_{SU(3) \ \text{SQCD}}(\vec c^{(1)}, \vec c^{(2)}, \frac{w^{\frac{1}{3}}}{r}, \frac{w^{- \frac{1}{3}}}{r})_{w \to q^{ - \frac{1}{2}}w}}{\theta(w^{ - 2})} \ ,
\end{align}
where $(wr, w^{-1}r, r^{-2})$ denotes an $SU(3)$ fugacity, and the theta function $\theta(z)$ is defined by
\begin{equation}
	\theta(z) \equiv \frac{\vartheta_1( \mathfrak{z})}{i z^{-\frac{1}{2}} q^{\frac{1}{8}} (q;q)} \ .
\end{equation}
At first sight, we can use this result for the index of the trinion theory $T_3$ to compute the indices of all theories of class $\mathcal S$ of type $\mathfrak a_2$. However, it turns out that one runs into integrals that cannot be evaluated with the integration formulas presented in section \ref{sec:integrals}. In particular, one encounters integrals whose integrand is an elliptic function multiplied by a product of Eisenstein series several of which contain the integration variable. Hence, a systematic computation of indices of theories of class $\mathcal{S}$ of type $\mathfrak{a}_2$ is beyond the scope of this paper.\footnote{We have been informed that \cite{beemetal} has obtained results for unflavored Schur indices of a number of higher-rank theories of class $\mathcal S.$}

Similarly, we can derive closed-form expressions for the indices of the theories of class $\mathcal S$ often denoted as $R_{0,N}$, and in particular, using these results, we can also evaluate the Schur index of the $E_7$ Minahan-Nemeschansky theory \cite{Agarwal:2018ejn}. As a theory of class $\mathcal{S}$, $R_{0, N}$ correspond to a sphere with two maximal punctures and one puncture associated with the partition $[N-2, 1^2]$. It arises in the strong-coupling limit of $SU(N)$ superconformal QCD in very much the same way as the $E_6$ theory appeared after applying Argyres-Seiberg duality: gauging the diagonal of an $SU(2)$ subgroup of the flavor symmetry of $R_{0, N}$ and the $SU(2)$ flavor symmetry of a hypermultiplet describes an S-duality frame of $SU(N)$ superconformal QCD.  Hence, using the Spiridonov-Warnaar inversion formula, one can obtain the $R_{0, N}$ indices from those of $SU(N)$ superconformal QCD. Finally, one derives the index of the $E_7$ Minahan-Nemeschansky theory by Higgsing the $R_{0,4}$ theory.

Another series of non-Lagrangian theories whose Schur indices are related to those of Lagrangian theories were discussed in \cite{Closset:2020afy,Kang:2021lic}. The theories in question are defined by conformally gauging different sets of $D_p(G)$ superconformal field theories \cite{Cecotti:2012jx,Cecotti:2013lda}.\footnote{The $D_p(G)$ theories can be given a class $\mathcal{S}$ description in terms of a regular full puncture (hence providing a $G$ flavor subgroup) and one irregular puncture, often denoted as $(G^b[p - h^\vee], F)$.} Due to the restrictions considered in \cite{Cecotti:2013lda,Kang:2021lic}, at most four $D_{p_i}(G)$ theories can be gauged along their common $G$-symmetry forming a quiver structure $\widehat{\Gamma}[G]$ with one gauge node and four (or less) $D_{p_i}(G)$-legs, where $\widehat{\Gamma} = D_4, E_{6,7,8}$. It was pointed out in \cite{Kang:2021lic} that for a set of $\Gamma$ and $G$ such that $\widehat{\Gamma}[G]$ is flavorless, the Schur index $\mathcal{I}_{\widehat{\Gamma}[G]}$ is actually related to the one of $\mathcal{N} = 4$ super Yang-Mills with gauge group $G$ as, 
\begin{equation}
	\mathcal{I}_{\widehat{\Gamma}[G]}(q) \sim \mathcal{I}_{\mathcal{N} = 4 \ G}(b = q^{\frac{\alpha_\Gamma}{2} - 1}, q \to q^{\alpha_\Gamma}) \ .
\end{equation}
Here $\alpha_\Gamma$ is the largest comark associated with the affine Dynkin diagram $\widehat{\Gamma}$, and more explicitly, $\alpha_{D_4} = 2$, $\alpha_{E_6} = 3$, $\alpha_{E_7} = 4$, $\alpha_{E_8} = 6$. Applying our closed-form expressions for the $\mathcal{N} = 4$ indices, we have for example
\begin{equation}
  \mathcal{I}_{\widehat{D_4}[SU(3)]}
  =  q \mathcal{I}_{\mathcal{N} = 4 \ SU(3)}(b = 1, q^2) = \frac{1}{24} + \frac{1}{2}E_2(2 \tau)\ .
\end{equation}
Denoting $\widehat{\vartheta}_i(z) \equiv \vartheta_i(z, 4\tau)$, one also finds
{\small\begin{align}
	\mathcal{I}_{\widehat{E_7}[SU(3)]}
	= & \ q^{-1}\mathcal{I}_{\mathcal{N} = 4 \ SU(3)} (b = q, q^4)\nonumber \\
	= & \ \frac{1}{12\pi}\frac{\widehat{\vartheta}_4(\tau)}{\widehat{\vartheta}_1(\tau)}
	\bigg[
	  - \frac{\widehat{\vartheta}'_4(0)}{\widehat{\vartheta}_4(0)}
	  - \frac{\widehat{\vartheta}'_4(\tau)}{\widehat{\vartheta}_4(\tau)}
	  - \frac{i}{\pi} \frac{\widehat{\vartheta}'_4(0)}{\widehat{\vartheta}_4(0)}\frac{\widehat{\vartheta}'_4(\tau)}{\widehat{\vartheta}_4(\tau)}
	  - \frac{i}{\pi} \frac{\widehat{\vartheta}'_4(\tau)^2}{\widehat{\vartheta}_4(\tau)^2}
	- \frac{i}{2\pi} \frac{\widehat{\vartheta}''_4(0)}{\widehat{\vartheta}_4(0)}
	+ \frac{i}{2\pi} \frac{\widehat{\vartheta}''_4(\tau)}{\widehat{\vartheta}_4(\tau)}
	\bigg]\ ,
\end{align}}%
and, with the notation $\widehat{E}_k\big[\substack{\phi \\ \theta}\big] \equiv E_k\big[\substack{\phi \\ \theta}\big](\frac{1}{2}\tau)$, 
\begin{footnotesize}
\begin{equation}
	\mathcal{I}_{\widehat{E_6}[SU(4)]} = \frac{\vartheta_4(3\tau|\frac{1}{2}\tau)}{\vartheta_4(12\tau|\frac{1}{2}\tau)}
	\Bigg(
	- \frac{i}{3} \widehat{E}_3\left[\begin{matrix}
		-1 \\ q^9
	\end{matrix}\right]
	+ \frac{i}{2}\widehat{E}_1\left[\begin{matrix}
		-1 \\ q^3
	\end{matrix}\right]
	\ \widehat{E}_1\left[\begin{matrix}
		-1 \\ q^6
	\end{matrix}\right]
	- \frac{i}{6} \widehat{E}_1\left[\begin{matrix}
		-1 \\ q^3
	\end{matrix}\right]  + \frac{i}{24}\widehat{E}_1\left[\begin{matrix}
		-1 \\ q^3
	\end{matrix}\right]
	+ \frac{i}{24}\widehat{E}_1\left[\begin{matrix}
		-1 \\ q^9
	\end{matrix}\right]
	\Bigg) \ .
\end{equation}
\end{footnotesize}%

\subsection{Modular properties}
Recall that to any four-dimensional $\mathcal N=2$ superconformal field theory one can associate a vertex operator algebra (VOA) \cite{Beem:2013sza}. The Schur index of the four-dimensional theory equals the vacuum character of the chiral algebra. The modular properties of these vacuum characters are of intrinsic interest. By showing that unflavored Schur indices must satisfy a modular differential equation, it was found in \cite{Beem:2017ooy} that the vacuum character of any vertex operator algebra associated with a four-dimensional sueprconformal field theory is an element of a vector-valued (quasi)-modular form. Establishing this fact directly, however, has so far been complicated due to a lack of closed-form expressions. In this paper, we have found precisely such expressions in terms of functions with well-understood modular properties and, even better, for the fully flavored indices. In this subsection we thus study the modular behavior of the Schur index in several examples.

\subsubsection{Small \texorpdfstring{$\mathcal N=4$}{N=4} VOA at \texorpdfstring{$c=-9$}{c=-9}}
Recall that the associated vertex operator algebra of $\mathcal{N} = 4$ super Yang-Mills theory with gauge group $SU(2)$ is the small $\mathcal{N} = 4$ vertex operator (super)algebra at $c=-9$. It contains an $\widehat{\mathfrak{su}}(2)_{k = - \frac{3}{2}}$ affine subalgebra. It is convenient to include the level $k = - \frac{3}{2}$ in the character as follows 
\begin{equation}
\mathcal{I}(\mathfrak{y},\mathfrak{b}) =  \frac{y^k}{2\pi} \frac{\vartheta'_4(\mathfrak{b})}{\vartheta_1(2 \mathfrak{b})}\ ,
\end{equation}
where we introduced a novel fugacity $y = e^{2\pi i \mathfrak{y}}$ and used our result of \eqref{N=4SU2}. Consider the following representation of the $S$ and $T$ modular transformations
\begin{equation}
	(\mathfrak{y} - \frac{\mathfrak{b}^2}{\tau}, \frac{\mathfrak{b}}{\tau}, - \frac{1}{\tau})\xleftarrow{~~S~~}(\mathfrak{y}, \mathfrak{b}, \tau) \xrightarrow{~~T~~} (\mathfrak{y}, \mathfrak{b}, \tau + 1)\ .
\end{equation}
One can easily check that $S^4 = (ST)^6 = 1$. Using the $S$ and $T$ transformation of $\vartheta_i$, it is easy to derive that
\begin{equation}\label{STSN=4}
	\mathcal{I}(\mathfrak{y},\mathfrak{b}) \xrightarrow{STS} \mathcal{I}_{\log}(\mathfrak{y},\mathfrak{b}) \colonequals - \mathfrak{b}\operatorname{ch}_{bc \beta \gamma}(\mathfrak{y},\mathfrak{b})  + (1 - \tau)\, \mathcal{I}(\mathfrak{y},\mathfrak{b})\ ,
\end{equation}
where we denoted
\begin{equation}
	\operatorname{ch}_{bc \beta \gamma}(\mathfrak{y},\mathfrak{b}) \colonequals i y^k\frac{\vartheta_4(\mathfrak{b})}{\vartheta_1(2 \mathfrak{b})}
	= y^k q^{\frac{3}{8}} \frac{(b^{-1}q^{- \frac{1}{2}};q)(b q^{\frac{3}{2}};q)}{(b^{-2};q)(b^2 q;q)} \ .
\end{equation}
As the notation suggests, this is simply a character of a free $(bc \beta \gamma)$-system of weights and $\mathfrak{u}(1)$-charges as follows,
\begin{center}
  \begin{tabular}{c|c|c}
    & $h$ & $m$\\
    \hline
    $(b, c)$ & $(\frac{3}{2}, -\frac{1}{2})$ & $(\frac{1}{2}, - \frac{1}{2})$\\ 
    $(\beta, \gamma)$ & $(1, 0)$ & $(1, - 1)$\\
  \end{tabular}
\end{center}

Note that we applied $STS$ rather than just $S$ as the index is acted on only by $\Gamma^0(2)$ rather than the full $SL(2,\mathbb Z)$.
Various comments are in order:
\begin{enumerate}
\item The transformation property in \eqref{STSN=4} shows that $\mathcal{I}$ is a quasi-Jacobi form \cite{Krauel:2013lra}.
\item The modularly transformed expression $\mathcal{I}_{\log}(\mathfrak{y},\mathfrak{b})$ has a smooth $b \to 1$ limit, unlike $\operatorname{ch}_{bc\beta \gamma}(\mathfrak{y},\mathfrak{b}) $ on its own. In fact, this limit precisely matches the logarithmic solution to the (unflavored) modular differential equations of \cite{Beem:2017ooy}. More precisely, as the logarithmic solution is ambiguous in that one can add an arbitrary multiple of the vacuum character to it, the limit $b \to 1$ reproduces the $STS$-transformation of the unflavored vacuum character: 
\begin{equation}
	\mathcal{I}(\mathfrak{y}) = \frac{y^k}{4\pi} \frac{\vartheta''_4(0)}{\vartheta'_1(0)} \xrightarrow{STS} - \frac{i}{2}y^k\frac{\vartheta_4(0)}{\vartheta_1'(0)}  + (1 - \tau)\, \mathcal{I}(\mathfrak{y})\ .
\end{equation}
\item The logarithmic expression $\mathcal{I}_\text{log}(\mathfrak{b})$ is a solution to all flavored modular differential equations that arise in the null-supermultiplet of the Sugawara condition, as is of course $\mathcal{I}(\mathfrak{y},\mathfrak{b})$ itself.
\item The character $\operatorname{ch}_{bc\beta \gamma}(\mathfrak{y},\mathfrak{b}) $ is proportional to the residue of the integrand of the contour integral defining the Schur index, see \eqref{resN=4}. Moreover, it precisely equals the character of the free fields used in  \cite{Adamovic:2014lra,Bonetti:2018fqz} to construct the vertex operator algebra. 
\item Finally, note that in \cite{Adamovic:2014lra} it was shown that  $\operatorname{ch}_{bc\beta \gamma}(\mathfrak{y},\mathfrak{b}) $ is reducible, and is given by the sum of the vacuum character and the character $\mathcal I_M$ of the (unique) irreducible non-vacuum module $M$ (in category $\mathcal O$):
\begin{equation}
\operatorname{ch}_{bc\beta \gamma}(\mathfrak{y},\mathfrak{b}) = \mathcal{I}(\mathfrak{y},\mathfrak{b}) + \mathcal{I}_{M}(\mathfrak{y},\mathfrak{b})\;.
\end{equation}
\end{enumerate}

\subsubsection{\texorpdfstring{$\mathfrak{so}(8)$}{so(8)} current algebra at \texorpdfstring{$k=-2$}{k=-2}}

Next we look at the modular properties of the vacuum character of $\widehat{\mathfrak{so}}(8)_{-2}$, which is the vertex operator algebra associated with $SU(2)$ superconformal QCD. We use the compact formula (\ref{Ign}) with $g = 0, n = 4$, and we also introduce some additional $y$ variables to define
\begin{align}
	\mathcal{I}_{0,4}(\mathfrak{y}, \mathfrak{b}) \colonequals y_1^{-2}y_2^{-2}y_3^{-2}y_4^{-2} \mathcal{I}_{0,4}(\mathfrak{b}) \colonequals \mathbf{y}^{-2} \mathcal{I}_{0,4}(\mathfrak{b}) \ .
\end{align}
Under the S-transformation, the variables $\mathfrak{y}_i$, $\mathfrak{b}_i$ and $\tau$ transform as
\begin{align}
	(\mathfrak{y}_i, \mathfrak{b}_i, \tau) \xrightarrow{S} (\mathfrak{y}_i - \frac{\mathfrak{b}^2_i}{\tau}, \frac{\mathfrak{b}_i}{\tau}, - \frac{1}{\tau})\ .
\end{align}
The index then transforms as (where the sums and products over $i$ run over $i = 1, \ldots, 4$, and $\alpha_i = \pm 1$, as before)
\begin{equation}
	\mathcal{I}_{0,4}(\mathfrak{y}, \mathfrak{b}) \xrightarrow{S}
	\frac{\log q}{2\pi} \mathcal{I}_{0,4}(\mathfrak{y}, \mathfrak{b})+ \frac{\eta(\tau)^2}{4\pi\prod_{i = 1}^{4}\vartheta_1(2 \mathfrak{b}_i)} \sum_{\vec\alpha = \pm} \left(\prod_i\alpha_i\right) \log(\prod_ib_i^{\alpha_i})E_1\left[
	\begin{matrix}
		1 \\ \prod_ib_i^{\alpha_i}
	\end{matrix}
	\right] \ . 
\end{equation}
This result can be rewritten as
\begin{equation}
	\mathcal{I}_{0,4}(\mathfrak{y}, \mathfrak{b})
	\xrightarrow{S}
	\frac{\log q}{2\pi} \mathcal{I}_{0,4}(\mathfrak{y}, \mathfrak{b}) + \frac{\mathbf{y}^{-2}}{\pi} \sum_{i = 1}^4 (\log m_i) R_i \ ,
\end{equation}
where $R_{j = 1,2,3,4}$ are the residues $R_{j +}$ in (\ref{SQCDresidues}) upon replacing the flavor fugacities there by those associated to the four punctures,
\begin{equation}
	m_1 = b_1 b_2, \qquad m_2 = \frac{b_1}{b_2}, \qquad m_3 =  b_3 b_4, \qquad
	m_4 = \frac{b_3}{b_4} \ .
\end{equation}
Further computing the $S$-transformations of the residues $R_j$, we conclude that $\{\mathcal{I}_{0,4}, R_j\}$ are closed under $S$-transformations
\begin{align}
	\mathcal{I}_{0, 4}\left(\mathfrak{y}, \mathfrak{b}\right)
	\xrightarrow{S} & \ \frac{\log q}{2\pi} \mathcal{I}_{0,4}(\mathfrak{y}, \mathfrak{b})
	  + \frac{\mathbf{y}^{-2}}{\pi}\sum_{j = 1}^{4}(\log m_j )R_j \\
	\mathbf{y}^{-2}R_j \xrightarrow{S} & \ i \mathbf{y}^{-2} R_j \ .
\end{align}
Let's make some comments:
\begin{enumerate}
\item One again observes that the flavored vacuum character transforms as a quasi-Jacobi form.
\item The $S$-transformation takes a similar form to the one we encountered in the $STS$-transformation of $\mathcal{I}_{\mathcal{N} = 4 \ SU(2)}$. What's more, the residues can once again be interpreted as the vacuum character of a system of free fields.\footnote{Also, these free fields can be used to build the vertex operator algebra, but, crucially differently from the $\mathcal N=4$ super Yang-Mills cases considered in detail in \cite{Bonetti:2018fqz}, they are still subjected to a BRST constraint \cite{Peelaers,Eager:2019zrc}.}
\item The four residues $R_j$ can be shown to be linear combinations of the characters of the modules of $\widehat{\mathfrak{so}}(8)_{-2}$. Apart from the vacuum module, it was shown in \cite{Arakawa:2015jya} that there are four nontrivial highest-weight modules. The finite part of their highest weights is given by $\lambda = w(\omega_1 + \omega_3 + \omega_4) - \rho$, where $w = 1, s_{1,3,4}$ are the basic Weyl reflections of $\mathfrak{so}(8)$, and $\omega_i$ are its fundamental weights. Their conformal weights are all equal to $h = - 1$.
\item Finally, and of course a consequence of the previous comment, one can show that all $R_j$ solve the full complement of flavored modular differential equations that follow from flavored null relations and which the vacuum character also satisfies \cite{Beem:2017ooy,Beem:202X}.
\end{enumerate}

\subsection{Defect index from Higgsing}

As a third application of our closed-form expressions, we illustrate the computation of indices of four-dimensional $\mathcal N=2$ superconformal field theories in the presence of BPS surface defects as engineered by the position-dependent Higgsing procedure of \cite{Gaiotto:2012xa}. In this procedure, when applied to theories of class $\mathcal S$ of type $\mathfrak a_1$, one starts with an IR theory $\mathcal{T}_{g, n}$ of genus $g$ with $n \ge 1$ punctures \footnote{In the following we focus on the cases with $n \ge 1$. Cases with $n = 0$ can be similarly computed by taking into account of the potential double poles when taking residues.}, and glues in an additional trinion theory by gauging the diagonal of the two $SU(2)$ flavor symmetries from both sides to obtain the UV theory $\mathcal{T}_{g, n + 1}$. Note that the UV index $\mathcal{I}_{g, n+1}(b)$ depends on one more flavor fugacity, denoted as $b$, than the IR index $\mathcal{I}_{g, n}$. One can give a position-dependent vacuum expectation value to a suitably chosen Higgs branch operator charged under the symmetry measured by $b$. This triggers a renormalization group flow, at the end of which one recovers the original IR theory $\mathcal{T}_{g, n}$ coupled to a surface operator. The vacuum expectation value of the Higgs branch operator depends on an integer $\kappa \ge 0$. At the level of the index the Higgsing operation is implemented by a residue computation \cite{Gaiotto:2012xa,Alday:2013kda,Cordova:2017mhb,Nishinaka:2018zwq}:
\begin{equation}\label{resprescr}
2 (-1)^\kappa q^{-\frac{1}{2}\kappa(\kappa+2)-\frac{1}{2}} \mathop{\operatorname{Res}}_{b \to q^{\frac{\kappa+1}{2}}}\frac{\eta(\tau)^2}{b}\mathcal{I}_{g, n + 1} = \mathcal{I}_{g, n}^{\text{defect}(\kappa)} \ .
\end{equation}
Here we already incorporated the normalization prefactor of \cite{Alday:2013kda}, $R_{(0,\kappa)} = (-1)^\kappa q^{-\frac{1}{2}\kappa(2+\kappa)},$ multiplied with a factor $q^{\frac{c_{\text{UV}}-c_{\text{IR}}}{24}} = q^{-\frac{5}{12}}$ to bridge the gap in central charge between the UV theory and the IR theory, and an extra factor $q^{-1/12}$ that allow us to write $(q,q)_{\infty}^2$ as the square of Dedekind eta functions. In particular, at $\kappa = 0$ the right-hand side is expected to be the original IR index. Armed with the compact expression (\ref{Ign}), we are able to take a closer look at these defect indices.

We first note that $b_i \to q^{\frac{\kappa+1}{2}}$ aren't poles of $\mathcal{I}_{0, 3}$, despite the presence of $\vartheta_1(2 \mathfrak{b}_i)$ in the denominator. This is due to the fact that
\begin{align}
	\sum_{\alpha_i} \left(\prod_{i = 1}^3 \alpha_i\right) E_1\left[\begin{matrix}
		-1 \\ b_1^{\alpha_1}  b_2^{\alpha_2} q^{\frac{\kappa+1}{2}\alpha_{3}}
	\end{matrix}\right] = 0 \ .
\end{align}
Instead, $\mathcal{I}_{0,3}$ has poles when $b_1b_2^{\alpha_2}b_3^{\alpha_3} = q^{\ell + \frac{1}{2}}$ reflecting the poles of the $E_1$'s. These poles are expected from the standard expression $\mathcal{I}_{0,3}(b) = \prod_{\pm\pm}\frac{\eta(\tau)}{\vartheta_4(\mathfrak{b_1} \pm \mathfrak{b}_2 \pm \mathfrak{b}_3)}$.

Besides the trinion index, all other indices with $n \ge 1$ do have simple poles at $b_i = q^{\frac{\kappa+1}{2}}$ arising from the $\vartheta_1$'s in the denominator. Let us consider the residue of $\mathcal{I}_{g, n + 1}(b)$ of the pole at $b \equiv b_{n + 1} \to q^{\frac{\kappa+1}{2}}$ with $\kappa \in \mathbb{N}$,
\begin{align}
	& \ \mathop{\operatorname{Res}}_{b \to q^{\frac{{\kappa+1}}{2}}} \frac{\eta(\tau)^2}{b}\mathcal{I}_{g, n + 1}(b) \nonumber\\
	= & \ (-1)^{{\kappa}} q^{\frac{{(\kappa+1)}^2}{2}} \frac{i^n}{2} \frac{\eta(\tau)^{n + 2g - 2}}{\prod_{i = 1}^{n}\vartheta_1(2 \mathfrak{b}_i)}\sum_{\vec\alpha = \pm} \prod_{i = 1}^{n}\alpha_i
	\sum_{\ell = 1}^{n + 1 + 2g - 2} \lambda_\ell^{(n + 1 + 2 g - 2)}
	E_\ell\left[
	\begin{matrix}
		(-1)^{n + 1} \\ b_1^{\alpha_1}\ldots b_n^{\alpha_n}q^{\frac{{\kappa+1}}{2}}
	\end{matrix}\right] \ .
\end{align}
Here we have performed the sum over $\alpha_{n + 1}$ using the symmetry property of $E_\ell$, and the fact that $\lambda^{(\text{even})}_\text{odd} = \lambda^{(\text{odd})}_\text{even} = 0$. The residue computation makes use of \eqref{theta-function-residue}, which introduces a $ \frac{ - i}{2} (-1)^\kappa q^{\frac{(\kappa + 1)^2}{2}}$ factor: the $ - i$ removes one $i$ from the original $i^{n + 1}$, while the sum over $\alpha_{n + 1}$ cancels the $\frac{1}{2}$.

When $\kappa = 0$, one has
\begin{equation}
	\mathop{\operatorname{Res}}_{b \to q^{\frac{1}{2}}} \frac{\eta(\tau)^2}{b}\mathcal{I}_{g, n + 1}(b)
	= \frac{i^n}{2}q^{\frac{1}{2}} \frac{\eta(\tau)^{n + 2 g - 2}}{\prod_{i = 1}^{n} \vartheta_1(2 \mathfrak{b}_1)}
	\sum_{\vec\alpha = \pm}\prod_{i = 1}^{n}\alpha_i
	  \sum_{\ell = 1}^{n + 1 + 2g - 2}\sum_{\ell' = 0}^\ell
	  \frac{\lambda_\ell^{(n + 1 + 2g - 2)}}{2^{\ell'} \ell'!}
	  E_{\ell - \ell'}\left[\begin{matrix}
	  	(-1)^n\\ \prod_{i=  1}^{n}b_i^{\alpha_i}
	  \end{matrix}\right] \ . 
\end{equation}
Here we applied the half-period shift properties of the Eisenstein series (\ref{Eisenstein-shift}). It turns out that the coefficients $\lambda$ satisfy an equation for any function $f$
\begin{equation}
	\sum_{\ell = 1}^{n + 1 + 2g - 2}\sum_{\substack{\ell' = 0 \\ \ell - \ell' = n \mod 2}}^\ell
	\frac{\lambda_\ell^{(n + 1 + 2g - 2)}}{2^{\ell'} \ell'!}
	f(\ell - \ell') = \frac{1}{2}\sum_{\substack{k = 1}}^{n + 2g - 2} \lambda^{(n + 2g - 2)}_kf(k) \ ,
\end{equation}
and when $(n,k) = (\text{even}, \text{odd})$ or $(\text{odd}, \text{even})$,
\begin{equation}\label{Eisenstein-alternating-sum}
	\sum_{\alpha_i = \pm 1} \left(\prod_{i = 1}^{n} \alpha_i\right) E_k\left[\begin{matrix}
		\pm 1 \\ \prod_{i = 1}^{n} b_i^{\alpha_i}
	\end{matrix}\right] = 0 \ .
\end{equation}
Consequently, the well-known result is recovered,
\begin{equation}
2q^{-\frac{1}{2}}\mathop{\operatorname{Res}}_{b \to q^{\frac{1}{2}}} \frac{\eta(\tau)^2}{b}\mathcal{I}_{g, n + 1}(b) =  \mathcal{I}_{g, n } \ .
\end{equation}

For $\kappa \ge 1$, by using (\ref{Eisenstein-shift}), one instead has
\begin{align}\label{A1-residue-1}
	\mathop{\operatorname{Res}}_{b \to q^{\frac{{\kappa+1}}{2}}} \frac{\eta(\tau)^2}{b}\mathcal{I}_{g, n + 1}(b)
	& \ =(-1)^{{\kappa}}q^{\frac{{(\kappa+1)}^2}{2}}
	  \frac{i^n}{2}\frac{\eta(\tau)^{n + 2g - 2}}{\prod_{i = 1}^{n}\vartheta_1(2 \mathfrak{b}_i)} \\
	  & \ \times \sum_{\vec\alpha = \pm} \Big(\prod_{i = 1}^{n} \alpha_i\Big)
	  \sum_{\ell = 1}^{n + 1 + 2 g - 2}\lambda_\ell^{(n + 1 + 2g - 2)}
	  \sum_{\ell' = 0}^{\ell} \bigg(\frac{{\kappa+1}}{2}\bigg)^{\ell'} \frac{1}{\ell'!}E_{\ell - \ell'}\left[\begin{matrix}
	  	(-1)^{n + {\kappa}} \\ b_1^{\alpha_1} \ldots b_n^{\alpha_n}
	  \end{matrix}\right] \ .\nonumber
\end{align}
After a bit of rewriting, the residues read
\begin{align}\label{A1-residue-2}
	& \ (-1)^{{\kappa}}q^{\frac{{(\kappa+1)}^2}{2}}
	  \frac{i^n}{2}\frac{\eta(\tau)^{n + 2g - 2}}{\prod_{i = 1}^{n}\vartheta_1(2 \mathfrak{b}_i)} 
	\sum_{\vec\alpha = \pm} \Big(\prod_{i = 1}^{n} \alpha_i\Big)
  \sum_{\ell = 1}^{n + 1 + 2 g - 2}
  \tilde \lambda_\ell^{(n + 1 + 2g - 2)}({\kappa+1}) E_\ell\left[\begin{matrix}
  	(-1)^{n + {\kappa}} \\ b_1^{\alpha_1} \ldots b_n^{\alpha_n}
  \end{matrix}\right] \ ,
\end{align}
where\footnote{Curiously, the entries $\tilde \lambda({\mathfrak K} = 2)$ have already appeared in the unflavored indices of $\mathcal{N} = 4$ $SU(N)$ gauge theories.}
\begin{align}
	\tilde \lambda_\ell^{(n + 1 + 2g - 2)}({\mathfrak K}) \colonequals \sum_{\ell' = \max(\ell, 1)}^{n + 1 + 2 g -2}\lambda_{\ell'}^{(n + 1 + 2g - 2)}\bigg(\frac{{\mathfrak K}}{2}\bigg)^{\ell' - \ell} \frac{1}{(\ell' - \ell)!}\ .
\end{align}

Due to (\ref{Eisenstein-alternating-sum}), in the simplest cases when $(g, n) = (0,4 \text{ or } 5)$, the sum over $\ell$ on the right hand side of the residue contains only one term, with only $\ell=1$ and coefficient $\tilde \lambda^{(2)}_1(\mathfrak K) =  \lambda^{(2)}_1 + \frac{{\mathfrak K}}{2}  \lambda^{(2)}_2 = \frac{{\mathfrak K}}{2}$ contributing for $n=4$ and only $\ell=2$ with coefficient $\tilde \lambda^{(3)}_2(\mathfrak K) = \lambda^{(3)}_2 + \frac{{\mathfrak K}}{2}  \lambda^{(3)}_3=\frac{{\mathfrak K}}{2}$ contributing when $n=5$. A simple consequence is that for these two cases
\begin{align}
2 (-1)^\kappa q^{-\frac{1}{2}\kappa(\kappa+2)-\frac{1}{2}}	\mathop{\operatorname{Res}}_{b \to q^{\frac{{\kappa+1}}{2}}} \frac{\eta(\tau)^2}{b}\mathcal{I}_{g, n}(b) = & \ (\kappa +1)\; \mathcal{I}_{g, n-1} \ , && {\kappa} \text{ is even ($n=4$ or $5$) } \ , \\
2 (-1)^\kappa q^{-\frac{1}{2}\kappa(\kappa+2)-\frac{1}{2}}	\mathop{\operatorname{Res}}_{b \to q^{\frac{{\kappa+1}}{2}}} \frac{\eta(\tau)^2}{b}\mathcal{I}_{g, n}(b) = & \ \frac{{\kappa+1}}{2}\;  \widetilde{\mathcal{I}}_{g, n-1} \ , && {\kappa} \text{ is odd  ($n=4$ or $5$) } \ .
\end{align}

Here,
\begin{align}
	\widetilde{\mathcal{I}}_{0,3} = & \  \frac{- i\eta(\tau)}{\prod_{i = 1}^{3}\vartheta_1(2 \mathfrak{b}_i)} \sum_{\vec\alpha = \pm} \left(\prod_{i = 1}^3 \alpha_i\right)E_1\left[\begin{matrix}
		+1 \\ \prod_{i = 1}^{3} b_i^{\alpha_i}
	\end{matrix}\right] \ ,\\
	\widetilde{\mathcal{I}}_{0,4} = & \ \frac{\eta(\tau)^2}{\prod_{i = 1}^{4}\vartheta_1(2 \mathfrak{b}_i)} \sum_{\vec\alpha = \pm} \left(\prod_{i = 1}^4 \alpha_i\right)E_2\left[\begin{matrix}
		-1 \\ \prod_{i = 1}^{4} b_i^{\alpha_i}
	\end{matrix}\right] \ ,
\end{align}
which are simply twice $\mathcal{I}_{0,3}$ ($\mathcal{I}_{0,4}$) with $E_1\Big[\substack{- 1 \\ \#}\Big]$ ($E_2 \Big[  \substack{ + 1\\ \# } \Big]$ ) replaced by $E_1\Big[\substack{ + 1 \\ \#}\Big]$ ($E_2 \Big[  \substack{- 1 \\ \#} \Big]$). 

In \cite{Gaiotto:2012xa} it was shown that the residue prescription \eqref{resprescr} can be equivalently implemented by the action of a difference operator $\mathfrak{S}_{(0,\kappa)}$. See also \cite{Alday:2013kda}.\footnote{We have included the normalizing prefactor of the difference operator already in the definition of the residue \eqref{resprescr}.} For $\kappa= 1$, it reads
\begin{equation}
\mathfrak{S}_{(0,1)} f(a) =  -q^{-\frac{1}{2}} \left(a^2 f(a q^{-\frac{1}{2}}) + a^{-2} f(a q^{\frac{1}{2}})\right)\;.
\end{equation}
Note that $E_2$ enjoys the shift property \eqref{Eisenstein-shift}, and it is thus straightforward to verify that indeed
\begin{equation}\label{shiftopact}
\tilde{\mathcal{I}}_{0,4} = \mathfrak{S}_{(0,1)} \mathcal{I}_{0,4}(b_{1,2,3}, b_4) =  - q^{-\frac{1}{2}}b_4^{2}\ \mathcal{I}_{0,4}(b_{1,2,3}, b_4 q^{ - \frac{1}{2}}) - q^{-\frac{1}{2}} b_4^{-2}\ \mathcal{I}_{0,4}(b_{1,2,3}, b_4 q^{\frac{1}{2}})\ .
\end{equation}
Note that this equality holds at the level of analytic functions. In other words, the replacements $b_4 \to b_4 q^{\pm \frac{1}{2}}$ should be performed before expanding $\mathcal{I}_{0,4}$ in a $q$-series. What's more, the difference operator acts equivalently on any of the other fugacities.

The two terms on the right-hand side of \eqref{shiftopact} admit a nice interpretation from the point of view of the SCFT/VOA correspondence. Recalling that the $\mathfrak{su}(2)$ algebras associated with the punctures are critical, \ie{}, their affine level equals $k=-2$, one can easily convince oneself that these terms equal exactly, including prefactors, the character of the spectrally flowed module by  plus one and minus one unit respectively \cite{Creutzig:2012sd}. Altogether, we see that the defect with $\kappa = 1$ corresponds to the thus-defined twisted module of the associated vertex operator algebra.\footnote{More generally, in theories of class $\mathcal S$ of type $\mathfrak a_1$, defects with even $\kappa$ correspond to untwisted modules of the vertex operator algebra, while those with odd $\kappa$ are associated with twisted modules. }


\section{Discussion}
In this paper we introduced techniques to evaluate the contour integrals defining the Schur limit of the superconformal index of a large class of Lagrangian four-dimensional $\mathcal{N} = 2$ superconformal field theories. Our methods heavily rely on the multivariate ellipticity of their integrand, which, in particular, guarantees that also their residues are elliptic. We have shown that after a single contour integration, ellipticity of the full integrand is lost, which is why we have developed integral formulas to deal with integrals like \eqref{integral-formula-1}. Armed with these formulas, we have evaluated in closed form the fully flavored Schur indices of all theories of class $\mathcal S$ of type $\mathfrak a_1$, we have computed the indices of various low-rank $\mathcal N=4$ super Yang-Mills theories and conjectured general expressions for the unflavored indices of $\mathcal N=4$ super Yang-Mills theories with gauge group $SU(N)$, we have analyzed superconformal QCD theories, and finally studied various applications of our closed-form, analytical expressions. 

While our method is highly successful, it's not omnipotent: the integral formulas introduced in \eqref{integral-formula-1} and \eqref{integral-formula-2} are insufficient to compute the Schur index of any and all Lagrangian theories. For example, to evaluate the Schur index of multi-node linear quivers with gauge group $SU(N)$, $N\geq 3$, we additionally need to be able to compute integrals of the form
\begin{align}
	\oint \frac{dz}{2\pi i z}f(z) E_{k}\left[\begin{matrix}
		\pm 1 \\ z a
	\end{matrix}\right]
	E_{\ell}\left[\begin{matrix}
		\pm 1 \\ z b
	\end{matrix}\right] \ldots
\end{align}
We hope to return to this in the future.

As alluded to in the introduction, the residue of a class of poles of the integrand of the flavored Schur index of four-dimensional $\mathcal{N} = 4$ super Yang-Mills theories with simple, simply-laced gauge groups is equal to the character of the free fields that can be used to realize the vertex operator algebra corresponding to the $\mathcal{N} = 4$ theory \cite{Bonetti:2018fqz}. At the level of the vertex operator algebra itself, the $\mathcal N=4$ algebra can be obtained as a subalgebra of said free field algebra $\mathbb{V}_{bc\beta \gamma}$ -- it is carved out as the kernel of a screening charge. Denoting the projection operator onto this kernel as $P$, we thus have
\begin{equation}
\mathcal{I}_{\mathcal{N} = 4}(b) = \operatorname{str}_{\mathbb{V}_{bc\beta \gamma} } P\, q^{L_0 - \frac{c}{24}} b^f\ .
\end{equation}
Our results, see for example (\ref{N=4SU2}) and (\ref{N4SU3index}), seem to indicate that one can pull the projection operator out of the trace. It would be very interesting to understand this operation better and to generalize it to the $\mathcal N=3$ theories also considered in \cite{Bonetti:2018fqz}. What's more, it is conceivable that our closed-form expressions for the index arise by applying the flavored version of Zhu's recursion formula \cite{zhu1996modular,Mason:2008zzb,Gaberdiel:2008pr,Beem:2017ooy,Beem:202X} to the one-point function of the projection operator $P$.

Some more future lines of inquiry are as follows. Besides evaluating the Schur index itself, one can consider the index of the theory in the presence of various local or non-local operators compatible with the supercharges defining the index. For example, in \cite{Pan:2019bor}, a localization computation is carried out to compute correlation functions of Schur operators, and the results are given once again in terms of contour integrals. The methods developed in this paper easily carry over to this case and allow one to compute Schur correlation functions in closed form (possibly with the help of the more general integral formulas mentioned above).\footnote{As a next step, it may be of interest to revisit the dimensional reduction of these correlators to three dimensions, where they are related to the deformation quantizations of \cite{Beem:2016cbd}. See \cite{Pan:2020cgc,Dedushenko:2019mzv,Dedushenko:2019mnd}. } Furthermore, BPS surface operators engineered by 4d/2d coupled systems \cite{Gadde:2013dda,Cordova:2017mhb}, and BPS line operators can also be inserted \cite{Gang:2012yr}. At least in simple cases, the resulting index can be written as a modified contour integral which can be evaluated with our methods. What's more, these setups have an interesting interpretation in terms of modules when viewed through the lens of the VOA/SCFT correspondence \cite{Cordova:2017mhb,Cordova:2016uwk,Pan:2017zie,Bianchi:2019sxz}. It would therefore be of great interest to compute the closed-form indices of these systems and to analyze their modular properties.

As mentioned above, we have derived a compact formula for the Schur indices of all the theories of class $\mathcal{S}$ of type $\mathfrak a_1$. Similarly, it would be useful to also obtain compact formulas for the flavored indices of $\mathcal{N} = 4$ super Yang-Mills theories and superconformal QCD. Such results would, for example, allow one to analyze their large $N$ behavior and its gravity dual interpretation. See, for example, \cite{Arai:2020qaj,Gaiotto:2021xce} for a recent discussion on the correspondence between the $\mathcal{N} = 4$ $U(N)$ Schur index and the theory of D3 branes wrapping the $S^5$ in $AdS_5 \times S^5$.

Finally, in the recent literature, a ``BAE approach'' to compute the superconformal index of four-dimensional $\mathcal{N} =  1$ and $\mathcal{N} = 4$ theories has been developed. In this approach, the index is written as a sum over solutions to some Bethe-Ansatz-like equations. See for example \cite{Closset:2017bse,Benini:2018mlo,Benini:2021ano,Lezcano:2021qbj}. For $\mathcal{N} = 4$ theories, such Bethe-Ansatz expansion can be interpreted holographically as accounting for the contributions of wrapped D3-branes \cite{Aharony:2021zkr}. While this approach does not seem to be naively applicable to the various limits of enhanced supersymmetry (Macdonald, Hall-Littlewood, Schur) of $\mathcal{N} = 2$ superconformal indices, it could be of interest to compare our results with a carefully performed limit of the ``BAE'' final result.\footnote{An older approach aims to express the fully refined index in terms of vortex partition functions/holomorphic blocks. See, for example, \cite{Peelaers:2014ima,Nieri:2015yia}. It would also be of interest to re-analyze these results armed with our closed-form expressions.}

\section*{Acknowledgments}
The authors would like to thank Chris Beem, Minxin Huang, Yongchao L\"{u}, Carlo Meneghelli, and Yufan Wang for useful discussions. Y.P. is supported by the National Natural Science Foundation of China (NSFC) under Grant No. 11905301, the Fundamental Research Funds for the Central Universities, Sun Yat-sen University under Grant No. 2021qntd27. The work of W.P. was partially supported by grant \#{}494786 from the Simons Foundation.

\appendix

\section{Special Functions}\label{specialfuctions}

The techniques introduced in this paper to evaluate Schur indices in closed form heavily rely on various families of elliptic functions, in particular Jacobi theta functions, the Eisenstein series, and the family of Weierstrass functions. In this appendix we first collect their definitions and basic properties, and then we list several useful identities in the last subsection.

\subsection{The Weierstrass family}

An elliptic function with respect to the complex structure $\tau$ can be viewed as a meromorphic function on $\mathbb{C}$ with double periodicity
\begin{align}
	f(z) = f(z + \tau) = f(z + 1) \ ,
\end{align}
where $\tau \in \mathbb{C}$ with positive imaginary part. One may therefore restrict the domain to be the \emph{fundamental parallelogram} in $\mathbb{C}$ with vertices $0$, $1$, $\tau$, $1 + \tau$. Alternatively, one may view an elliptic function as a meromorphic function on the torus $T^2_\tau$ with complex structure $\tau$. In this appendix and in the main text we often omit the specification of the complex structure $\tau$ in our notations.

One may visualize or construct basic elliptic functions by starting with functions on $\mathbb{C}$ of the form $f(z) \equiv z^{-k}$ and subsequently try to enforce periodicity by summing over all shifts by the periods $1$ and $\tau$, schematically, $P_k(z) \equiv \sum_{m, n} (z - m - n \tau)^{-k}$. After subtracting divergences, one arrives at the following set of (almost) elliptic functions.
\begin{itemize}
	\item The Weierstrass $\zeta$-function is defined by
	\begin{align}\label{weierstrasszetadef}
		\zeta(z) \colonequals \frac{1}{z} + \sum'_{\substack{(m, n) \in \mathbb{Z}^2\\(m, n) \ne (0, 0)}}
		\left[\frac{1}{z - m - n \tau} + \frac{1}{m + n \tau} + \frac{z }{(m + n \tau)^2} \right]\ .
	\end{align}
	In the following and in the main text we will often abbreviate
	\begin{align}
		\sum'_{\substack{(m, n) \in \mathbb{Z}^2\\(m, n) \ne (0, 0)}} \to \ \ \sum'_{m, n} \ , \qquad \sum_{\substack{m \in \mathbb{Z}\\m \ne 0}} \to \sum_m '\ .
	\end{align}
	The $\zeta$ function is not quite elliptic, but instead it satisfies
	\begin{align}\label{shift-formula-zeta}
	  \zeta(z + 1 | \tau) - \zeta(z| \tau) = & \ 2\eta_1(\tau)\\
	  \zeta(z + \tau |\tau) - \zeta(z|\tau) = & \ 2 \eta_2(\tau) \equiv 2\tau \eta_1(\tau) - 2\pi i\ ,
	\end{align}
	where $\eta_1$ and $\eta_2$ are independent of $z$ and are both related to the Eisenstein series $E_2$. We will come back to this in Appendix \ref{app:usefulidentities}. Note that $\zeta$ has a simple pole at each lattice point $m + n \tau$ with unit residue. The fact that $\zeta$ fails to be fully elliptic is tied to the fact that meromorphic functions on $T^2$ with a single simple pole don't exist. In this sense $\zeta(z)$ is the best one can do in terms of double periodicity.

	\item The Weierstrass $\wp$-function
	\begin{align}
		\wp(z) \colonequals & \ \frac{1}{z^2} + \sum_{(m,n) \ne (0,0)} \left[\frac{1}{(z - m - n \tau)^2} - \frac{1}{(m + n \tau)^2}\right] \ .
	\end{align}
	This function is elliptic,
	\begin{align}
		\wp(z) = \wp(z + 1) = \wp(z + \tau) \ .
	\end{align}
	Following from the simple fact that $\partial_z z^{-1} = - z^{-2}$, one has
	\begin{align}
		\wp(z) = - \partial_z \zeta(z)\ .
	\end{align}
	By definition, $\wp$ has only one double pole on $T^2_\tau$.

	\item The descendants $\partial_z^n \wp(z)$ are all elliptic functions, all with a single $n + 2$-th order pole on $T^2_\tau$.
\end{itemize}

\subsection{Jacobi theta functions}

The standard Jacobi theta functions are defined as
\begin{align}
	\vartheta_1(\mathfrak{z}|\tau) \colonequals & \ -i \sum_{r \in \mathbb{Z} + \frac{1}{2}} (-1)^{r-\frac{1}{2}} e^{2\pi i r \mathfrak{z}} q^{\frac{r^2}{2}} ,
	& \vartheta_2(\mathfrak{z}|\tau) \colonequals & \sum_{r \in \mathbb{Z} + \frac{1}{2}} e^{2\pi i r \mathfrak{z}} q^{\frac{r^2}{2}} \ ,\\
	\vartheta_3(\mathfrak{z}|\tau) \colonequals & \ \sum_{n \in \mathbb{Z}} e^{2\pi i n \mathfrak{z}} q^{\frac{n^2}{2}},
	& \vartheta_4(\mathfrak{z}|\tau) \colonequals & \sum_{n \in \mathbb{Z}} (-1)^n e^{2\pi i n \mathfrak{z}} q^{\frac{n^2}{2}} \ .
\end{align}
In the main text and these appendices we will often omit $|\tau$ in the notation. It is well-known that the Jacobi-theta functions can be rewritten as triple product of the $q$-Pochhammer symbol, for example,
\begin{align}\label{theta1-product-formula}
	\vartheta_1(\mathfrak{z}) = - i z^{\frac{1}{2}}q^{\frac{1}{8}}(q;q)(zq;q)(z^{-1};q) \ ,\qquad (z;q) \colonequals \prod_{k = 0}^{+\infty}(1 - zq) \ .
\end{align}

The functions $\vartheta_i(z)$ behave nicely under full-period shifts,
\begin{align}
	\vartheta_{1,2}(\mathfrak{z} + 1) = & - \vartheta_{1,2}(\mathfrak{z}) , & 
	\vartheta_{3,4}(\mathfrak{z} + 1) = & + \vartheta_{3,4}(\mathfrak{z}) , & \\
	\vartheta_{1,4}(\mathfrak{z} + \tau) = & - \lambda \vartheta_{1,4}(\mathfrak{z}), & 
	\vartheta_{2,3}(\mathfrak{z} + \tau) = & + \lambda \vartheta_{2,3}(\mathfrak{z}) , & 
\end{align}
where $\lambda \equiv e^{-2\pi i \mathfrak{z}}e^{- \pi i \tau}$. In particular, one can derive
\begin{align}
	\vartheta_1(\mathfrak{z} + m \tau + n) = (-1)^{m + n} e^{-2\pi i m \mathfrak{z}} q^{ - \frac{1}{2}m^2}\vartheta_1(\mathfrak{z})\ .
\end{align}
Moreover, the four Jacobi theta functions are related by half-period shifts which can be summarized as in the following diagram,
\begin{center}
	\includegraphics[height=100pt]{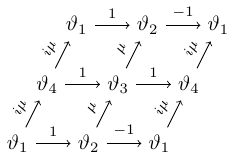}
\end{center}
where $\mu = e^{- \pi i \mathfrak{z}} e^{- \frac{\pi i}{4}}$, and $f \xrightarrow{a} g$ means
\begin{align}
	\text{either}\qquad  f\left(\mathfrak{z} + \frac{1}{2}\right) = a g(\mathfrak{z}) \qquad \text{or} \qquad
	f\left(\mathfrak{z} + \frac{\tau}{2}\right) = a g(\mathfrak{z}) \ ,
\end{align}
depending on whether the arrow is horizontal or (slanted) vertical respectively.

The functions $\vartheta_i(z | \tau)$ transform nicely under the modular $S$ and $T$ transformations, which act, as usual, on the nome and flavor fugacity as $(\frac{\mathfrak{z}}{\tau}, - \frac{1}{\tau})\xleftarrow{~~S~~}(\mathfrak{z}, \tau) \xrightarrow{~~T~~} (\mathfrak{z}, \tau + 1).$ In summary
\begin{center}
	\includegraphics[height=0.2\textheight]{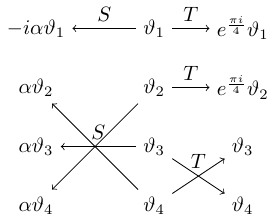}
\end{center}
where $\alpha = \sqrt{-i \tau}e^{\frac{\pi i z^2}{\tau}}$.

The $\tau$-derivative of the Jacobi theta functions is related to the double $z$-derivative as
\begin{align}
	4\pi i \partial_\tau \vartheta_i(z|\tau) = \vartheta''_i(z|\tau)\ .
\end{align}

Finally, we will frequently encounter residues of the $\vartheta$ functions. In particular,
\begin{align}\label{theta-function-residue}
	\mathop{\operatorname{Res}}\limits_{a \to b^{\frac{1}{n}}q^{\frac{k}{n} + \frac{1}{2n}}e^{2\pi i \frac{\ell}{n}}} \frac{1}{a} \frac{1}{\vartheta_4(n\mathfrak{a} - \mathfrak{b})} = & \ \frac{1}{n} \frac{1}{(q;q)^3} (-1)^k q^{\frac{1}{2} k (k + 1)} \ , \\
	\mathop{\operatorname{Res}}\limits_{a \to b^{\frac{1}{n}}q^{\frac{k}{n}}e^{2\pi i \frac{\ell}{n}}} \frac{1}{a} \frac{1}{\vartheta_1(n\mathfrak{a} - \mathfrak{b})} = & \ \frac{1}{n} \frac{i }{\eta(\tau)^3} (-1)^{k + \ell} q^{\frac{1}{2}k^2}\ .
\end{align}
Note that the $(-1)^\ell$ in the second line is related to the presence of a branch point at $z = 0$ according to (\ref{theta1-product-formula}). Let us quickly derive the second formula,
\begin{align}
	\mathop{\operatorname{Res}}\limits_{a \to b^{\frac{1}{n}}q^{\frac{k}{n}}e^{2\pi i \frac{\ell}{n}}} \frac{1}{a} \frac{1}{\vartheta_1(n\mathfrak{a} - \mathfrak{b})} \nonumber
	\colonequals & \ \oint_{b^{\frac{1}{n}}q^{\frac{k}{n}}e^{2\pi i \frac{\ell}{n}}} \frac{da}{2\pi i a} \frac{1}{\vartheta_1(n \mathfrak{a} - \mathfrak{b})} 
	= \oint_{1} \frac{dz}{2\pi i z} \frac{1}{\vartheta_1(n \mathfrak{z} + k \tau + \ell)}\nonumber\\
	= & \ \oint_{1} \frac{dz}{2\pi i z} \frac{(-1)^{k + \ell}z^{nk} q^{\frac{1}{2}k^2}}{\vartheta_1(n \mathfrak{z})}\nonumber\\
	= & \ \oint_{1} \frac{dz}{2\pi i z} \frac{(-1)^{k + \ell}z^{nk} q^{\frac{1}{2}k^2}}{-i q^{\frac{1}{8}} z^{\frac{1}{2}} (q;q) (z^nq;q)(z^{-n}q;q)(1 - z^{-n})} \nonumber\\
	= & \  \frac{1}{n}\frac{i}{\eta(\tau)^3} (-1)^{k + \ell}q^{\frac{k^2}{2}}\ . 
\end{align}
Here we used the shift property of $\vartheta_1$ and \eqref{theta1-product-formula}.

\subsection{Eisenstein series}

The twisted Eisenstein series are defined as
\begin{align}
	E_{k \ge 1}\left[\begin{matrix}
		\phi \\ \theta
	\end{matrix}\right] \colonequals & \ - \frac{B_k(\lambda)}{k!} \\
	& \ + \frac{1}{(k-1)!}\sum_{r \ge 0}' \frac{(r + \lambda)^{k - 1}\theta^{-1} q^{r + \lambda}}{1 - \theta^{-1}q^{r + \lambda}}
	+ \frac{(-1)^k}{(k-1)!}\sum_{r \ge 1} \frac{(r - \lambda)^{k - 1}\theta q^{r - \lambda}}{1 - \theta q^{r - \lambda}} \ ,
\end{align}
where $\phi \equiv e^{2\pi i \lambda}$ with $0 \le \lambda < 1$, $B_k(x)$ denotes the $k$-th Bernoulli polynomial, and the prime in the sum indicates that the $r = 0$ should be omitted when $\phi = \theta = 1$. Additionally, we also define
\begin{align}
	E_0\left[\begin{matrix}
		\phi \\ \theta
	\end{matrix}\right] = -1 \ .
\end{align}
When $k = 2n$ is even, the $\theta = \phi = 1$ limit reproduces the usual Eisenstein series $E_{2n}$, while when $k$ is odd, $\theta = \phi = 1$ is a vanishing limit except for $k = 1$ where it is singular,\footnote{See appendix \ref{app:usefulidentities}.}
\begin{align}
	E_{2n}\left[\begin{matrix}
		+1 \\ +1
	\end{matrix}\right] = E_{2n} \ , \qquad E_1\left[\begin{matrix}
		+ 1 \\ z
	\end{matrix}\right] = \frac{1}{2\pi i }\frac{\vartheta'_1(\mathfrak{z})}{\vartheta_1(\mathfrak{z})}, \qquad
	E_{2n + 1 \ge 3}\left[\begin{matrix}
		+1 \\ +1
	\end{matrix}\right] = 0 \ .
\end{align}
As a result, among all the $E_k\big[\substack{\pm 1 \\ z}\big]$, only $E_1\big[{\substack{\pm 1 \\ z}}\big]$ has a pole at $z = 1$.

A closely related property is the symmetry of the Eisenstein series
\begin{align}\label{Eisenstein-symmetry}
	E_k\left[\begin{matrix}
	  \pm 1 \\ z^{-1}
	\end{matrix}\right] = (-1)^k E_k\left[\begin{matrix}
	  \pm 1 \\ z
	\end{matrix}\right] \ .
\end{align}
The twisted Eisenstein series of neighboring weights are related by
\begin{align}\label{EisensteinDerivative}
	q \partial_q E_k\left[\begin{matrix}
		\phi \\ b
	\end{matrix}
	\right] = (- k) b \partial_b E_{k + 1}\left[\begin{matrix}
		\phi \\ b
	\end{matrix}
	\right]\ .
\end{align}

When shifting the argument $\mathfrak{z}$ of the Eisenstein series by several half or full periods $\tau$, one has for any non-zero $n \in \mathbb{Z}$
\begin{align}\label{Eisenstein-shift}
	E_k\left[\begin{matrix}
		\pm 1\\ z q^{\frac{n}{2}}
	\end{matrix}\right]
	=
	\sum_{\ell = 0}^{k} \left(\frac{n}{2}\right)^\ell \frac{1}{\ell !}
	E_{k - \ell}\left[\begin{matrix}
		(-1)^n(\pm 1) \\ z
	\end{matrix}\right] \ .
\end{align}
To prove these equalities recursively, one can start with the identification (\ref{EisensteinToTheta}) between Eisenstein series and the Jacobi-theta functions, where the periodicity of the latter is clear, and then apply (\ref{EisensteinDerivative}). A similar discussion can also be found in \cite{2012arXiv1209.5628O,Krauel:2013lra}. A natural consequence is that\footnote{In fact, these equalities remain true even after replacing $1$ by $e^{2\pi i \lambda}$ and $- 1$ by $e^{2\pi i (\lambda + \frac{1}{2})}$.}
\begin{align}\label{Eisenstein-shift-1}
	\Delta_k \left[\begin{matrix}
		\pm 1 \\ z
	\end{matrix}\right]
	\equiv E_k\left[\begin{matrix}
		\pm 1 \\ zq^{\frac{1}{2}}
	\end{matrix}\right]
	- E_k\left[\begin{matrix}
		\pm 1 \\ zq^{ - \frac{1}{2}}
	\end{matrix}\right]
	= & \ \sum_{m = 0}^{\floor{\frac{k - 1}{2}}} \frac{1}{2^{2m}(2m+1)!}E_{k - 1 - 2m}\left[\begin{matrix}
		\mp 1\\z
	\end{matrix}\right] \ ,
\end{align}
or more generally
\begin{align}
	E_k\left[\begin{matrix}
		\pm 1 \\ zq^{\frac{1}{2} + n}
	\end{matrix}\right]
	- E_k\left[\begin{matrix}
		\pm 1 \\ zq^{ - \frac{1}{2} - n}
	\end{matrix}\right]
	= & \ 2\sum_{m = 0}^{\floor{\frac{k - 1}{2}}} \left(\frac{2n+1}{2}\right)^{2m + 1}\frac{1}{(2m+1)!}E_{k - 1 - 2m}\left[\begin{matrix}
		\mp 1\\z
	\end{matrix}\right] \ .
\end{align}

The Eisenstein series are often reorganized into twisted Elliptic-$P$ functions, generalizing the Weierstrass $\wp$-family. In particular \cite{Mason:2008zzb},
\begin{align}\label{P1}
	P_{k = 1}\left[\begin{matrix}
		\phi \\ \theta
	\end{matrix}\right](y) \colonequals - \frac{1}{y}\sum_{m \ge 0}E_m\left[\begin{matrix}
		\phi \\ \theta
	\end{matrix}\right] y^m \ ,
\end{align}
while the remaining twisted-$P_k$ with higher $k$ are obtained by taking $y$-derivatives. In particular, we will later use
\begin{align}\label{P2}
	P_2(y) \colonequals - \sum_{n = 1}^{\infty} \frac{1}{2n} E_{2n}(\tau)y^{2n} \ ,
\end{align}
whose derivative reproduces $P_1\big[\substack{+ 1 \\ + 1}\big](y)$ up to a $y^{-1}$ term.

With $P$, the difference equations can be further reorganized into the more compact formula
\begin{align}\label{Delta-Eisenstein}
	\Delta_k \left[\begin{matrix}
		\pm 1 \\ z
	\end{matrix}\right]
	= - 2\oint_0 \frac{dy}{2\pi i} \frac{1}{y^k} \sinh \left(\frac{y}{2}\right) P_1\left[\begin{matrix}
		\mp 1 \\ z
	\end{matrix}\right](y) \ .
\end{align}
where the $y$-contour goes around the origin. Conversely, the individual twisted Eisenstein series can be rewritten in terms of the above differences $\Delta_k$. Let us define $\mathcal{S}_\ell$ by
\begin{align}\label{S2k}
	\frac{1}{2}\frac{y}{\sinh \frac{y}{2}}
	\equiv \sum_{\ell \ge 0} \mathcal{S}_\ell\, y^\ell .
\end{align}
It is straightforward to show that
\begin{align}\label{Eisenstein-from-Delta}
	E_k\left[\begin{matrix}
		\pm 1 \\ z
	\end{matrix}\right]
	= \sum_{\ell = 0}^{k} \mathcal{S}_\ell\, \Delta_{k - \ell + 1}\left[\begin{matrix}
		\mp 1 \\ z
	\end{matrix}\right]\ .
\end{align}

\subsubsection{Constant terms}

The constant terms in $z$ of the Eisenstein series play an important role in the main text when writing down the integration formulas. These numbers are given by
\begin{align}\label{const-terms}
	& \ \text{const. term of }E_{2n + 1}\left[\begin{matrix}
		\pm 1 \\ z
	\end{matrix}\right] = 0\ , \quad \text{ except for } \quad \text{const. term of } E_1\left[\begin{matrix}
		+ 1 \\ z
	\end{matrix}\right] = - \frac{1}{2} \ , \nonumber\\
	& \ \text{const. term of }E_{2n}\left[\begin{matrix}
		+ 1 \\ z
	\end{matrix}\right] = - \frac{B_{2n}}{(2n)!} = - \bigg[\frac{y}{2}\coth \frac{y}{2}\bigg]_{2n} \ , \\
	& \ \text{const. term of }E_{2n}\left[\begin{matrix}
		- 1 \\ z
	\end{matrix}\right] = - \mathcal{S}_{2n} = - \bigg[\frac{y}{2}\frac{1}{\sinh \frac{y}{2}}\bigg]_{2n}\ , \nonumber
\end{align}
and their differences are given by
\begin{align}\label{D2k}
	\mathcal{D}_{2n} \equiv \mathcal{S}_{2n} - \frac{B_{2n}}{(2n)!} = \text{const. term of } \left(E_{2n}\left[\begin{matrix}
		+ 1 \\ z
	\end{matrix}\right]
	- E_{2n}\left[\begin{matrix}
				- 1 \\ z
			\end{matrix}\right]
	\right) = \left[- \frac{y}{2} \tanh \frac{y}{4}\right]_{2n} \ .
\end{align}
In the above, $[f(y)]_k$ denotes the $k$-th coefficient of the Tayler expansion in $y$ around $y = 0$, and $B_{2n}$ are simply the Bernoulli numbers. For the reader's convenience, we collect here the first few values of these numbers,
\begin{center}
	\begin{tabular}{c|c|c|c|c|c|c}
		$n=$  & $1$ & $2$ & $3$ & $4$ & $5$ & $6$\\
		\hline
		$\frac{1}{(2n)!}B_{2n}$ & $\frac{1}{12}$ & $- \frac{1}{720}$ & $\frac{1}{30240}$ & $- \frac{1}{1209600}$ & $\frac{1}{47900160}$ & $ - \frac{691}{1307674368000}$ \\ 
		$\mathcal{S}_{2n}$ & $-\frac{1}{24}$ & $\frac{7}{5760}$ & $- \frac{31}{967680}$ & $\frac{127}{154828800}$ & $- \frac{73}{3503554560}$ & $\frac{1414477}{2678117105664000}$\\
		$\mathcal{D}_{2n}$ & $- \frac{1}{8}$ & $\frac{1}{384}$ & $- \frac{1}{15360}$ & $\frac{17}{10321920}$ & $ - \frac{31}{743178240}$ & $\frac{691}{653996851200}$
	\end{tabular}
\end{center}

\subsection{Useful identities}\label{app:usefulidentities}

The Jacobi theta functions satisfy a collection of \emph{duplication formulas}, for example,
\begin{align}\label{duplication}
	\vartheta_1(2 \mathfrak{z})\vartheta_1'(0)
	= & \ 2\pi\prod_{i = 1}^{4}\vartheta_i(\mathfrak{z})
	= \pi \vartheta_1(2 \mathfrak{z}) \prod_{i = 2}^{4}\vartheta_i(0) \ ,\\
  \vartheta_4(2 \mathfrak{z}) \vartheta_4(0)^3
  = & \ \vartheta_4(\mathfrak{z})^4 - \vartheta_1(\mathfrak{z})^4
  = \vartheta_3(\mathfrak{z})^4 - \vartheta_2(\mathfrak{z})^4 \ .
\end{align}
The $\mathfrak{z} \to 0$ limit of the first line gives the well-known identity $\vartheta'_1(0) = \pi \vartheta_2(0)\vartheta_3(0)\vartheta_4(0)$. The derivatives of $\vartheta_i$ satisfy, among a few other relations,
\begin{align}\label{theta-derivative-formula}
	\frac{d}{d \mathfrak{z}} \left[\frac{\vartheta_1(\mathfrak{z})}{\vartheta_4(\mathfrak{z})}\right] = \vartheta_4(0)^2 \frac{\vartheta_2(\mathfrak{z})\vartheta_3(\mathfrak{z})}{\vartheta_4(\mathfrak{z})^2} \quad \Rightarrow
	\quad
	\frac{\vartheta'_4(\mathfrak{z})}{\vartheta_4(\mathfrak{z})}
	- \frac{\vartheta'_1(\mathfrak{z})}{\vartheta_1(\mathfrak{z})}
	= - \pi \vartheta_4(0)^2 \frac{\vartheta_2(\mathfrak{z})\vartheta_3(\mathfrak{z})}{\vartheta_1(\mathfrak{z})\vartheta_4(\mathfrak{z})}\ .
\end{align}

One can express both the Weierstrass family and the Eisenstein series in terms of the Jacobi theta functions. For example,
\begin{align}\label{zeta-thetap}
	\zeta(\mathfrak{z}) = \frac{\vartheta'_1(\mathfrak{z})}{\vartheta_1(\mathfrak{z})} - 4\pi^2 \mathfrak{z} E_2 \ .
\end{align}
The quasi-periodicity of $\zeta$ now follows and one can express the $\eta_i(\tau)$ in \eqref{shift-formula-zeta} as
\begin{align}
	\eta_1(\tau) = - 2\pi^2 E_2, \qquad \eta_2(\tau) = \tau \eta_1(\tau) - \pi i \ .
\end{align}

The schematic relation between the Eisenstein series and the Jacob-theta functions can be summarized in the diagram
\begin{center}
	\includegraphics[width=0.7\textwidth]{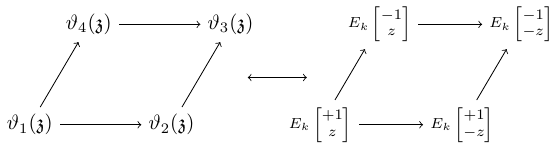}
\end{center}
In more details, the Eisenstein series can be rewritten in terms of ratios of $\vartheta$ functions and their derivatives,
\begin{align}\label{EisensteinToTheta}
	E_k\left[\begin{matrix}
		+ 1 \\ z
	\end{matrix}\right] = - \left[e^{ - \frac{y}{2\pi i }\mathcal{D}_\mathfrak{z} - P_2(y) }\right]_k \vartheta_1(\mathfrak{z})
\end{align}
where $P_2$ is a Weierstrass elliptic-$P$ function (\ref{P2}), $[f(y)]_k$ denotes the $k$-th coefficient of the Taylor series of $f(y)$ around $y=0$, and we define an abstract differential operators $\mathcal{D}_\mathfrak{z}^n$ by
\begin{align}
	\underbrace{\mathcal{D}_\mathfrak{z} \ldots \mathcal{D}_\mathfrak{z}}_{n \text{ copies}} \vartheta_i(\mathfrak{z}) = \mathcal{D}_\mathfrak{z}^n \vartheta_i(\mathfrak{z}) \equiv \frac{\vartheta^{(n)}_i(\mathfrak{z})}{\vartheta_i(\mathfrak{z})} \ .
\end{align}
More explicitly, we have
\begin{align}\label{EisensteinToTheta-2}
	E_k\left[\begin{matrix}
		+ 1 \\ z
	\end{matrix}\right] = \sum_{\ell = 0}^{\floor{k/2}}  \frac{(-1)^{k + 1}}{(k - 2\ell)!}\left(\frac{1}{2\pi i}\right)^{k - 2\ell} \mathbb{E}_{2\ell} \frac{\vartheta_1^{(k - 2\ell)}(\mathfrak{z})}{\vartheta_1(\mathfrak{z})} \ ,
\end{align}
where we define
\begin{align}\label{Ebold}
	& \mathbb{E}_{2} \colonequals E_2, \qquad \mathbb{E}_4 \colonequals E_4 + \frac{1}{2}(E_2)^2, \qquad
	\mathbb{E}_6 \colonequals E_6 + \frac{3}{4}E_4E_2 + \frac{1}{8}(E_2)^3 \ , \qquad \ldots\\
	& \mathbb{E}_{2\ell} \colonequals \sum_{\substack{\{n_p\} \\ \sum_{p \ge 1} (2p)n_p = 2\ell}} \prod_{p\ge 1} \frac{1}{n_p !} \left(\frac{1}{2p}E_{2p}\right)^{n_p}\ .
\end{align}
The conversion from $E_k\left[\substack{- 1 \\ \pm z}\right]$ can be obtained by replacing $\vartheta_1$ with $\vartheta_{2,3,4}$ according to the previous diagram. One can show these relations by observing that both sides satisfy the same difference equations. (Those of the Eisenstein series have been discussed in the previous subsection.) For the reader's convenience we list the first few conversions here.
\begin{align}\label{Ek-thetap}
	E_1\left[\begin{matrix}
		+1 \\ z
	\end{matrix}
	\right] = & \ \frac{1}{2\pi i} \frac{\vartheta'_1(\mathfrak{z})}{\vartheta_1(\mathfrak{z})}\ ,  \\
	E_2\left[\begin{matrix}
		+1 \\ z
	\end{matrix}
	\right] = & \ \frac{1}{8\pi^2}\frac{\vartheta_1''(\mathfrak{z})}{\vartheta_1(\mathfrak{z})} - \frac{1}{2} E_2 \ , \\
	E_3\left[\begin{matrix}
		+1 \\ z
	\end{matrix}
	\right] = & \ \frac{i}{48\pi^3} \frac{\vartheta'''_1(\mathfrak{z})}{\vartheta_1(\mathfrak{z})}
	  - \frac{i}{4\pi}\frac{\vartheta'_1(\mathfrak{z})}{\vartheta_1(\mathfrak{z})} E_2,  \\
	E_4\left[\begin{matrix}
		+1 \\ z
	\end{matrix}\right] = & \ - \frac{1}{384\pi^4} \frac{\vartheta''''_1(\mathfrak{z})}{\vartheta_1(\mathfrak{z})} + \frac{1}{16\pi^2}E_2 \frac{\vartheta''_1(\mathfrak{z})}{\vartheta_1(\mathfrak{z})} - \frac{1}{4} \left(E_4 + \frac{1}{2}(E_2)^2\right) \\
	E_5\left[\begin{matrix}
		+1 \\ z
	\end{matrix}\right]
	= & \ - \frac{i}{3840 \pi^5} \frac{\vartheta^{(5)}_1(\mathfrak{z})}{\vartheta_1(\mathfrak{z})} + \frac{i}{96\pi^3}E_2 \frac{\vartheta_1^{(3)}(\mathfrak{z})}{\vartheta_1(\mathfrak{z})} - \frac{i}{8\pi}\left(E_4 + \frac{1}{2}(E_2)^2\right)\frac{\vartheta_1'(\mathfrak{z})}{\vartheta_1(\mathfrak{z})} \\
	E_6\left[\begin{matrix}
		+ 1 \\ z
	\end{matrix}\right]
	= & \ \frac{1}{46080\pi^6} \frac{\vartheta^{(6)}_1(\mathfrak{z})}{\vartheta_1(\mathfrak{z})} - \frac{1}{768\pi^4}E_2 \frac{\vartheta_1^{(4)}(\mathfrak{z})}{\vartheta_1(\mathfrak{z})} + \frac{1}{32\pi^2} \left(E_4 + \frac{1}{2}(E_2)^2\right) \frac{\vartheta_1^{(2)}(\mathfrak{z})}{\vartheta_1(\mathfrak{z})} \nonumber\\
	& \ - \frac{1}{6}\left(E_6 + \frac{3}{4}E_4 E_2 + \frac{1}{8}E_2^3\right) \ .
\end{align}

From the above conversion one computes the residues of Eisenstein series,
\begin{align}
	\mathop{\operatorname{Res}}_{z \to 1}\frac{1}{z}E_k\left[\begin{matrix}
		+ 1 \\ z
	\end{matrix}\right] = \delta_{k1} \ ,
	\qquad
	\mathop{\operatorname{Res}}_{z \to q^{\frac{1}{2} + n}}\frac{1}{z}E_k\left[\begin{matrix}
		- 1 \\ z
	\end{matrix}\right] = \frac{1}{2^{k - 1} (k - 1)!} \ .
\end{align}

Moreover, the Eisenstein series satisfy the following relations
\begin{align}\label{duplication-Eisenstein}
	\sum_{\pm}E_k\left[\begin{matrix}
		\phi \\ \pm z
	\end{matrix}\right](\tau) = & \ 2 E_k\left[\begin{matrix}
		\phi \\ z^2
	\end{matrix}\right](2\tau) \ , \nonumber \\
	\sum_{\pm} \pm E_k\left[\begin{matrix}
		\phi \\ \pm z
	\end{matrix}\right](\tau)
	= & \ -2 E_k\left[\begin{matrix}
		\phi \\ z^2
	\end{matrix}\right](2\tau)
	 + 2 E_k\left[\begin{matrix}
	 	\phi \\ z
	 \end{matrix}\right](\tau)\ , \nonumber
	\\
	E_k\left[\begin{matrix}
		+ 1\\z
	\end{matrix}\right](2\tau)
	+ E_k\left[\begin{matrix}
		- 1\\z
	\end{matrix}\right](2\tau) = & \ 
	\frac{2}{2^k}E_k\left[\begin{matrix}
		+ 1 \\ z
	\end{matrix}\right] \ ,\\
	E_k\left[\begin{matrix}
		+ 1\\z
	\end{matrix}\right](2\tau)
	- E_k\left[\begin{matrix}
		- 1\\z
	\end{matrix}\right](2\tau) = & \
	- \frac{2}{2^k}E_k\left[\begin{matrix}
		+ 1 \\ z
	\end{matrix}\right](\tau)
	+ 2 E_k\left[\begin{matrix}
		+ 1 \\ z
	\end{matrix}\right](2\tau)\ , \nonumber
	\\
	\sum_{\pm \pm} E_k\left[\begin{matrix}
		\pm 1 \\ \pm z
	\end{matrix}\right](\tau) = & \ \frac{4}{2^k}E_k\left[
	\begin{matrix}
		+ 1 \\ z^2
	\end{matrix}\right](\tau)\ . \nonumber
\end{align}
Applying the shift $z \to z q^{\frac{1}{2}}$, one can also generate similar formulas with $E_1\Big[\substack{ \pm 1 \\ *}\Big] \rightarrow E_1\Big[\substack{ \mp 1 \\ *}\Big]$. These formulas are generalizations of the duplication formulas, for instance, the last identity at $k = 1$ reduces to the duplication formula (\ref{duplication}). Combining the duplication formulas and (\ref{theta-derivative-formula}), one finds the useful identity
\begin{align}\label{Eisenstein-identity-1}
	E_1\left[\begin{matrix}
		+ 1 \\ z
	\end{matrix}\right]
	- E_1\left[\begin{matrix}
		- 1 \\ z
	\end{matrix}\right]
	= \frac{\eta(\tau)^3}{2i} \frac{\vartheta_1(2 \mathfrak{z})\vartheta_4(0)^2}{\vartheta_1(\mathfrak{z})^2 \vartheta_4(\mathfrak{z})^2}\ .
\end{align}


\section{Fourier series}\label{app:fourier}

In this paper, the Schur index of a Lagrangian theory is computed by directly evaluating the contour multi-integral of a multivariate elliptic function, one integral after another. Expanding the elliptic function in terms of a sum of $\zeta$-functions allows us to perform the first integral, but unfortunately, the result is in general non-elliptic with respect to the remaining integration variables due to the presence of the Eisenstein series $E_1$.

Luckily, in all cases that we will be dealing with, albeit lacking ellipticity, each summand in the result is always a product of an elliptic function (with respect to the remaining integration variables) -- the residues $R_i$ --, and some Eisenstein series. A powerful tool to compute integrals of such almost elliptic functions is the Fourier series of the Eisenstein series.

Let us start by defining the Fourier series
\begin{equation}
	S_k(\mathfrak{z}) \colonequals \sum_{n}' \frac{1}{\sin ^k n \pi \tau} e^{2\pi i n \mathfrak{z}} \ , \quad \text{for } k \in \mathbb{N}_{\ge 1}, \qquad
	\text{and}, \qquad S_0(\mathfrak{z}) = -1 \ .
\end{equation}
This series is a Taylor series in $q$ provided the imaginary part of $\mathfrak{z}$ is not too large. Concretely, let $\mathfrak{z} = \mathfrak{z}_\mathbb{R} + \lambda \tau$, then the summand reads
\begin{equation}
	\frac{1}{\sin^k n \pi \tau}q^{ n \lambda} \propto \frac{q^{n\lambda}}{(q^{\frac{n}{2}} - q^{- \frac{n}{2}})^k}
	= \left\{ \begin{gathered}
  \frac{q^{(\lambda + \frac{k}{2})n}}{(1 - q^{+ n})^k} \sim q^{(\lambda + \frac{k}{2})n} (1 + q^{+n} + \ldots)^k , \qquad n > 0 \hfill \\
  \frac{q^{(\lambda - \frac{k}{2})n}}{(1 - q^{- n})^k} \sim q^{(\lambda - \frac{k}{2})n} (1 + q^{-n} + \ldots)^k, \qquad n < 0 \hfill \\ 
\end{gathered}  \right. \ .
\end{equation}
When $- \frac{k}{2}\le \lambda \le \frac{k}{2}$, $S_k$ can be expanded in non-negative poweres in $q$. In the following and in the main text, we will always operate under the assumption that the entire argument of $S_k$ sits well within this range.

The simplest Eisenstein series can be Fourier expanded as
\begin{align}
	E_1\left[\begin{matrix}
		-1 \\ z
	\end{matrix}\right] = & \ \frac{1}{2i} S_1(\mathfrak{z}), 
	& E_1\left[\begin{matrix}
		+1 \\ z
	\end{matrix}\right]  = & \ - \frac{1}{2} + \frac{1}{2i} S_1(\mathfrak{z} - \frac{\tau}{2})\ ,\\
	E_2\left[\begin{matrix}
	  -1 \\ z
	\end{matrix}\right]
	= & \ - \frac{1}{4} S_2(\mathfrak{z} + \frac{\tau}{2}) + \frac{i}{4}S_1(\mathfrak{z})
	  + \frac{1}{24}
	   \ ,
	& E_2\left[\begin{matrix}
	  +1 \\ z
	\end{matrix}\right]
	= & \  - \frac{1}{4} S_2(\mathfrak{z})
	  - \frac{1}{12} \ .
\end{align}
The first line can be seen by first translating the $E_1$ to Jacobi-theta functions and applying the well-known Fourier expansion of $\vartheta'_i(\mathfrak{z})/\vartheta_i(\mathfrak{z})$ \cite{Wittaker}. The second line follows by analyzing the $\tau$-derivative of the first. From these results, the Fourier expansion of $\zeta(\mathfrak{z})$ can also be determined. Recall that both $\zeta$ and $E_1$ are related to Jacobi theta functions by \eqref{zeta-thetap}, \eqref{Ek-thetap}, and therefore when $\zeta = \zeta_\mathbb{R} + \lambda \tau$ with $\lambda \in [0, 1)$
\begin{equation}\label{zetaFourier}
	\zeta(\mathfrak{z}) = 2\pi i E_1\left[\begin{matrix}
		+ 1 \\ z
	\end{matrix}\right] - 4\pi^2 \mathfrak{z} E_2 = \pi S_1(\mathfrak{z} - \frac{\tau}{2}) - \pi i - 4\pi^2 \mathfrak{z} E_2 \ .
\end{equation}

To obtain the Fourier expansions for higher Eisenstein series, we make the following Ansatz,
\begin{equation}
	E_{2n}\left[\begin{matrix}
		+1 \\ z
	\end{matrix}
	\right] = \sum_{m = 0}^n c_{2n}(2m)S_{2m}(\mathfrak{z}) \ ,\qquad
	E_{2n + 1}\left[\begin{matrix}
		-1 \\ z
	\end{matrix}\right] = \sum_{m = 0}^n c_{2n+1}(2m+1)S_{2m + 1}(\mathfrak{z}) \ .
\end{equation}
In particular, the known Fourier series for low weight Eisenstein series then imply
\begin{equation}
	c_0(0) = 1, \qquad
	c_1(1) = \frac{1}{2i} \ ,\qquad
	c_2(2) = - \frac{1}{4}, \qquad
	c_2(0) = + \frac{1}{12} \ .
\end{equation}

These data initiate a recursion for the coefficients. Concretely, Using (\ref{Eisenstein-shift-1}), we see that
\begin{align}
	\sum_{m = 0}^n c_{2n + 1}(2m + 1)2i\ S_{2m}(\mathfrak{z})
	= & \ \sum_{\ell = 0}^n \frac{1}{2^{2\ell}(2\ell + 1)!} E_{2n - 2\ell}\left[\begin{matrix}
		-1 \\ z
	\end{matrix}\right] \nonumber\\
	= & \ \sum_{\ell = 0}^n \frac{1}{2^{2\ell}(2\ell + 1)!} \sum_{m = 0}^{n - \ell}c_{2n - 2\ell}(2m)\ S_{2m}(\mathfrak{z})\\
	= & \ \sum_{m = 0}^n\sum_{\ell = 0}^{n - m} \frac{1}{2^{2\ell}(2\ell + 1)!} c_{2n - 2\ell}(2m)\ S_{2m}(\mathfrak{z})\ . \nonumber
\end{align}
Such analysis provides recursion relations for $c$'s,
\begin{align}
	2i\, c_{2n + 1}(2m + 1) = & \ \sum_{\ell = 0}^{n - m} \frac{1}{2^{2\ell}(2\ell+1)!}c_{2n - 2\ell}(2m) \ , & \ m \in \mathbb{N}\ , \\
	2i\, c_{2n + 2}(2m + 2) = & \ \sum_{\ell = 0}^{n - m} \frac{1}{2^{2\ell}(2\ell+1)!}c_{2n + 1 - 2\ell}(2m + 1) \ , & \ m \in \mathbb{N}\ ,
\end{align}
with $c_{2n + 2}(0)$ undetermined since the constant term in $E_{2n + 2}$ does not contribute to the difference equation. However, the constant terms of the twisted Eisenstein series $E_{2n}$ are of course well-known in terms of the Bernoulli numbers, and thus (note the minus sign in $S_{0}(\mathfrak{z}) = -1$)
\begin{align}
	c_{2n}(0) = (- 1) \text{ const. term of } E_{2n}\left[\begin{matrix}
		+ 1 \\ z
	\end{matrix}\right] = + \frac{B_{2n}}{(2n)!} \ .
\end{align}

\clearpage

{
\bibliographystyle{utphys}
\bibliography{ref}
}

\end{document}